# Inverse-IMPRESSION: A Graph-based Platform for Molecular Structure Elucidation from Experimental NMR Spectroscopic Properties

Zheqi Jin[a], Grace Armitage[a], Richard Cox[a], Ben Honoré[a], Mohammad Golbabaee[b] and Craig Butts[a,*]

[a]School of Chemistry, University of Bristol, Cantock's Close, Bristol, BS8 1TS, U.K.

[b]School of Engineering Mathematics and Technology, University of Bristol, Ada Lovelace Building, Tankard's Cl, Bristol BS8 1TW

*Correspondence to: Craig.Butts@bristol.ac.uk

**Abstract:** Here, we present a platform built on our inverted Graph Transformer Network, IMPRESSION-G2, which can accurately and rapidly reconstruct molecular bonding directly from experimental nuclear magnetic resonance (NMR) spectroscopic information. It comprises three interconnected stages: a one-shot model that predicts bond connectivity between atoms; a structure-correction stage that corrects the predicted structures by removing uncertain bonds and iteratively reassigning them; noise-augmented multi-shot prediction, generating an ensemble of candidate structures, which are ranked to identify the best-fit structure. By integrating a range of $^{1}$H and $^{13}$C NMR data, including two-dimensional (2D) experiments such as COSY, HSQC, and HMBC, the inverse-IMPRESSION platform correctly identifies the structures of 77.8% of molecules with up to 30 heavy atoms (H, C, N, O and F) using simulated NMR data, and 10 of 19 (53%) molecules using experimental NMR data. The experimental structures solved have molecular weights of up to 480 Da and are representative of the complex structures in synthetic and natural products that routinely challenge chemists. The inverse-IMPRESSION framework thus provides the first effective approach for automated molecular structure elucidation using graph-based machine learning on experimental data.

## Introduction

Accurate and rapid molecular structure elucidation is a fundamental task across fields such as organic synthesis, materials science, and drug discovery. Among available analytical techniques, nuclear magnetic resonance (NMR) spectroscopy plays a central role due to its ability to provide detailed information on local chemical environments, including atomic connectivity, functional groups, and stereochemistry. However, interpreting NMR spectra remains a challenging and time-consuming process, with the primary bottleneck arising from the combinatorial explosion of possible molecular structures. As molecular size increases, the number of feasible candidate structures grows exponentially, rendering exhaustive enumeration and manual selection impractical on the order of $10^{60}$ molecules.[1] Therefore, computer-assisted approaches are essential to systematically manage this combinatorial complexity and enable practical and scalable molecular structure elucidation.

Traditional Computer-Assisted Structure Elucidation (CASE) systems[2-5] that determine molecular structure elucidation from spectroscopic data, especially NMR data, typically rely on logical-combinatorial algorithms to derive molecular fragments from experimental spectra and assemble them into candidate structures, which are subsequently ranked based on their agreement with the input data. Modern CASE systems have incorporated density functional theory (DFT),[6] statistical tools[7-9] and genetic algorithms,[10] to improve the accuracy and efficiency of structure generation and ranking. However, the reliance on exhaustive structure generation and filtering often results in the ranking of highly similar structures at high computational cost, limiting full automation for complex molecules.

Recent advances in artificial intelligence (AI) and machine learning have driven the emergence of data-driven approaches for molecular structure elucidation. These approaches are at an early stage and require large training datasets to effectively capture structure-spectrum relationships, however they have the potential to enable fully automated and substantially faster predictions compared to traditional rule-based methods. Current data-driven methods of this type can be broadly categorised into large language model (LLM)-based methods[11-16] and graph-based methods.[17-20]

Graph-based methods represent molecules explicitly as graphs, in which atoms correspond to nodes and bonds or atom pairs correspond to edges. This representation enables atom- and pair-specific information, such as chemical shifts and coupling constants, to be naturally incorporated as node and edge features, providing a flexible and chemically intuitive framework for representing both molecular structure and spectroscopic data. Graph neural networks (GNNs) iteratively aggregate information from neighbouring nodes and edges to update the molecular graph, enabling the property prediction in structure elucidation tasks, including bond orders and interatomic distances. Several representative efforts highlight both the promise and limitations of current graph-based approaches. Sridharan et al.[19] proposed a stepwise strategy combining Monte Carlo Tree Search (MCTS) with GNNs but largely restricted to small molecules containing fewer than ten heavy atoms. More recently, Yang et al.[20] integrated diffusion models with GNNs to iteratively refine molecular bonding patterns, conditioning the generation process on embedded NMR spectra. This approach achieved a Top-1 accuracy of 62% for simulated NMR data, but the performance decreased substantially to 11% for experimental data comprising molecules with up to 25 heavy atoms. We recently reported a proof-of-concept graph-based approach[21] using an inverted implementation of the IMPRESSION-G2 machine learning architecture, which demonstrated that graph-based methods can achieve near 100% success if idealised simulated NMR data are available, that drops to 33% with ablated data. The IMPRESSION-G2 model[22] was originally developed to predict chemical shifts and scalar coupling constants from 3D molecular structures represented by interatomic distances. In its inverted form,[21] IMPRESSION-G2 predicted interatomic distances (and hence molecular structure) directly from NMR data. This provides important evidence that molecular structure elucidation can be formulated as an edge-prediction task on molecular graphs and demonstrates that, in the hypothetical limit of complete NMR information, structures can be reconstructed reliably.

On the other hand, LLM-based approaches typically depend on Transformer encoder-decoder architectures and represent molecules using SMILES (Simplified Molecular-Input Line-Entry System) style strings, with NMR data encoded as text sequences. This compact representation facilitates efficient storage and enables training on large collections of molecular and spectroscopic data which can overcome the lack of the more task-relevant molecular graph representation. While it is challenging to directly compare performance of models due to variability of methods, representations and training and test datasets reported in the literature, the reported performances of these LLM models are broadly comparable to those achieved by GNN-based methods. For example, the recently reported NMRMind system[13] achieved 92% accuracy on "comprehensive" simulated 1D and 2D NMR data and 71% on ablated NMR data. On a set of 12 experimental molecules, it achieved Top-1 accuracy of 33%, with the largest solved molecule having a mass of 399.5 Da. The recently reported NMR-Solver[14] achieved a Top-1 accuracy of 60% on experimentally synthesised molecules, however it incorporated prior knowledge of reactant structures as additional constraints which thus explores a substantially reduced chemical space relative to methods that rely solely on spectroscopic data.

Herein, we present an alternative framework for our inverse-IMPRESSION platform in which edge prediction can be expressed as prediction of bond probabilities, i.e., the likelihood that two atoms are bonded given a set of NMR data. This simplified expression more directly maps on the skeletal molecular structure prediction task, where the primary objective is to infer atomic connectivity, without explicitly resolving 3D features such as stereochemistry or conformation. This inverse-IMPRESSION platform comprises interconnected models based on the IMPRESSION-G2 architecture, each employing distinct node and edge features and responsible for a specific component of the structure elucidation process. A noise-augmentation multi-shot strategy is applied to generate multiple candidate structures for each prediction event, which are subsequently ranked to identify the best-fit structure. Together, this framework operates as a unified platform to produce 2D molecular structures with assigned bond orders and demonstrates reliability and robustness across both simulated and experimental NMR datasets. Notably, it successfully predicts complex organic molecules including experimental examples that represent real-world challenging tasks for expert chemists.

## Methods

### Datasets

A dataset comprising 122,000 pharmaceutically relevant molecules containing H, C, N, O, and F atoms, each with no more than 30 non-hydrogen atoms, was derived from the public PubChem[23] (72,998 molecules) and ChEMBL[24] (49,002 molecules) databases. For each molecule, 100 conformers were generated and optimised using the Merck Molecular Force Field[25] (MMFF) as implemented in RDKit,[26] and their corresponding MMFF energies were computed. NMR parameters, including chemical shifts and scalar coupling constants, were subsequently predicted for each conformer using the IMPRESSION-G2 model,[22] which provides DFT-level accuracy at substantially lower computational cost. The final molecular NMR parameters were obtained as Boltzmann-weighted averages over all conformers using their MMFF energies (Supplementary Fig. 3).

Molecular structures were encoded as fully-connected graphs (Supplementary Fig. 2), with atomic numbers assigned as node features and atom pair labels, e.g. H-C or C-O, assigned as edge features. Spectroscopic properties were incorporated by assigning chemical shifts as node features and scalar coupling constants or "correlations" as edge features. Coupling correlations were assigned a value of 0 when the corresponding scalar coupling constants were below 2 Hz, as such couplings would give rise to very weak or unobservable NMR correlation peaks under typical experimental conditions. The remaining coupling constants were converted into categorical labels representing correlations (peaks) arising from different 2D NMR experiments: 1 for $^{1}$H-$^{1}$H COSY, 2 for $^{1}$H-$^{13}$C HSQC and 3 for $^{1}$H-$^{13}$C HMBC.

This dataset was split into a training set of 120,000 molecules (4,974,382 nodes and 107,317,458 edges) and a test set of 2,000 molecules (86,645 nodes and 1,923,615 edges). To distinguish these datasets from the fragment training subset, they are referred to as the full-molecule training set and full-molecule test set, respectively. From the full-molecule training set, 60,000 molecules containing at least two oxygen atoms were randomly sampled and used to construct a fragment training subset for the stepwise model. For each selected molecule, two molecular fragments were generated using a breadth-first search (BFS) algorithm, yielding a total of 120,000 molecular fragments for training (see Supplementary Section 1.3.2 for details).

**NMR Datasets**

Two sets of NMR parameters were constructed to train the one-shot and stepwise models in the inverse-IMPRESSION platform, each mimicking a different experimental NMR scenario:

- NMR Dataset A represents an idealised control scenario where NMR chemical shifts and scalar coupling constants (rather than the coupling correlations used in NMR Dataset B) are available for all nuclei, including those that are not experimentally observable or routinely measurable. Although unrealistic, this setting provides an estimate of the upper bound of reconstruction accuracy.

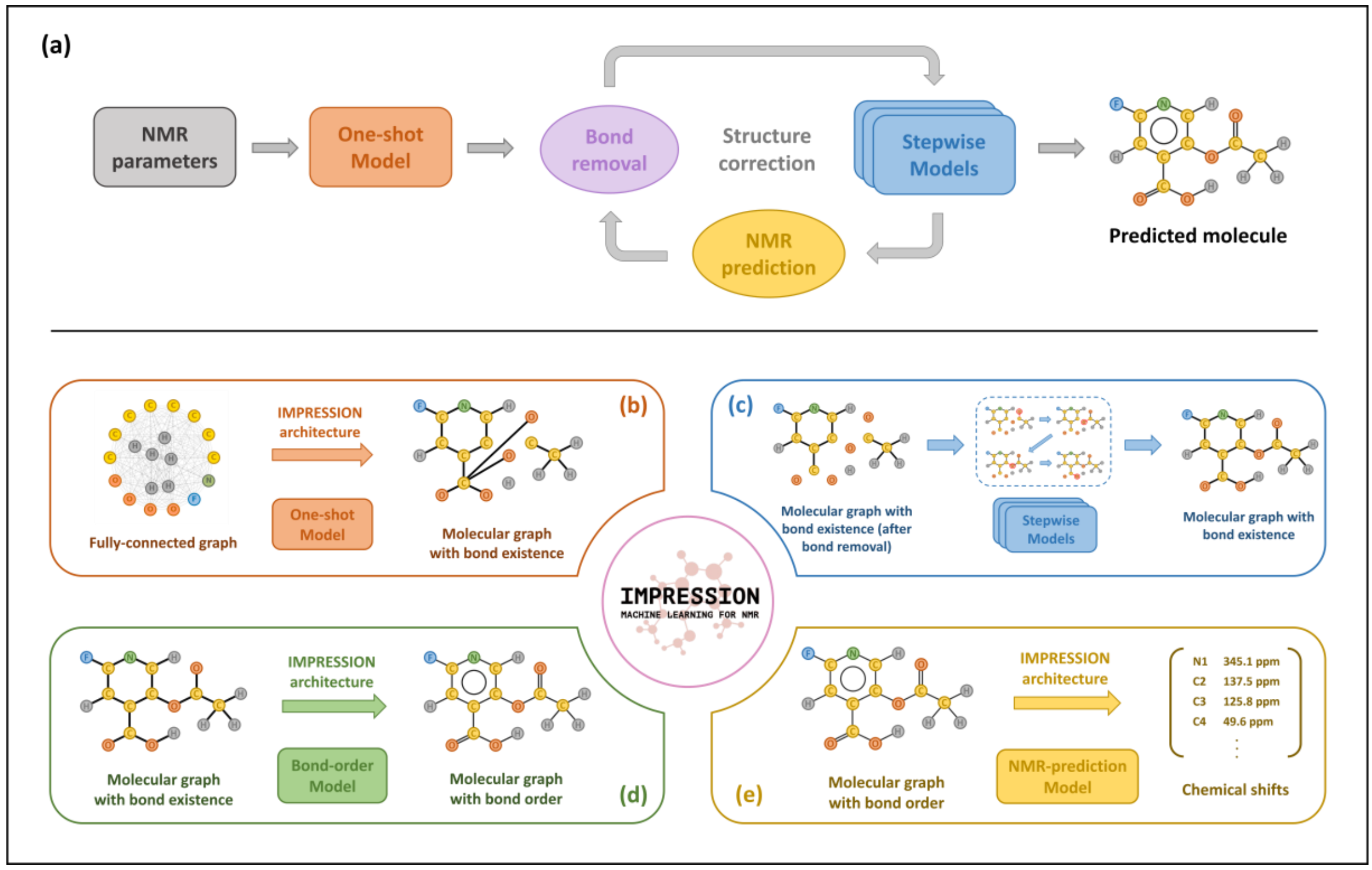

**Fig. 1 | Structure elucidation using the inverse-IMPRESSION framework and the roles of its four component models. a,** Overview of the structure elucidation workflow. **b,** One-shot model, which predicts bond connectivity between atoms simultaneously. **c,** Stepwise model, which iteratively assigns bonds to partially constructed structures. **d,** Bond-order model, which assigns bond orders to structures generated by the one-shot or stepwise models. **e,** NMR-prediction model, which predicts chemical shifts for the resulting structures and identifies atoms with incorrect bonding assignments.

- NMR Dataset B represents an experimentally relevant scenario in which only the most commonly used NMR experiments are available. It includes chemical shifts for $^{1}$H and $^{13}$C nuclei and scalar coupling correlations for H-H and C-H atom pairs, corresponding to COSY, HSQC, and HMBC spectra.

## Models in the inverse-IMPRESSION platform

The inverse-IMPRESSION platform (Fig. 1a) comprises four Graph Transformer Network (GTN) models, which share the same neural network architecture (Supplementary Section 1.2.1 and 1.2.3) as the previously reported IMPRESSION-G2 model.[22] Each model was trained separately using task-specific node and edge features (details below). Full details on training methodology can be found in Supplementary Section 1.2.4-1.2.6.

### One-shot Model

The one-shot model (Fig. 1b) predicts all bonds within a molecule simultaneously from NMR spectroscopic inputs. Trained on the full-molecule training set, it employs atomic numbers and chemical shifts concatenated as node features, and scalar coupling correlations or constants as edge features. These input features are processed through the message-passing scheme of the IMPRESSION graph transformer-based architecture, followed by linear layers applied to each edge and a final sigmoid activation layer, producing a graph in which the updated edge features represent the probabilities of bond existence between atom pairs. A threshold of 0.5 is applied to convert these probabilities into binary bond predictions to balance assignment between presence and absence of bonds.

**Stepwise Model**

The stepwise model (Fig. 1c) performs iterative bond prediction on molecular fragments generated from earlier-stage outputs. These fragments are obtained either by removing bonds to all heteroatoms following the one-shot model, or by removing bonds around carbon atoms whose predicted chemical shifts (see 'NMR-Prediction Model' section below) show large deviations from the experimental input data. At each iteration, the model focuses on a sub-valent atom and predicts bond probabilities to all other atoms, thereby identifying the most likely attachment atom within the current fragment(s). Bonds are then added sequentially to the focussed atom until no further suitable attachment is found, after which the model proceeds to the next sub-valent atom. This process is repeated until all sub-valent atoms have been considered, which is equivalent to the generation process of the fragment training subset visualised in Supplementary Fig. 4.

The stepwise model incorporates the same node and edge features as the one-shot model. As it operates on molecular fragments, additional labels are concatenated to the node and edge features, respectively, to indicate which atoms need to be focussed on during the stepwise iterations. Following the IMPRESSION architecture and a sigmoid activation layer, the model outputs bond probabilities between the focused atom and all other atoms in the fragment. The atom with the highest bond probability exceeding a predefined threshold of 0.5 is selected as the attachment atom, and a bond is formed accordingly. If no probabilities exceed this threshold, the model proceeds to focus on the next sub-valent atom.

**Bond-Order and NMR-Prediction Models**

The bond-order model (Fig. 1d) assigns bond orders to structures generated by the one-shot or stepwise models, enabling their use in the subsequent NMR-prediction step. It is applied after each generation step, but is not explicitly shown in Fig. 1a for clarity. The NMR-prediction model (Fig. 1e) then predicts chemical shifts for the resulting structures, and by comparing to the given NMR measurements, it identifies atoms with incorrect bonding assignments to be corrected in the later structure correction stage. Both models were trained on the full-molecule dataset; their architectures and input features are described in detail in Supplementary Section 1.2.5.

**Multi-Shot Noise-Augmented Molecule Reconstruction**

During testing, structural candidates were generated using a multi-shot noise-augmentation strategy to enable Top-k sampling across a distribution of experimental variability in the NMR spectroscopic chemical shift and scalar coupling data. Gaussian noise was added to the input chemical shifts and scalar coupling constants, with standard deviations corresponding to realistic uncertainties between experiment and our simulation method (0.3 ppm for $^{1}H$, 2.5 ppm for $^{13}C$, and 0.4 Hz for all coupling constants between NMR-active nuclei, unless otherwise stated). Following noise augmentation, edges associated with coupling constants greater than 2 Hz were assigned the corresponding coupling correlation type (see 'Dataset' section), whereas all others were assigned as uncorrelated (value of 0). Each noise-perturbed input produced a single predicted structure, yielding an ensemble of candidate structures for a given set of NMR parameters. These candidates were subsequently ranked using performance metrics described below.

### Performance metrics

The performance of the inverse-IMPRESSION framework was evaluated using two metrics: (1) Single-prediction accuracy, defined as the fraction of test molecules which were correctly predicted from the input NMR parameters without noise augmentation; (2) Top-k accuracy, defined as the fraction of test molecules in which the correct structure is ranked within the Top-k predictions generated via noise augmentation. For experimental data, candidate structures were ranked using the mean absolute error (MAE) between the experimental $^{13}C$ chemical shifts and those predicted for each candidate using forward IMPRESSION-G2.[22] For simulated data, candidate structures were ranked by the frequency with which a given structure was predicted amongst all noise-augmented predictions. A more informed ranking, such as MAE between simulated chemical shifts for candidates and the true structure, was not used for simulated data in order to avoid bias.

## Results and discussion

### One-shot Prediction and Node Degeneracy

The first stage of the prediction pipeline employed a "one-shot" model (Fig. 1b) that aimed to predict all bonds within a molecule simultaneously. When trained and tested on the complete set of theoretically available NMR parameters (NMR Dataset A), comprising all $^{1}H$, $^{13}C$, $^{15}N$, $^{17}O$ and $^{19}F$ chemical shifts together with all pairwise scalar coupling constants, the model achieved excellent performance, correctly reconstructing 99.4% of molecules in the test set (Table 1, entry 1) in single prediction events. This result is consistent with our recent proof-of-concept work[21] on predicting interatomic distances and demonstrates the ability of the inverse-IMPRESSION graph transformer network (GTN) architecture to effectively learn the relationship between NMR spectroscopic data encoded in graph-based representations and underlying molecular structure – in this case reflected by bond probabilities.

**Table 1 | Performance of inverse-IMPRESSION models as a function of the composition of NMR dataset composition**. $\delta^1H$, $\delta^{13}C$, $\delta^{15}N$, $\delta^{17}O$ and $\delta^{19}F$ are chemical shifts of the corresponding nuclei. $J_{HCNOF\text{-}HCNOF}$ refers to scalar coupling constants between all pairs of the five nuclei, and $J_{CH\text{-}H}$ correlations are coupling correlations between C-H and H-H atoms pairs.

| Entry | Model | NMR dataset | NMR node input features | NMR edge input features | Single prediction accuracy | Top-1 accuracy | Top-3 accuracy | Top-10 accuracy | Top-200 accuracy |
|---|---|---|---|---|---|---|---|---|---|
| **1** | One-shot | A | $\delta^1H$, $\delta^{13}C$, $\delta^{15}N$, $\delta^{17}O$, $\delta^{19}F$ | $J_{HCNOF\text{-}HCNOF}$ | 99.4% | - | - | - | - |
| **2** | One-shot | B | $\delta^1H$, $\delta^{13}C$ | $J_{CH\text{-}H}$ correlations | 5.3% | - | - | - | - |
| **3** | Stepwise | B | $\delta^1H$, $\delta^{13}C$ | $J_{CH\text{-}H}$ correlations | 49.6% | 74.2% | 81.2% | 82.7% | 82.8% |
| **4** | Stepwise with structure correction | B | $\delta^1H$, $\delta^{13}C$ | $J_{CH\text{-}H}$ correlations | 53.4% | 77.8% | 86.4% | 88.7% | 89.2% |

However, NMR Dataset A represents an experimentally unrealistic scenario, as many of the included NMR parameters such as $^{17}O$ chemical shifts or very small scalar couplings (below 2 Hz), are generally impractical to measure using routine solution-state NMR methods. To address this, a model was trained under more experimentally realistic conditions, using only commonly measured parameters: $^1H$ and $^{13}C$ chemical shifts, together with $^1H$-$^1H$ and $^1H$-$^{13}C$ scalar couplings greater than 2 Hz, which were then converted to coupling correlation labels (NMR Dataset B). Under these conditions, the model achieved a single-prediction accuracy of only 5.3% (Table 1, entry 2).

Analysis of the one-shot model performance as a function of molecular composition (Supplementary Section 2.3.1) using NMR Dataset B, reveals that molecules containing two or more nitrogen atoms or two or more oxygen atoms consistently failed to yield any correct structures. This behaviour is also in line with that previously observed[21] and can be rationalised by the degeneracy of the node/edge information for such heteroatoms i.e., all atoms of these types contain identical (null) node/edge features. This is demonstrated by a representative failure case in Fig. 2. The ground-truth structure, 2-hydroxy-2-methylbutanoic acid, contains three oxygen atoms (hydrogens and bond orders omitted for clarity), each bonded to one of two adjacent carbons (C1 and C2; Fig. 2e). While the one-shot model correctly predicts the carbon framework (Fig. 2a), it incorrectly assigns all three oxygen atoms to C1, and leaves C2 unbound to any oxygen (Fig. 2d). This behaviour arises because the corresponding bond existence probabilities show identical values for all three oxygen atoms (66% for bonding to C1 and 33% to C2; Fig. 2b and 2c). This uniformity reflects the absence of NMR information for all oxygen (as well as N and F) atoms in NMR Dataset B, resulting in identical (null) node and edge features for these atoms in the input graph. Consequently, the model assigns indistinguishable and chemically unrealistic bonding probabilities across the predicted structure.

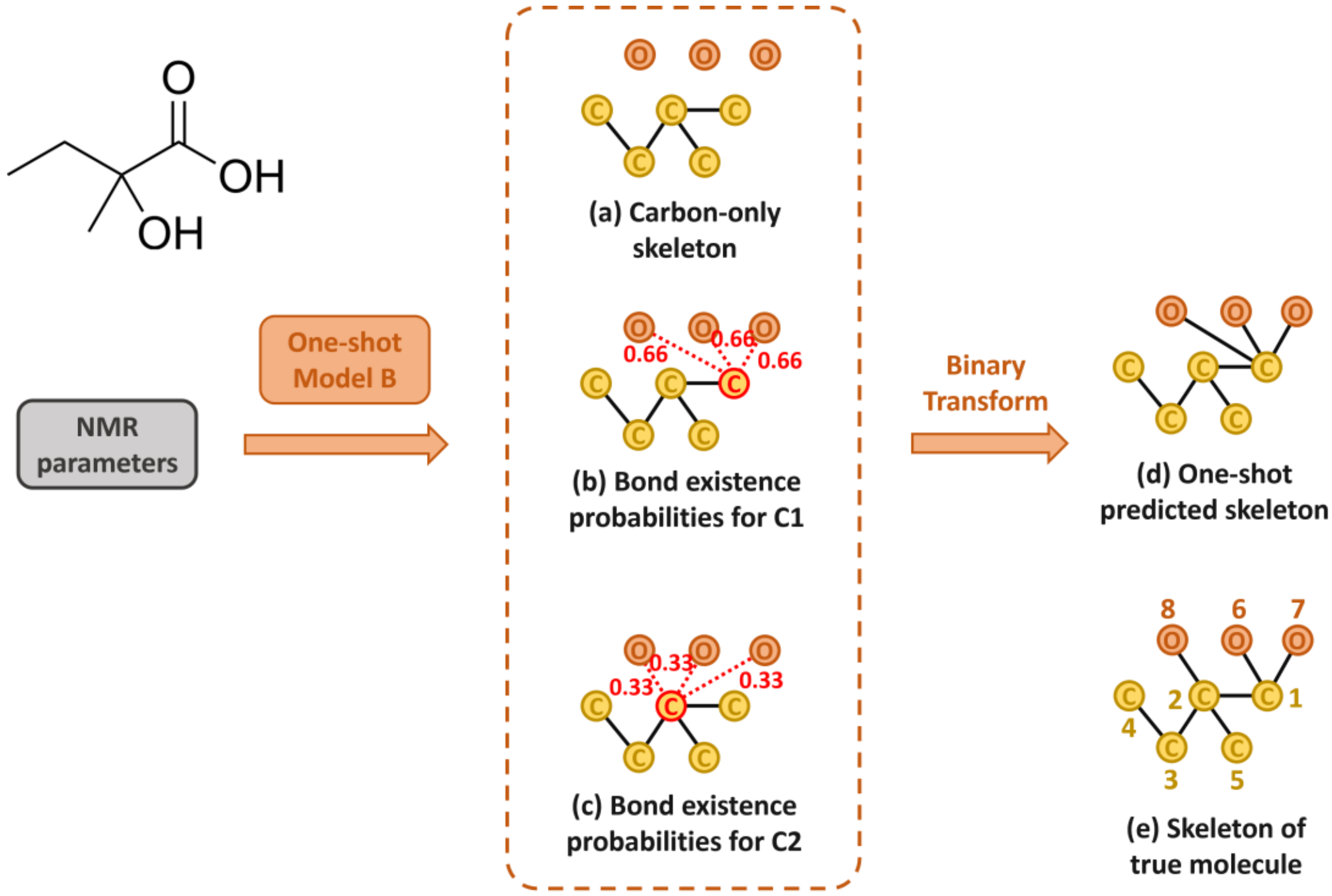

**Fig. 2 | Incorrect prediction of 2-hydroxy-2-methylbutanoic acid by the one-shot model trained on NMR Dataset B.** Predicted bond existence probabilities and corresponding molecular structures are shown (hydrogens and bond orders omitted for clarity). **a,** Predicted carbon skeleton, with high bond existence probabilities for C-C and C-H bonds. **b,c,** Predicted bond existence probabilities between the oxygen atoms and C1/C2. **d,** Predicted molecular structure obtained by binarising the bond existence probabilities from (**b**) and (**c**). **e,** Carbon-oxygen skeleton of the ground-truth molecule, with carbon and oxygen atoms indexed.

**Structure Correction**

To address node degeneracy, a second stage is introduced to refine the structures generated by the initial one-shot model. Following the initial prediction, all bonds involving degenerate atoms (N, O and F in NMR Dataset B) are removed. A “stepwise” model (Fig. 1c) then iteratively focuses on one sub-valent atom at a time, and predicts bonding probabilities for that focussed atom to each other atom in the graph. A bond is then assigned to the edge with the highest predicted probability, and the process is repeated until there is no bond probability predicted to be greater than 0.5, at which point the next focussed atom is selected.

This stepwise procedure substantially improves performance, increasing the single-prediction accuracy to 49.6% across the simulated full-molecule test set (Table 1, Entry 3). Further improvement was achieved through multi-shot noise-augmentation, in which each prediction was repeated 200 times with small perturbations applied to the input spectroscopic data. The resulting candidate structures were then ranked according to their frequency of occurrence across the ensemble of 200 structures.

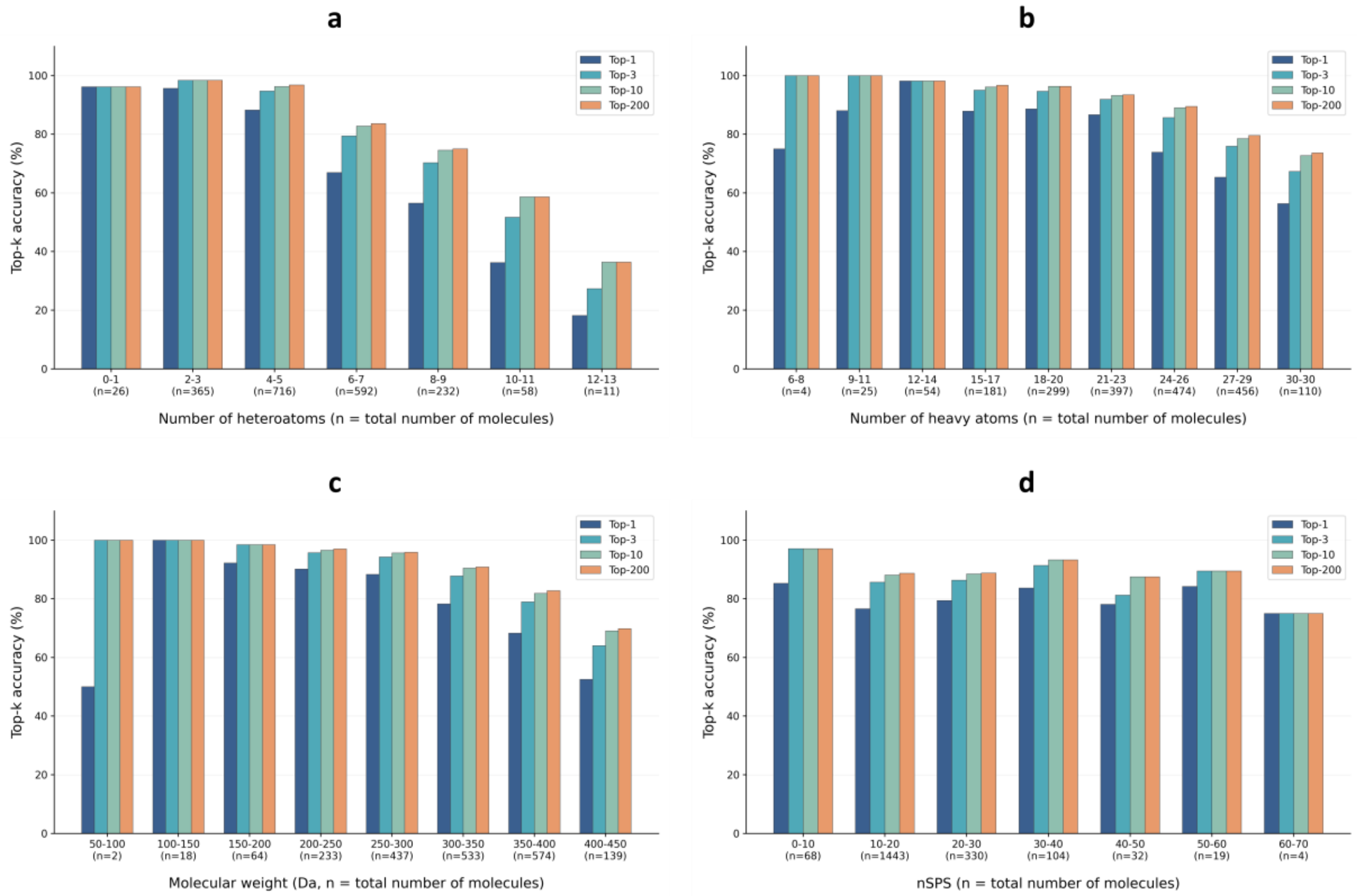

**Fig. 3 | Effect of molecular properties on Top-k (k = 1, 3, 10 and 200) accuracy after structure correction with NMR Dataset B.** The total number of molecules within each property range is indicated below the corresponding bars. **a,** Dependence on heteroatom count (N, O and F). **b,** Dependence on heavy atom count (C, N, O and F). **c,** Dependence on molecular weight. **d,** Dependence on molecular complexity, quantified by normalised spacial score (nSPS).

Under these conditions, the combined "one-shot + stepwise" workflow achieved a Top-1 accuracy of 74.2%, with the correct structure appearing within the Top-3 and Top-10 in 81.2% and 82.7% of cases, respectively, and among all the 200 generated candidates in 82.8% of cases.

To further improve prediction performance, an NMR prediction stage of structure correction was added. $^{13}$C chemical shifts are predicted for any fully valent candidate structure output by stepwise model, and these are compared to the input chemical shift values. Carbon atoms exhibiting large prediction errors (mean absolute error (MAE) >10 ppm), together with their neighbouring N, O and F atoms, have their bonds removed, after which the stepwise model is again applied to reassign bonds for the resulting sub-valent atoms. Incorporation of this further structure-correction procedure increased the Top-1 accuracy to 77.8% for NMR Dataset B (Table 1, entry 4), with corresponding Top-3, Top-10 and Top-200 accuracies of 86.4%, 88.7% and 89.2%, respectively.

Notably, the overall performance of this platform on simulated datasets is robust with increasing molecular size (Fig. 3b,c) and provides consistently high Top-k accuracy as a function of molecular complexity, as quantified by the normalised spacial score (nSPS)[27] (Fig. 3d). The ranking method in particular performs well, with Top-3 to Top-200 accuracies exceeding 95% for molecules containing five or fewer heteroatoms, suggesting that node degeneracy has been substantially mitigated.

1-(4-ethylphenyl)ethenone (1)

aloperine (2)

4-piperidone ethylene acetal (3)

tadalafil (4)

abiraterone (5)

betulin (6)

24-epibrassinolide (7)

brucine (8)

2-ethoxy-5-methylpyridin-4-amine (9)

1-Boc-4-(4-aminophenyl)piperazine (10)

**Fig. 4 | Structures of 10 successfully elucidated molecules from experimental NMR data.**

However, these accuracies decrease to below 60% for molecules with more than ten heteroatoms (Fig. 3a). This is attributed to the low density of NMR data available for molecules rich in nitrogen, oxygen, and fluorine, which reduces the information available to inform the bond probability prediction around these atoms, thereby limiting model performance even when node degeneracy is addressed. This sensitivity to heteroatoms can be reduced further by inclusion of $^{15}N$ and $^{19}F$ NMR data, which are experimentally accessible but rarely used in complex structure elucidation. This is readily demonstrated by retraining the platform to include these nuclei, resulting in over 80% of Top-3 to Top-200 accuracy on structures containing up to 13 heteroatoms (structure correction model with NMR Dataset C, Supplementary Information 2.3.4)

**Performance on Experimental Data**

Having established robust performance on simulated data, the complete inverse-IMPRESSION platform was then evaluated on an experimental dataset comprising 19 molecules ($M_R$ = 143.2-608.7 g/mol, nSPS = 9.6-59.2), representative of the types of complex and challenging organic molecular structures which chemists working in pharmaceutical and natural products chemistry may be interested in solving. The experimental NMR data were obtained from the literature[28] or generated in this work. As shown in Fig. 4, ten structures were correctly identified by the platform. Compounds **1-7** were recovered using 2000-shot predictions with the same noise parameters as in the test on simulated data above. Compound **8** was identified after an additional 5000 shots. Two further compounds (**9-10**) were found by increasing the standard deviation of the coupling constant noise to explore a greater region of spectroscopic data space. The largest and most complex successful examples include the polycyclic steroids 24-epibrassinolide **7** ($M_R$ = 480.7 g/mol, nSPS = 48.1) and betulin **6** ($M_R$ = 442.7 g/mol, nSPS = 56.5), which to our knowledge, represent among the most challenging and structurally complex molecules solved using machine learning approaches to date.

In our experimental testing, correct structures were consistently and reliably identified as Top-1 by ranking according to the mean absolute error (MAE) between experimental and predicted $^{13}C$ chemical shifts, computed using IMPRESSION-G2 (Supplementary Section 2.4.1). Using this criterion, the correct structure was ranked first in all cases where it was generated among the noise-augmented candidates. In each case, the corresponding MAE values for the correct structure was below 4.4 ppm, which is in line with the benchmarked performance of the NMR prediction tool, IMPRESSION-G2, against experimental data. By contrast, in all nine unsuccessful cases, the incorrect structures exhibited substantially higher MAE values (≥ 5.3 ppm), enabling clear discrimination from correct predictions (Supplementary Section 2.4.2) and highlighting that the MAE ranking criteria offers a substantial indicator of confidence for incorrect predictions made by this machine learning platform.

In line with the performance on simulated data described in Fig. 3, accuracy on experimental data was relatively, but not completely, insensitive to molecular size and complexity. Successful cases spanned $M_R$ = 143.2-480.7 g/mol and nSPS = 9.6-56.5, while the unsuccessful cases spanned $M_R$ = 221.3-608.7 g/mol and nSPS = 11.8-59.2. Additional factors influencing prediction success are summarised in Supplementary Table 4. As expected, performance decreased when less 2D NMR data were available, and when the target structure contained more quaternary-quaternary carbon bonds and heteroatoms. These factors highlight the importance of both the quantity and, importantly, the distribution of NMR spectroscopic information across the molecular structure that is being solved i.e., molecules that contain regions described by minimal NMR information, such as quaternary centres and heteroatom-rich substructures, are inherently more difficult to resolve with NMR data alone. This limitation is well recognised in NMR-based structure elucidation: when insufficient spectroscopic connectivity is available for a substructure, accurate structure prediction becomes challenging for human analysts and unsurprisingly also for this machine learning system as well.

## Conclusions

Herein we demonstrate that the GTN-based inverse-IMPRESSION platform can be adapted to the task of molecular structure elucidation directly from measurable NMR parameters, including chemical shifts and scalar coupling correlations in 2-dimensional NMR spectra. We overcome the limitations of degeneracy for nodes/atoms that have no associated NMR data with an iterative bond-prediction strategy. When combined with noise-augmented multi-shot prediction and frequency ranking, the resulting workflow achieves strong performance on simulated datasets containing only $^{1}$H and $^{13}$C NMR information, achieving a Top-1 accuracy exceeding 77%. This simulation-trained model is effectively transferred to testing on experimentally measured NMR data without additional fine-tuning, correctly identifying 10 out of 19 molecules (52.6%), including complex natural product-like and synthetic structures that challenge expert chemists to solve. These cutting-edge results demonstrate that graph neural network models can learn sufficient structural information from routine experimental NMR measurements to enable automated molecular structure elucidation, opening new opportunities for applications across fields where NMR-based structure elucidation is critical.

**Data availability**

All data supporting the findings of this study are available within the article and the Supporting Information. Code associated with this work is publicly available via Zenodo at https://doi.org/10.5281/zenodo.20432239

**Acknowledgements**

Z. J. and R. C. acknowledge funding from the EPSRC Technology Enhanced Chemical Synthesis Centre for Doctoral Training (EP/Y03483X/1) to support their work.

**Author contributions**

Z. J. conceived the entire workflow used in this programme, led the architecture and code improvements, generated, curated and enhanced the majority of the training sets used for inverse-IMPRESSION and conducted the testing against computational and experimental datasets. G. A. provided the experimental NMR measurements and curated the experimental NMR data used for experimental testing, supported by R. C. B. H. developed the early code base from which inverse-IMPRESSION was developed, and provided some portion of the training sets. M. G. co-supervised Z. J., provided insight on workflow development. C. B. led project conception and development, supervising all researchers in this programme. All authors contributed to manuscript development, which was led by C. B. and Z. J.

**Competing interests**

The authors declare no competing interests.

**Supporting Information**

# Inverse-IMPRESSION: A Graph-based Platform for Molecular Structure Elucidation from Experimental NMR Spectroscopic Properties

Zheqi Jin[a], Grace Armitage[a], Richard Cox[a], Ben Honoré[a], Mohammad Golbabaee[b] and Craig Butts[a,*]

[a]School of Chemistry, University of Bristol, Cantock's Close, Bristol, BS8 1TS, U.K.

[b]School of Engineering Mathematics and Technology, University of Bristol, Ada Lovelace Building, Tankard's Cl, Bristol BS8 1TW

*Correspondence to: Craig.Butts@bristol.ac.uk

# Table of Contents

# 1 Methods

## 1.1 Hardware and Software Availability

All computations were performed on a single NVIDIA GH200 Grace Hopper Superchip hosted on the Isambard-AI Phase 2 high-performance computing system, running SUSE Linux Enterprise Server 15 SP6. The superchip integrates a 72-core NVIDIA Grace CPU and an NVIDIA Hopper H100 Tensor Core GPU, providing 115 GB of CPU memory and 96 GB of GPU memory.

The computational environment was managed using Conda (v25.7.0) with Python 3.12.11. GPU acceleration was provided by NVIDIA CUDA (v12.6). The main software dependencies included Numpy (v2.3.3), Pandas (v2.3.2), RDKit (v2025.03.6), Scipy (v1.16.2), PyTorch (v2.6.0) and PyTorch Geometric (v2.6.1).

Three in-house Python packages were employed in this study. IMPRESSION-G2,[1] previously developed and reported in our published work, was used to predict chemical shifts and scalar coupling constants for 3D molecular structures, thus generating the training and test datasets used in this work. Efficient Molecular Storage (EMS) was developed for molecular and nuclear magnetic resonance (NMR) data processing and storage and was additionally used for conformer generation and geometry optimization via RDKit.[2] Inverse-IMPRESSION was developed based on IMPRESSION-G2, aiming at extending the original framework to multiple learning tasks. The four models investigated in this study (one-shot, stepwise, bond-order, and NMR-prediction models) were all trained using the Inverse-IMPRESSION platform.

The code for EMS and inverse-IMPRESSION, and the SDF files of 19 molecules with experimental NMR parameters are available at: https://github.com/Buttsgroup and https://doi.org/10.5281/zenodo.20432239

# 1.2 Inverse-IMPRESSION platform

## 1.2.1 Graph Transformer Network architecture

The Graph Transformer Network (GTN) architecture of inverse-IMPRESSION is an adapted version of the solution presented in the CHAMMPS Kaggle competition[3] and deployed in IMPRESSION-G2.[1] All models used in inverse-IMPRESSION platform share the same GTN architecture but differ in their input and output node/edge features, and therefore have distinct learned weights and biases. The layers of the GTN architecture are described as follows (Supplementary Fig. 1):

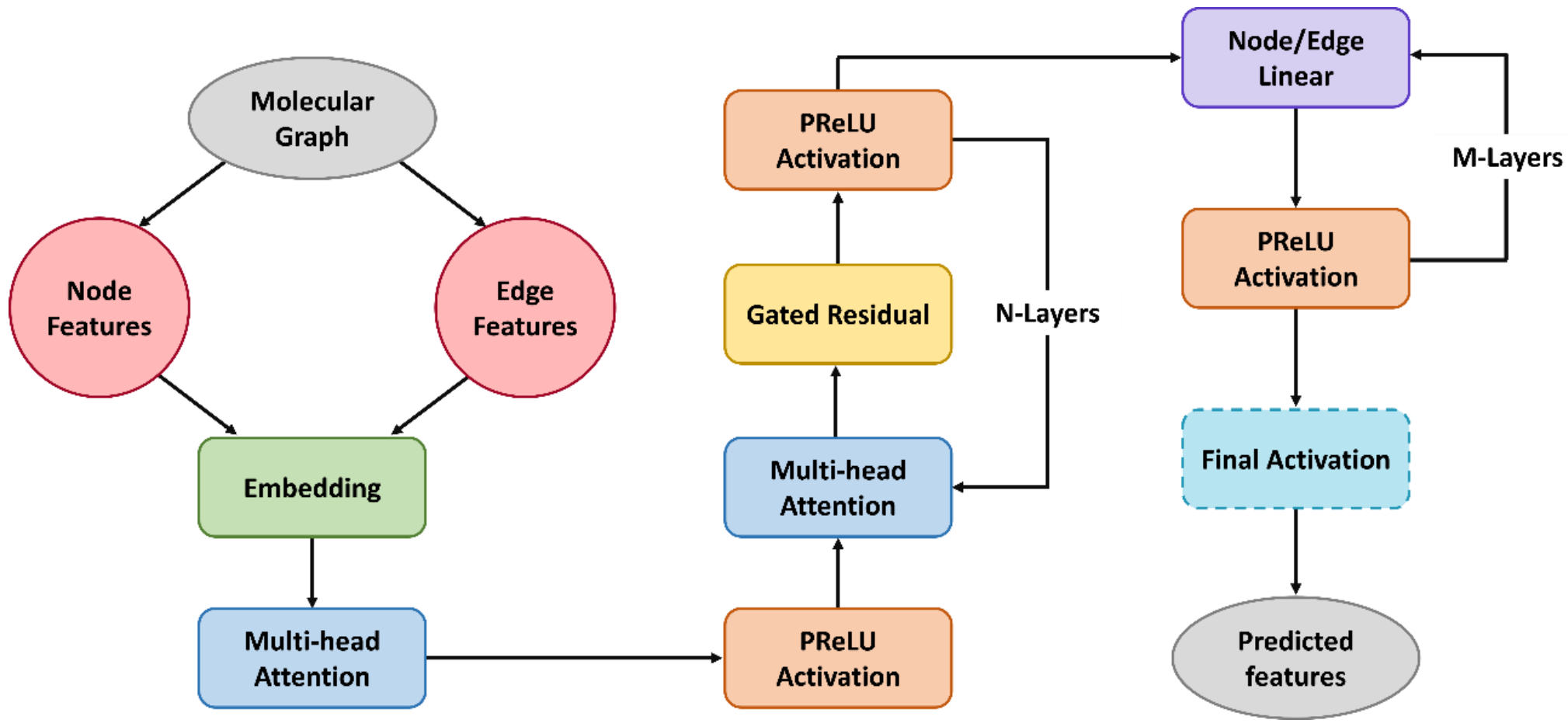

**Supplementary Fig. 1** The GTN architecture of the inverse-IMPRESSION platform

1. **Molecular Graph:** The input molecular structure or NMR spectroscopic data is represented as a fully-connected graph (Supplementary Fig. 2), in which atoms correspond to nodes and atom pairs correspond to edges. Each node and edge is associated with a feature vector encoding the relevant atomic or pairwise information. The input and output node/edge features for each model are discussed in Supplementary Section 1.2.5.

2. **Embedding Layer:** Embeddings provide continuous vector representation for discrete data such as categorical labels and words, which are not directly usable by neural networks. These representations are learnable and can be updated during training to capture task-specific relationships. The node and edge features that are transformed into embeddings for each model are discussed in Supplementary Section 1.2.5.

3. **Multi-head Attention Layer:** After categorical features are transformed into embeddings and all node and edge features are concatenated, they are passed to the multi-head attention layer, which is described in Supplementary Section 1.2.3.

4. **PReLU Activation Layer:** An activation layer introduces non-linearity to the model, enabling it to learn complex patterns and relationships within the data. In inverse-IMPRESSION, PReLU[4] (Parametric Rectified Linear Unit) is used as the activation function following the multi-head attention layers, gated residual layers and linear layers. The PReLU function is defined as:

$$PReLU(x) = max(0, x) + \boldsymbol{a} \cdot min(0, x) \tag{S1}$$

where $\boldsymbol{a}$ is a learnable parameter that controls the slope of the negative part of the function.

5. **Gated Residual layer:** The concept of the gated residual layer builds on the gating mechanism introduced in gated recurrent units[5] (GRUs) and has been extended to graph neural networks (GNNs). These layers incorporate learnable gates to control the flow of information between representations from previous and current layers, enabling the model to adaptively determine which features should be retained or discarded. In inverse-IMPRESSION, the gated residual layer adopts an update-gate mechanism inspired by GRUs, which is defined as follows:

$$h_t = z_t \odot \tilde{h}_t + (1 - z_t) \odot h_{t-1} \tag{S2}$$

$$\tilde{h}_t = f(h_{t-1}) \tag{S3}$$

$$z_t = \sigma(\boldsymbol{W}_t h_{t-1} + \boldsymbol{U}_t \tilde{h}_t + \boldsymbol{b}_t) \tag{S4}$$

   where $h_t$ is the output node or edge representation of the current hidden layer, $h_{t-1}$ is the representation of the previous hidden layer, and $\odot$ denotes element-wise multiplication. The candidate representation $\tilde{h}_t$ is computed by applying a transformation function $f$ to $h_{t-1}$. The function $f$ represents a message passing neural network (MPNN), which is implemented as a multi-head attention layer in inverse-IMPRESSION. The update gate $z_t$ is calculated by applying a sigmoid function $\sigma$ to a linear transformation of both $h_{t-1}$ and $\tilde{h}_t$, producing a gating matrix with values between 0 and 1. Here, $\boldsymbol{W}_t$ and $\boldsymbol{U}_t$ are learnable weight matrices and $\boldsymbol{b}_t$ is a learnable bias vector.

6. **Node/Edge Linear layer:** The linear layer is a fundamental component of all neural networks. It performs a linear transformation on the input features:

$$y = \boldsymbol{W}x + \boldsymbol{b} \tag{S5}$$

   where $x$ is the input feature, $y$ is the output feature, $\boldsymbol{W}$ is a learnable weight, and $\boldsymbol{b}$ is a learnable bias vector.

7. **Final Activation Layer:** The final activation layer is the last layer in the inverse-IMPRESSION models and is used to map the output features to task-specific values. The sigmoid function is employed as the final activation layer in the one-shot and stepwise models to predict bond existence probabilities and atom connectivity probabilities. The softmax function is applied in the bond-order model to predict bond order probabilities across five classes: no bond, single bond, double bond, triple bond, and aromatic bond. No final activation layer is applied in the NMR-prediction model.

   The sigmoid function maps values to a range between 0 and 1 and can be converted into binary outputs using a threshold of 0.5. It is therefore commonly used in binary classification tasks. The sigmoid function is defined as:

$$\sigma(x) = \frac{1}{1 + e^{-x}} \tag{S6}$$

   The softmax function maps a vector of values to a probability distribution over multiple classes, such that the outputs are non-negative and sum to one. The resulting probability distribution can then be converted into a class prediction, typically by selecting the class with the highest probability. It is therefore commonly used in multi-class classification tasks. The softmax function ($K$ classes) is defined as:

$$softmax(x_i) = \frac{exp(x_i)}{\sum_{j=1}^{K} exp(x_j)} \tag{S7}$$

## 1.2.2 Graph Representation

In molecular modelling, graph representations are particularly useful as they naturally mirror the topological structure of molecules. Atoms are represented as nodes and are connected by edges, which can correspond to chemical bonds or any type of atomic interactions. Each node and edge is associated with a feature vector, allowing the encoding of relevant information such as atom type, chemical shift, bond order, and scalar coupling correlations. In the inverse-IMPRESSION platform, the graph representation is extended to a fully-connected graph, where every pair of atoms is connected by an edge (Supplementary Fig. 2). This design is motivated by the structure elucidation task, in which the molecular structure is unknown. Inverse-IMPRESSION aims to infer which edges correspond to true chemical bonds.

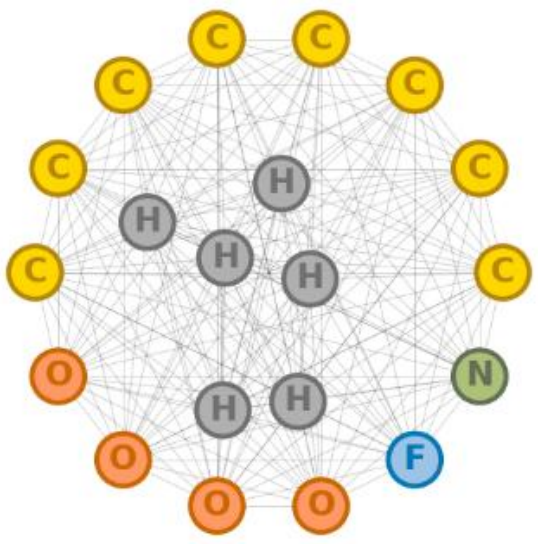

**Supplementary Fig. 2** A fully-connected graph with 20 nodes (6 H atoms, 8 C atoms, 1 N atom, 4 O atoms, and 1 F atom). Each pair of atoms (nodes) is connected by an edge (grey lines).

### 1.2.3 Multi-head Attention Mechanism

All four models in the inverse-IMPRESSION platform are built according to the Graph Transformer Network (GTN) architecture,[6] which integrates the Transformer framework[7] into GNNs. The Transformer architecture relies on the attention mechanism, allowing the model to weight the importance of information aggregated from neighbouring nodes and edges, and thus capture relational information within the graph. The multi-head attention mechanism extends this capability by employing multiple independent attention heads, enabling different aspects of node and edge features, as well as their relationships, to be captured by different heads.

The multi-head attention mechanism in inverse-IMPRESSION is implemented as follows:

$$z_{uv}^{(h)} = \left[ h_u^{(t-1,h)} \| e_{uv}^{(t-1,h)} \| h_v^{(t-1,h)} \right] \quad \text{(S8)}$$

$$\alpha_{uv}^{h} = softmax_u\left( \boldsymbol{w}^{(h)^{\mathrm{T}}} z_{uv}^{h} \right) \quad \text{(S9)}$$

$$\tilde{h}_u^{(h)} = \alpha_{uv}^{(h)} \cdot h_u^{(t-1,h)} \quad \text{(S10)}$$

$$m_{u\to v}^{(h)} = \tilde{h}_u^{(h)} \odot e_{uv}^{(t-1,h)} \quad \text{(S11)}$$

$$m_v^{(h)} = \sum_{u\in\mathcal{N}(v)} m_{u\to v}^{(h)} \quad \text{(S12)}$$

$$h_v^{(t,h)} = h_v^{(t-1,h)} + m_v^{(h)} \quad \text{(S13)}$$

$$e_{uv}^{(t,h)} = \boldsymbol{W}_E^{(h)} \left[ h_u^{(t,h)} \| e_{uv}^{(t-1,h)} \| h_v^{(t,h)} \right] \quad \text{(S14)}$$

1. **Attention calculation (Supplementary Eqs. 8-9):** The source node features $h_u^{(t-1,h)}$, edge features $e_{uv}^{(t-1,h)}$ and target node features $h_v^{(t-1,h)}$ in the $(t-1)$-th layer and $h$-th attention head are first concatenated to form a triplet vector $z_{uv}^{(h)}$ for each edge $(u, v)$. The attention weight $\alpha_{uv}^{h}$ is then computed by applying a linear transformation with a learnable weight vector $\boldsymbol{w}^{(h)}$ to $z_{uv}^{h}$, followed by a softmax operation. The softmax operation is applied over the source nodes $u$ for each target node $v$, ensuring that the attention weights sum to 1 across all neighboring nodes of $v$.

2. **Message function (Supplementary Eqs. 10-11):** The source node features $h_u^{(t-1,h)}$ are weighted by the attention weights $\alpha_{uv}^{h}$ to form a weighted node feature vector $\tilde{h}_u^{(h)}$. The message $m_{u\to v}^{(h)}$ is then computed by multiplying the weighted node feature vector $\tilde{h}_u^{(h)}$ with the edge feature $e_{uv}^{(t-1,h)}$.

3. **Aggregation function (Supplementary Eq. 12):** The messages $m_{u\to v}^{(h)}$ from all neighboring nodes $\mathcal{N}(v)$ are aggregated by sum operation to form a single message $m_v^{(h)}$ for the target node $v$.

4. **Node update function (Supplementary Eq. 13):** The target node feature $h_v^{(t,h)}$ is updated by adding the aggregated message $m_v^{(h)}$ to the target node feature $h_v^{(t-1,h)}$ in the previous layer.

5. **Edge update function (Supplementary Eq. 14):** The edge feature $e_{uv}^{(t,h)}$ is updated by applying a linear transformation with a learnable weight matrix $\boldsymbol{W}_E^{(h)}$ to the concatenated vector. This allows the edge features to be updated based on the new node features after the attention mechanism.

## 1.2.4 Training elements

### *1.2.4.1 Noise-Augmented Training*

Random Gaussian noise was added to the chemical shifts and scalar coupling constants at each training epoch for both the one-shot and stepwise models to improve robustness. The standard deviations applied to the chemical shifts were 0.3 ppm for $^{1}H$, 2.5 ppm for $^{13}C$, 7.5 ppm for $^{15}N$, and 7.5 ppm for $^{19}F$ nuclei. For all scalar coupling constants, a standard deviation of 0.4 Hz was used. Coupling constants were subsequently converted into integer-encoded scalar coupling correlations using a threshold of 2 Hz. The encoding scheme was defined as follows: $^{1}H$-$^{1}H$ COSY = 1, $^{1}H$-$^{13}C$ HSQC = 2, $^{1}H$-$^{13}C$ HMBC = 3, $^{1}H$-$^{15}N$ HSQC = 4, $^{1}H$-$^{15}N$ HMBC = 5, $^{19}F$-$^{13}C$ HSQC = 6, and $^{19}F$-$^{13}C$ HMBC = 7.

### *1.2.4.2 LAMBW Optimiser*

The LAMBW (Layer-wise Adaptive Moments optimiser for Batching training with Weight decay) optimiser[8] is used to improve the convergence speed and training stability when the inverse-IMPRESSION models are trained on large datasets. The weight decay mechanism is incorporated in the optimiser to prevent overfitting by penalizing large weights, which helps the model generalise better to unseen data.

### *1.2.4.3 CyclicLR scheduler*

The CyclicLR (Cyclic Learning Rate) scheduler[9] is used to vary the learning rate between a lower and an upper bound during training. The learning rate periodically increases and decreases over a fixed number of epochs, which helps the models escape local minima and saddle points in the loss landscape. In the inverse-IMPRESSION platform, the CyclicLR scheduler is applied for a single cycle, after which the learning rate is fixed at the lower bound.

### *1.2.4.4 Hyperparameters*

Hyperparameters are predefined settings that control the training process and model architecture. The four inverse-IMPRESSION models share the same tuneable hyperparameters:

- **Batch Size**: Number of molecular graphs per training iteration, set to**15**.
- **Epochs**: Total number of times the entire training dataset passes over the model, set to **100**.
- **Attention Heads**: Number of attention heads in the multi-head attention layer, set to **16**.
- **Gated Residual Layers**: Number of gated residual layers to update node and edge features, set to **6**.
- **Embedding Dimension**: Dimension of the atom and edge embeddings, set to **48**.
- **Weight Decay**: Weight decay factor in the LAMBW optimiser, set to **0.05**.
- **Minimum Learning Rate**: Lower bound of learning rate in the CyclicLR scheduler, set to **0.001**.
- **Maximum Learning Rate**: Upper bound of learning rate in the CyclicLR scheduler, set to **0.01**.
- **CyclicLR Up Epochs**: Number of epochs the learning rate increases in the CyclicLR scheduler, set to **15**.
- **CyclicLR Down Epochs**: Number of epochs the learning rate decreases in the CyclicLR scheduler, set to **15**.

## 1.2.5 Inverse-IMPRESSION models

The four models of inverse-IMPRESSION platform are introduced in the main text. Their key features are summarised in Supplementary Table 1.

**Supplementary Table 1** Summary of four inverse-IMPRESSION models

| Model | One-shot | Bond-order | Stepwise | NMR-prediction |
|---|---|---|---|---|
| **Input Node Feature** | • atom type<br>• chemical shift | • atom type | • atom type<br>• chemical shift<br>• atom-in-fragment label | • atom type |
| **Input Edge Feature** | • edge type<br>• scalar coupling correlation | • edge type<br>• bond existence | • edge type<br>• scalar coupling correlation<br>• bond-in-fragment label | • edge type<br>• bond order |
| **Output Node Feature** | / | / | • atom connectivity probability | • chemical shift |
| **Output Edge Feature** | • bond existence probability | • bond order probability | / | / |
| **Final Activation Layer** | Sigmoid | Softmax | Sigmoid | / |
| **Loss Function** | Binary cross-entropy | Categorical cross-entropy | Binary cross-entropy | L1 loss |

The features to be embedded are summarised below:

- **Atom type**: encoded by atomic number (H = 1, C = 6, N = 7, O = 8, F = 9)
- **Edge type**: integer-encoded by atom-pair category (H-H = 0, C-H = 1, C-C = 2, N-H = 3, N-C = 4, N-N = 5, O-H = 6, O-C = 7, O-N = 8, O-O = 9, F-H = 10, F-C = 11, F-N = 12, F-O = 13, F-F = 14)
- **Scalar coupling correlation**: integer-encoded by 2D NMR peaks ($^{1}H$-$^{1}H$ COSY = 1, $^{1}H$-$^{13}C$ HSQC = 2, $^{1}H$-$^{13}C$ HMBC = 3, $^{1}H$-$^{15}N$ HSQC = 4, $^{1}H$-$^{15}N$ HMBC = 5, $^{19}F$-$^{13}C$ HSQC = 6, and $^{19}F$-$^{13}C$ HMBC = 7)
- **Bond existence**: binary-encoded bond indicator (0 = non-bond, 1 = bond)
- **Atom-in-fragment label**: categorical indicator of atom state (0 = outside the molecular fragment, 1 = within the molecular fragment, 2 = focused atom)
- **Bond-in-fragment label**: binary-encoded bond indicator (0 = outside the molecular fragment, 1 = within the molecular fragment)
- **Bond order**: categorical bond type label (0 = non-bond, 1 = single, 2 = double, 3 = triple, 4 = aromatic)

## 1.2.6 Loss Function

The loss function is used to quantify the discrepancy between the predicted values and the ground truth, thereby guiding the training process. The choice of loss function depends on the specific task and the final activation layer of each model.

### *1.2.6.1 Binary cross-entropy (BCE) loss*

Binary cross-entropy (BCE) loss is applied in the one-shot and stepwise models for the binary classification task. The BCE loss is defined as:

$$\mathcal{L} = -\frac{1}{N}\sum_{i=1}^{N}\left( y_i log(\hat{y}_i) + (1 - y_i)log(1 - \hat{y}_i)\right) \tag{S15}$$

where $N$ denotes the number of edges (one-shot model) or nodes (stepwise model) in a batch, $y_i$ is the ground-truth label (0 or 1) for edge or node $i$, and $\hat{y}_i$ denotes the predicted probability of bond existence (one-shot model) or atom connectivity (stepwise model). This loss function computes the average loss over all edges or nodes within a single batch.

### *1.2.6.2 Categorical cross-entropy loss*

Categorical cross-entropy loss is applied in the bond-order model for the multi-class classification task. It is defined as:

$$\mathcal{L} = -\frac{1}{N}\sum_{i=1}^{N}\sum_{k=1}^{K} y_{i,k} log\left(\hat{y}_{i,k}\right) \tag{S16}$$

where $N$ denotes the number of edges, $K$ is the number of bond types, $y_{i,k}$ is the ground-truth label of bond order (0 or 1) and $\hat{y}_{i,k}$ denotes the predicted probability for each bond type. This loss function computes the average loss over all edges within a single batch.

### *1.2.6.3 L1 loss*

L1 loss (mean absolute error) is applied in the NMR-prediction model to measure the absolute difference between the predicted and ground-truth chemical shift values. It is defined as

$$\mathcal{L} = \frac{1}{N}\sum_{i=1}^{N}| y_i - \hat{y}_i| \tag{S17}$$

where $N$ denotes the number of nodes, $y_i$ is the ground-truth chemical shift and $\hat{y}_i$ is the predicted chemical shift. This loss function computes the average loss over all nodes within a single batch.

# 1.3 Datasets

## 1.3.1 Full-Molecule Training and Test Set

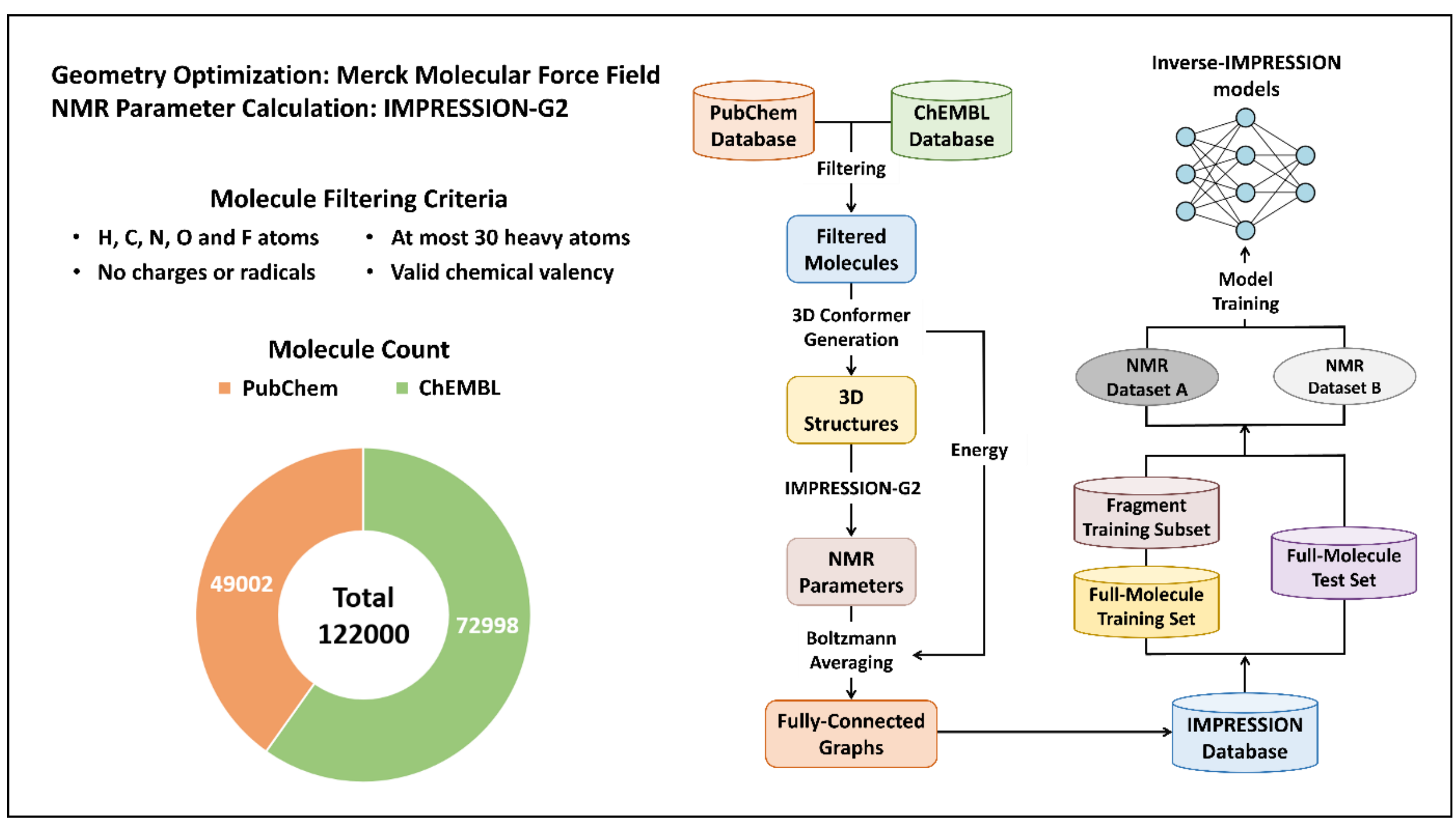

**Supplementary Fig. 3** Workflow of generating molecular graph datasets for training inverse-IMPRESSION models

The molecules of inverse-IMPRESSION dataset were derived from the public PubChem[10] and ChEMBL[11] databases, and filtered according to the following criteria: (1) containing only H, C, N, O and F atoms; (2) comprising at most 30 heavy atoms; (3) having no charges, radicals or disconnected fragments; and (4) satisfying standard chemical valency rules for all atoms. A final dataset of 122,000 organic molecules was obtained, including 72,998 molecules from the PubChem database and 49,002 molecules from the ChEMBL database. Details are shown in Supplementary Fig. 3.

A total of 100 conformers were generated and optimised for each molecule using the Merck Molecular Force Field[12] (MMFF) as implemented in RDKit with default settings. The conformer energies were also calculated using MMFF. The NMR parameters (chemical shifts and scalar coupling constants) of each conformer were predicted using the IMPRESSION-G2 model.[1] The final NMR parameters for each molecule were obtained as Boltzmann-weighted averages over all conformers based on their MMFF energies.

The molecules together with their NMR parameters were processed using our in-house public package, Efficient Molecular Storage (EMS), and stored in DataFrames.[13] They were then transformed into fully-connected graphs using the PyTorch Geometric package,[14] in which each atom was represented as a node and every pair of atoms was connected by an edge. These graph representations were used as inputs to the inverse-IMPRESSION models with customised node and edge features.

The inverse-IMPRESSION dataset was split into a training set of 120,000 molecules (4,974,382 nodes and 107,317,458 edges) and a test set of 2,000 molecules (86,645 nodes and 1,923,615 edges). To distinguish these datasets from the fragment training subset used for training the stepwise model, the inverse-IMPRESSION training and test sets are referred to as the full-molecule training and test sets. The diversity of the molecules can be approximated by visualizing the distributions of molecular weight, atom counts, ring counts, cLogP value, nSPS value and $^{1}H/^{13}C/^{15}N/^{19}F$ chemical shifts, as shown in Supplementary

Section 3. The cLogP (calculated logarithm of partition coefficient) value measures molecule's lipophilicity, which is calculated using the Wildman–Crippen fragment-based model[15] as implemented in RDKit. The nSPS[16] (normalised spacial score) value reflects the 3D structural complexity of a molecule, which is calculated using the SpacialScore module implemented in RDKit.

## 1.3.2 Fragment Training Subset

**Supplementary Fig. 4** The fragment generation process according to the BFS ordering of O atoms. The atom with a red circle is the focused atom in each step.

To train the stepwise model, which predicts the next attachment atom given the current molecular fragment, a subset of 60,000 molecules with no less than 2 oxygen atoms was randomly sampled from the full-molecule training dataset. Two molecular fragments were randomly generated for each molecule using breadth-first search (BFS) algorithm, yielding a fragment training subset of 120,000 molecular fragments.

The BFS algorithm orders graph nodes based on their path distance from a starting node. In the inverse-IMPRESSION platform, BFS is extended to rank NMR-silent atoms by their distance from the NMR-active molecular fragment predicted by the one-shot model. For NMR Dataset B, the BFS algorithm orders the N, O and F atoms by their path distance from the NMR-active fragments composed of H and C atoms. For NMR Dataset C, the BFS algorithm orders the O atoms by their path distance from the NMR-active fragments composed of H, C, N, and F atoms.

Supplementary Fig. 4 illustrates a molecule with one BFS ordering of O atoms and the corresponding fragment generation process. The fragment generation starts from the molecular fragment consisting of H, C, N, and F atoms and proceeds by iteratively adding bonds to the growing fragment following the BFS order. A *focus* mechanism is introduced to iterate over all equivalent atoms (O atoms in this example) and determine whether to continue adding bonds to the currently focused atom. For example, Fragment 1 is the initial molecular fragment with focus on the first oxygen atom O1. Fragment 2 is generated by adding a bond between O1 and a carbon atom according to the ground truth. Since O1 has no bonds to add, Fragment 3 retains the same structure but shifts the focus to the next oxygen atom O2. Fragment 4 is then generated by adding a bond between O2 and a carbon atom. The bond-adding and re-focusing steps are repeated until all oxygen atoms have been processed.

# 2 Results and Discussion

## 2.1 Prediction Performance of Inverse-IMPRESSION Models

This section summarises the prediction performance of three inverse-IMPRESSION models: (1) one-shot model; (2) stepwise model; and (3) stepwise model with structure correction. Three NMR datasets were used to train the models, representing different scenarios. In addition to NMR Datasets A and B discussed in the main text, NMR Dataset C was included to incorporate NMR data for $^{1}$H, $^{13}$C, $^{15}$N, and $^{19}$F. The details of the three NMR datasets are summarised below:

- **NMR Dataset A**: represents a control scenario where NMR chemical shifts and scalar coupling constants (rather than the coupling correlations used in NMR Dataset B and C) are available for all $^{1}$H, $^{13}$C, $^{15}$N, $^{17}$O and $^{19}$F nuclei, including those that are not experimentally observable and measurable. Although idealised and unrealistic, this setting provides an estimate of inverse-IMPRESSION platform's upper bound of reconstruction accuracy.

- **NMR Dataset B**: represents a scenario in which only the most commonly used NMR experiments are available. It includes chemical shifts for $^{1}$H and $^{13}$C nuclei and scalar coupling correlations for H-H and C-H atom pairs, corresponding to COSY, HSQC, and HMBC spectra. A threshold of 2 Hz was applied to filter the weak peaks.

  For the stepwise model trained with NMR Dataset B, all bonds involving N, O, and F atoms are removed for the structures generated by the initial one-shot model. The stepwise model then reassigns bonds for these atoms to reconstruct the molecular structure. For the stepwise model with structure correction, additional bond removal is performed for carbon atoms exhibiting large prediction errors (mean absolute error, MAE > 10 ppm) and their neighbouring N, O, and F atoms. The stepwise model subsequently reassigns bonds for these atoms to correct and complete the structure.

- **NMR Dataset C**: represents a scenario with comprehensive experimentally accessible NMR data. It includes chemical shifts for $^{1}$H, $^{13}$C, $^{15}$N, and $^{19}$F nuclei, as well as scalar coupling correlations for H-H, C-H, N-H, and F-C atom pairs, corresponding to peaks from *$^{1}$H-$^{1}$H* COSY, *$^{1}$H-$^{13}$C* HSQC, *$^{1}$H-$^{13}$C* HMBC, *$^{1}$H-$^{15}$N* HSQC, *$^{1}$H-$^{15}$N* HMBC, *$^{19}$F-$^{13}$C* HSQC, and *$^{19}$F-$^{13}$C* HMBC spectra. A threshold of 2 Hz was applied to filter the weak peaks.

  For the stepwise model trained with NMR Dataset C, all bonds involving O atoms are removed for the structures generated by the initial one-shot model. The stepwise model then reassigns bonds for these atoms to reconstruct the molecular structure. For the stepwise model with structure correction, additional bond removal is performed for carbon atoms exhibiting large prediction errors (MAE > 10 ppm) and their neighbouring N, O, and F atoms. The stepwise model subsequently reassigns bonds for these atoms to correct and complete the structure.

A multi-shot noise-augmented strategy was applied to inverse-IMPRESSION models trained on NMR Datasets B and C. For each NMR input, 200 structural candidates are generated by adding gaussian noise to the input chemical shifts and scalar coupling constants, with standard deviations of 0.3 ppm for $^{1}$H, 2.5 ppm for $^{13}$C, and 7.5 ppm for $^{15}$N and $^{19}$F, and 0.4 Hz for all coupling constants between NMR-active nuclei. After noise augmentation, the resulting edges with coupling constants of over 2 Hz are assigned as the corresponding coupling correlation type (see Supplementary Section 1.2.5 for details), with the remainder assigned as uncorrelated (value of 0). Each noise-perturbed input yields a single predicted structure, producing an ensemble of candidate structures for a given set of NMR parameters. The chemical validity of the generated candidates was assessed using the sanitization function implemented in the RDKit package.

The prediction performance of inverse-IMPRESSION models are evaluated by the following metrics:

- **Single prediction accuracy**: the fraction of tests in which the correct structure is predicted using the original NMR parameters without noise augmentation.
- **Validity**: the fraction of tests where at least one chemically valid structure appears among the 200 generated candidates.
- **Top-200 accuracy**: the fraction of tests where the correct structure appears among the 200 generated candidates prior to ranking, which indicates the upper bound of the ranking's ability to recover the correct structure.
- **Top-10 frequency accuracy**: the fraction of tests where the correct structure appears within the 10 most frequent candidates.
- **Top-3 frequency accuracy**: the fraction of tests where the correct structure appears within the 3 most frequent candidates.
- **Top-1 frequency accuracy**: the fraction of tests where the correct structure appears as the most frequent candidate.

**Supplementary Table 2** The performance metrics of inverse-IMPRESSION models trained with different NMR datasets

| Model | NMR dataset | Single prediction accuracy | Validity | Top-200 accuracy | Top-10 frequency accuracy | Top-3 frequency accuracy | Top-1 frequency accuracy |
|---|---|---|---|---|---|---|---|
| One-shot | A | 99.4% | - | - | - | - | - |
| | B | 5.3% | 6.8% | 6.5% | 6.5% | 6.5% | 6.5% |
| | C | 23.4% | 33.4% | 31.1% | 31.1% | 31.1% | 30.6% |
| Stepwise | B | 49.6% | 95.5% | 82.8% | 82.7% | 81.2% | 74.2% |
| | C | 59.4% | 94.4% | 90.0% | 90.0% | 89.5% | 86.2% |
| Stepwise with structure correction | B | 53.4% | 99.2% | 89.2% | 88.7% | 86.4% | 77.8% |
| | C | 64.4% | 99.5% | 93.6% | 93.3% | 91.7% | 86.2% |

## 2.2 Edge-level Analysis on One-Shot Models

An edge-level analysis was performed to evaluate the prediction accuracy for each edge type in the one-shot models. To ensure a concise and clear analysis, no noise was applied during prediction, and only one candidate was generated for each molecule. Two metrics are used to quantify the prediction accuracy:

- **False Negative Rate (FNR):** defined as FN/(TP+FN), where FN is the number of false negative edges among one edge type, and TP is the number of true positive edges. The FNR indicates the proportion of true bonds of each bond type that are not identified in ground-truth molecules.
- **False Discovery Rate (FDR):** defined as FP/(TP+FP), where FP is the number of false positive edges among on edge type, and TP is the number of true positive edges. The FDR indicates the proportion of predicted bonds of each bond type that are incorrectly identified, i.e., bonds predicted by the model that do not exist in the true molecule.

**Supplementary Table 3** False negative rates (FNR) and false discovery rates (FDR) associated with each edge type from the one-shot models

| Edge type | NMR Dataset A | | NMR Dataset B | | NMR Dataset C | |
|---|---|---|---|---|---|---|
| | FNR (%) | FDR (%) | FNR (%) | FDR (%) | FNR (%) | FDR (%) |
| **C-C** | 0.00 | 0.00 | 3.26 | 4.21 | 2.59 | 2.87 |
| **C-H** | 0.00 | 0.00 | 0.00 | 0.00 | 0.00 | 0.00 |
| **N-H** | 0.00 | 0.00 | 71.46 | 35.68 | 0.00 | 0.00 |
| **O-H** | 0.00 | 0.00 | 82.82 | 40.88 | 85.88 | 37.61 |
| **N-C** | 0.01 | 0.01 | 69.16 | 34.51 | 4.45 | 3.19 |
| **C-O** | 0.00 | 0.00 | 73.22 | 34.55 | 76.16 | 30.84 |
| **C-F** | 0.00 | 0.00 | 21.82 | 17.23 | 0.00 | 0.00 |

## 2.3 Top-k Accuracy Trends for Inverse-IMPRESSION Models

### 2.3.1 One-Shot Model B

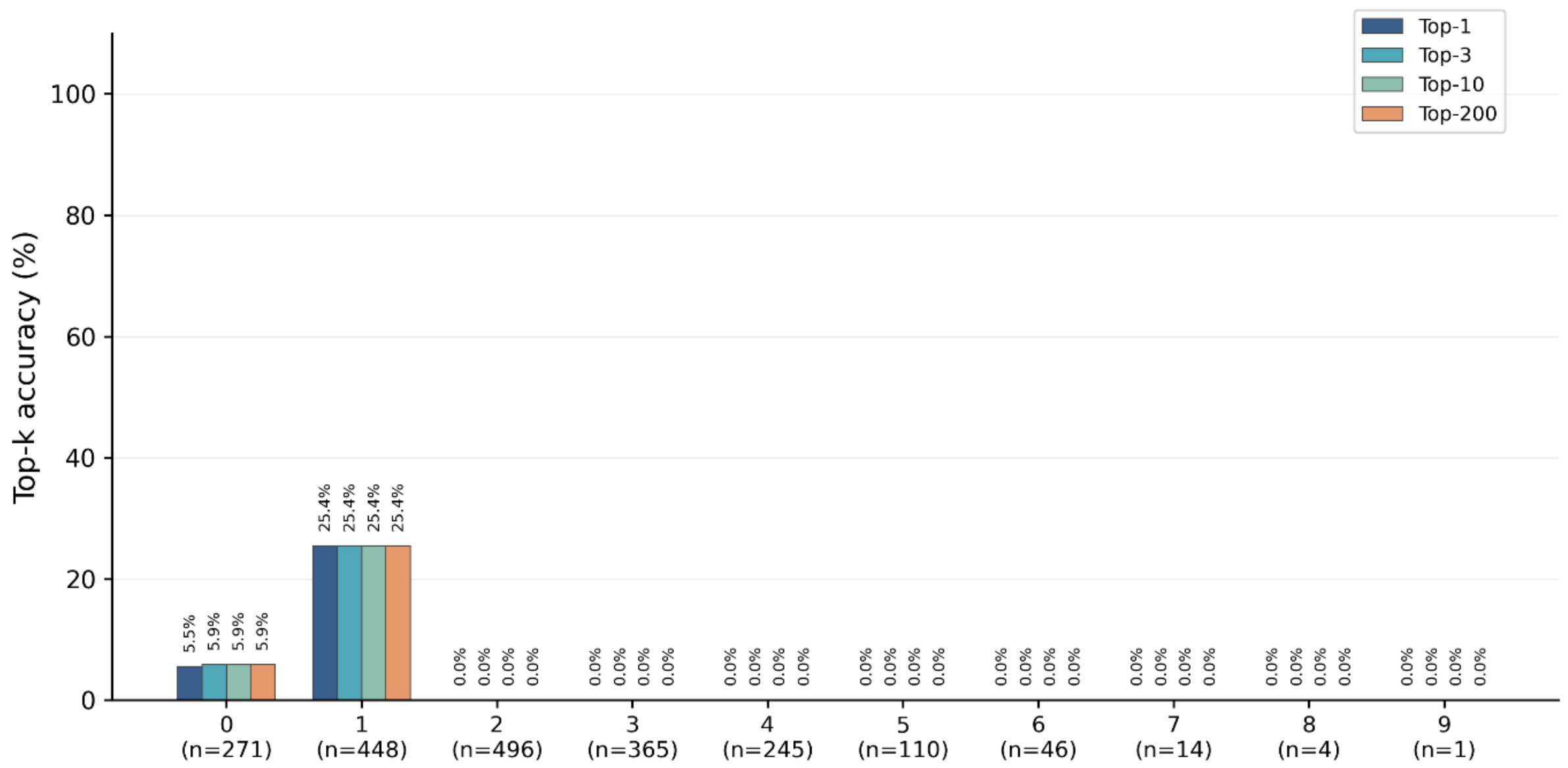

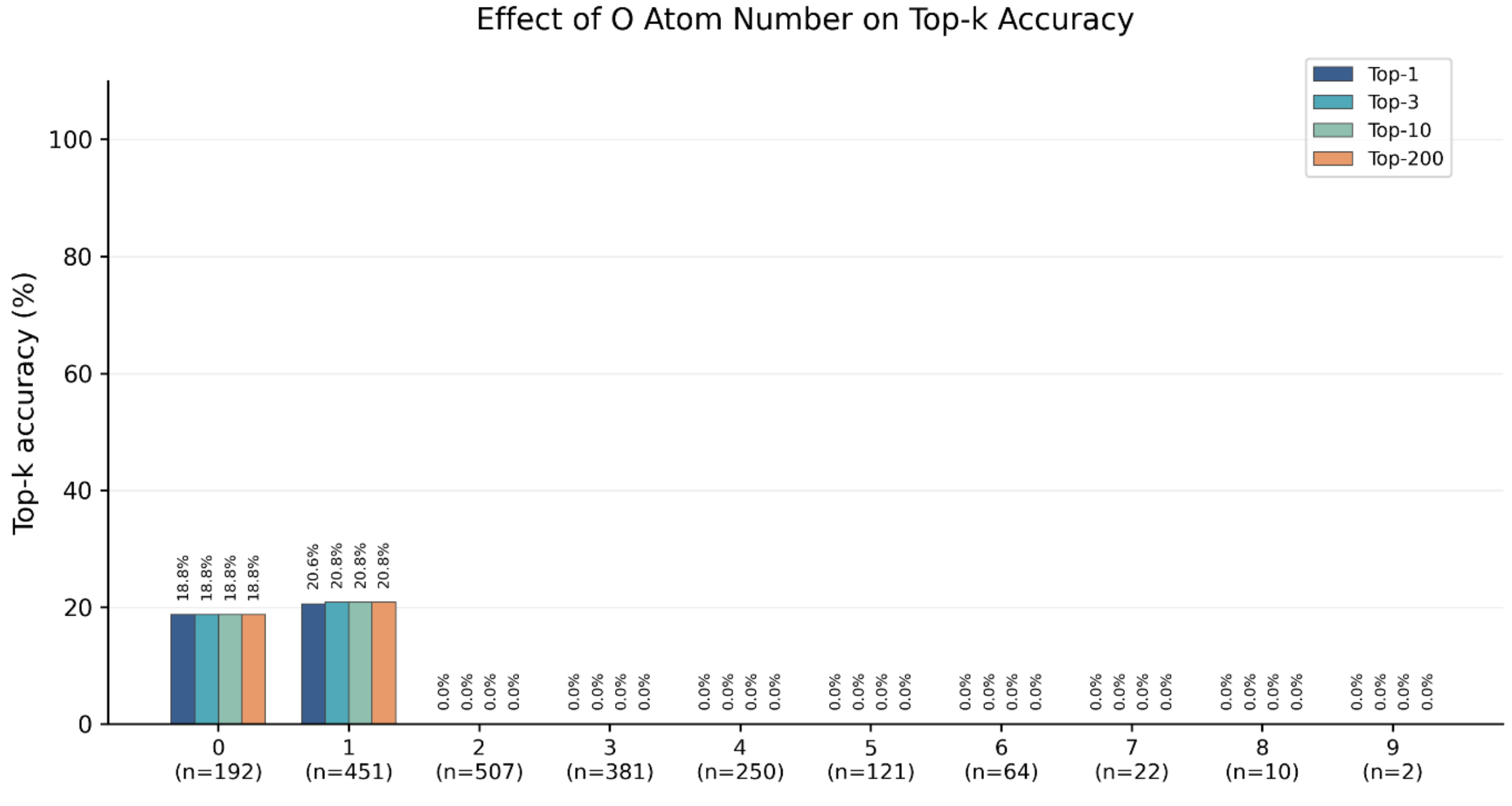

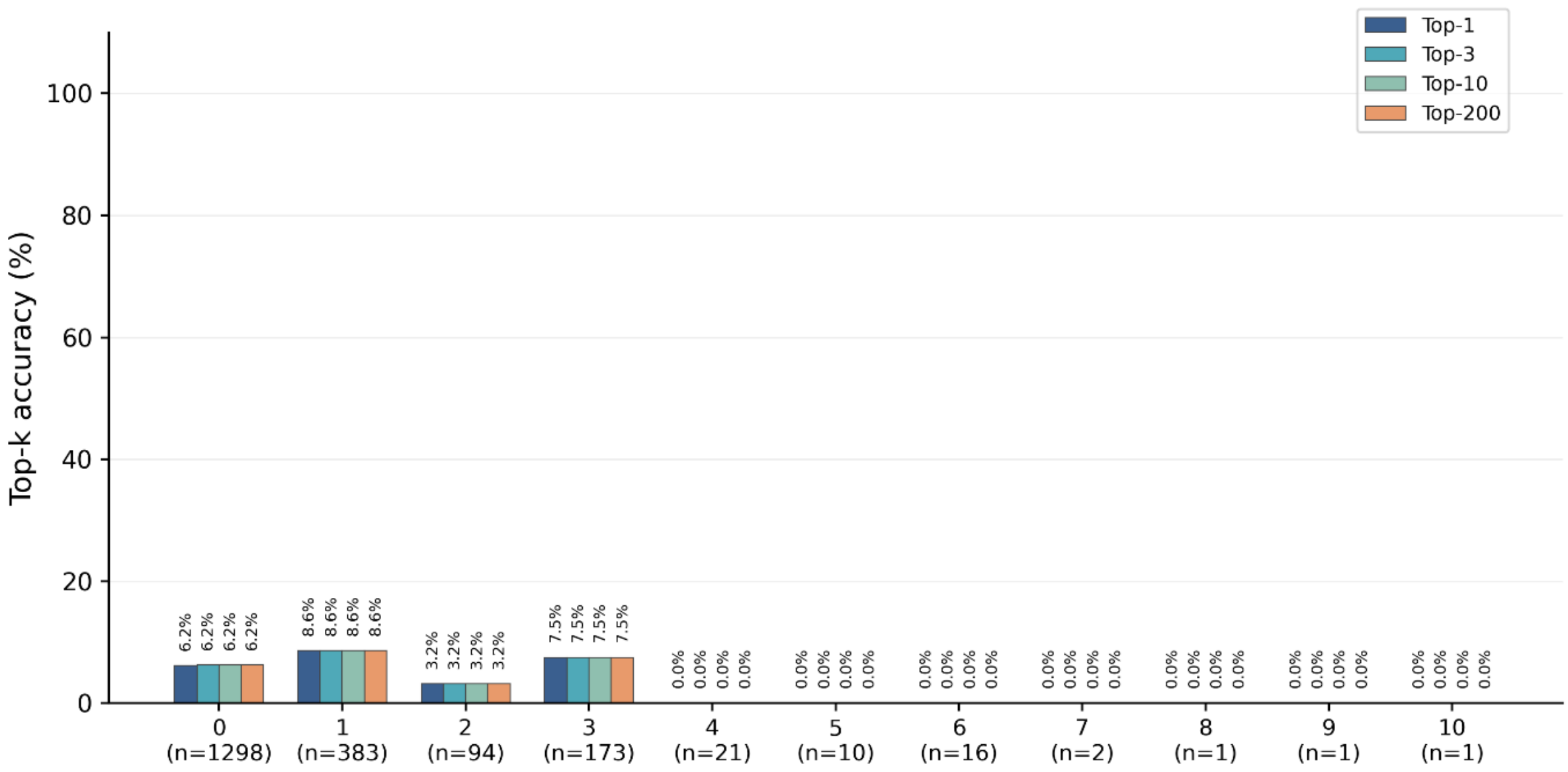
One-shot Model B:
Effect of F Atom Number on Top-k Accuracy
Top-1
Top-3
Top-10
Top-200
Top-k accuracy (%)
6.2% 6.2% 6.2% 6.2%
8.6% 8.6% 8.6% 8.6%
3.2% 3.2% 3.2% 3.2%
7.5% 7.5% 7.5% 7.5%
0 (n=1298)
1 (n=383)
2 (n=94)
3 (n=173)
4 (n=21)
5 (n=10)
6 (n=16)
7 (n=2)
8 (n=1)
9 (n=1)
10 (n=1)
Number of F atoms (n = total number of molecules)

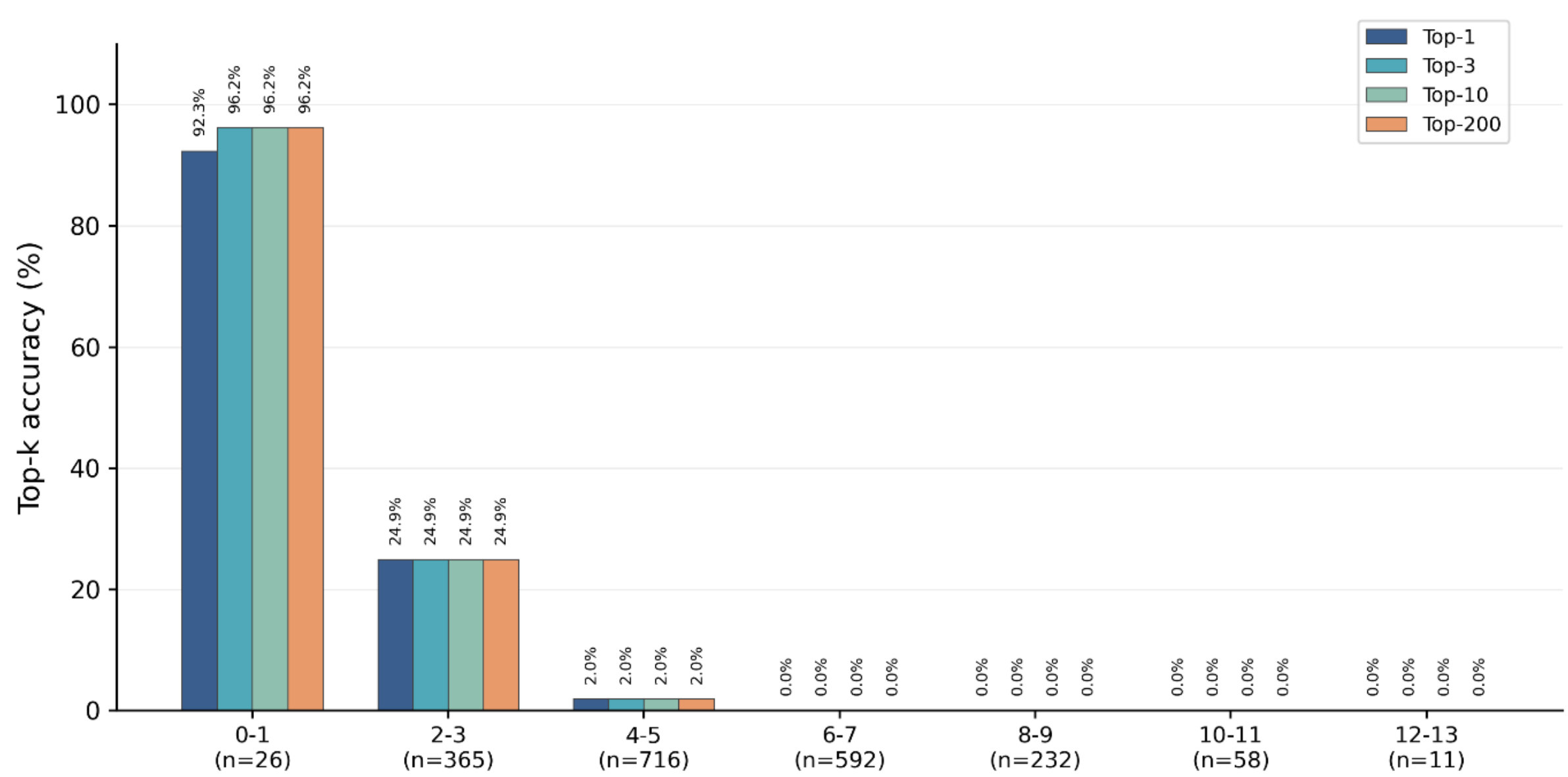
One-shot Model B:
Effect of Heteroatom (N, O and F) Number on Top-k Accuracy
Top-1
Top-3
Top-10
Top-200
Top-k accuracy (%)
92.3% 96.2% 96.2% 96.2%
24.9% 24.9% 24.9% 24.9%
2.0% 2.0% 2.0% 2.0%
0-1 (n=26)
2-3 (n=365)
4-5 (n=716)
6-7 (n=592)
8-9 (n=232)
10-11 (n=58)
12-13 (n=11)
Number of heteroatoms (n = total number of molecules)

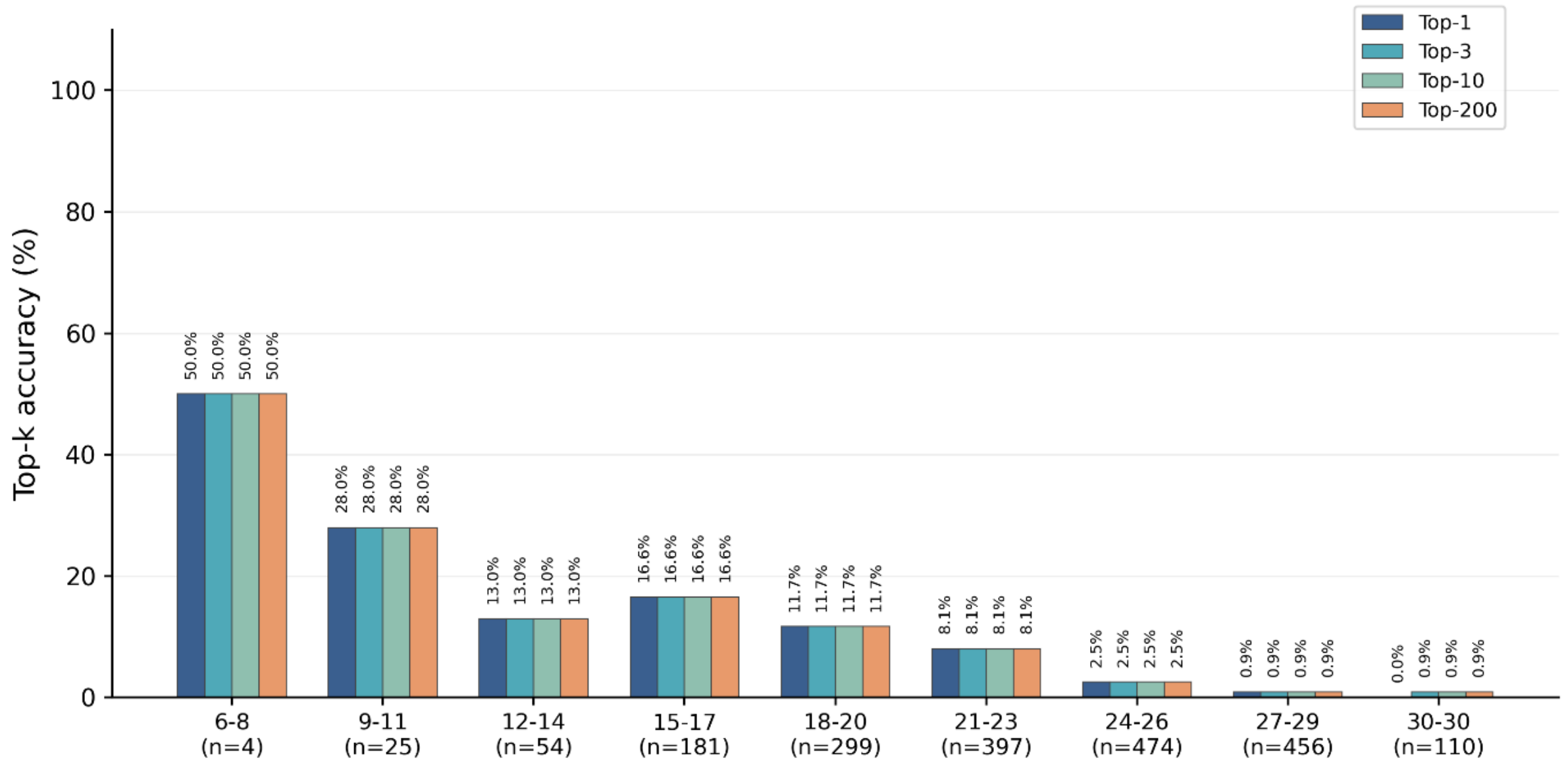
One-shot Model B:
Effect of Heavy Atom (C, N, O and F) Number on Top-k Accuracy
Top-1
Top-3
Top-10
Top-200
Top-k accuracy (%)
0
20
40
60
80
100
50.0%
28.0%
13.0%
16.6%
11.7%
8.1%
2.5%
0.9%
0.0%
6-8 (n=4)
9-11 (n=25)
12-14 (n=54)
15-17 (n=181)
18-20 (n=299)
21-23 (n=397)
24-26 (n=474)
27-29 (n=456)
30-30 (n=110)
Number of heavy atoms (n = total number of molecules)

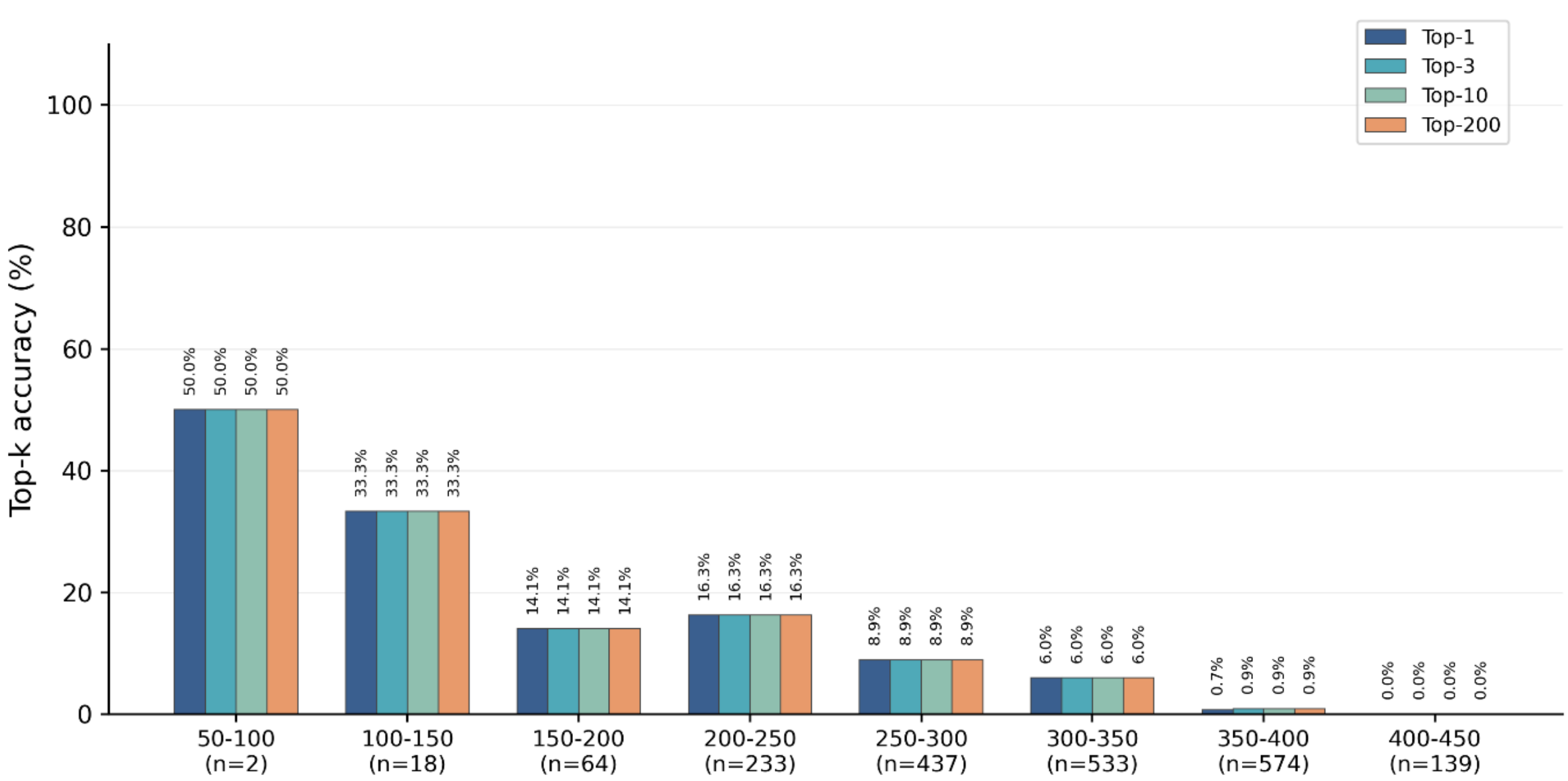
One-shot Model B:
Effect of Molecular Weight on Top-k Accuracy
Top-1
Top-3
Top-10
Top-200
Top-k accuracy (%)
0
20
40
60
80
100
50.0%
33.3%
14.1%
16.3%
8.9%
6.0%
0.7%
0.9%
0.0%
50-100 (n=2)
100-150 (n=18)
150-200 (n=64)
200-250 (n=233)
250-300 (n=437)
300-350 (n=533)
350-400 (n=574)
400-450 (n=139)
Molecular weight (Da, n = total number of molecules)

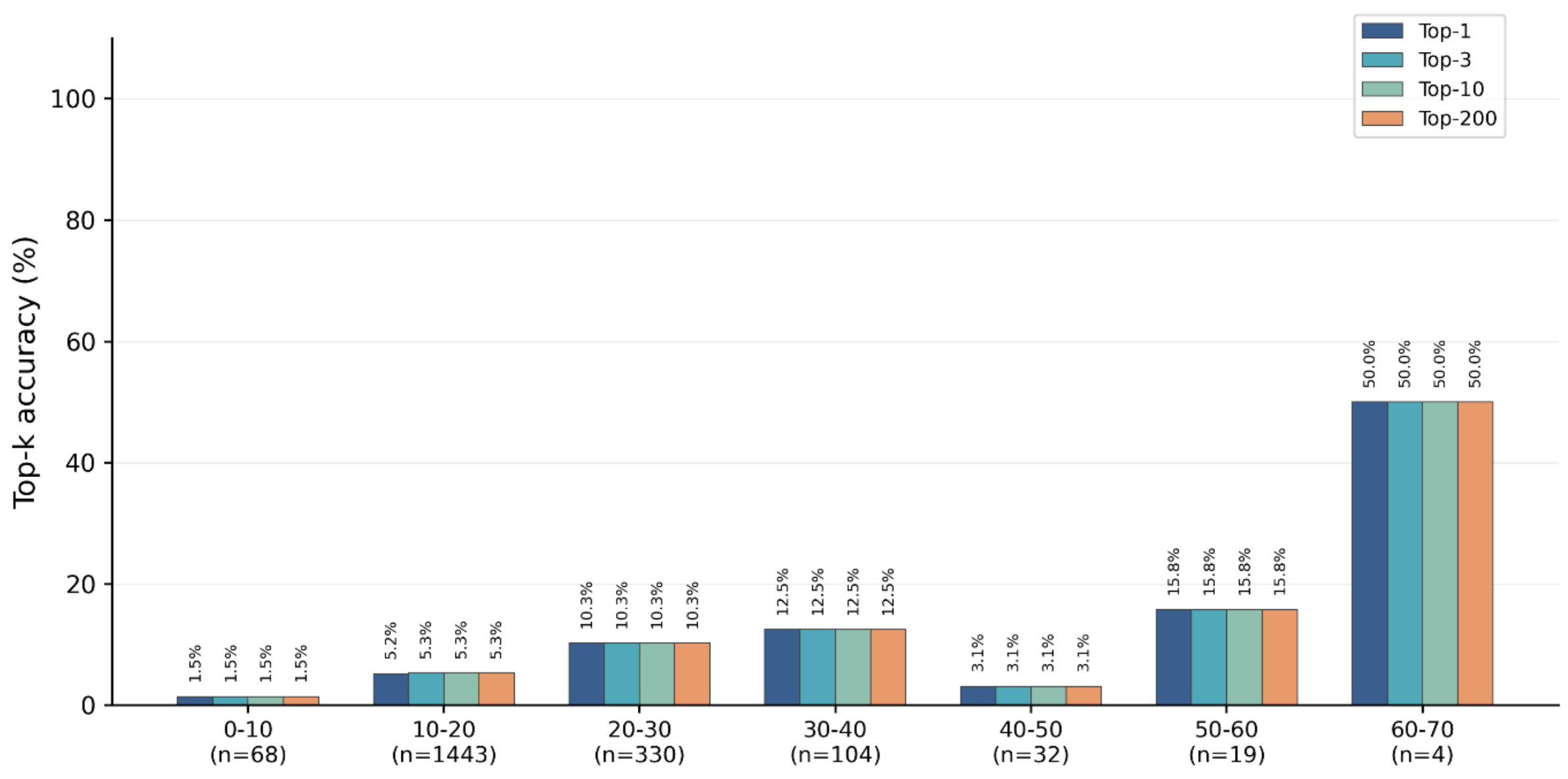
One-shot Model B:
Effect of Molecular Complexity (nSPS) on Top-k Accuracy
Top-1
Top-3
Top-10
Top-200
Top-k accuracy (%)
0
20
40
60
80
100
1.5%
5.2%
5.3%
10.3%
12.5%
3.1%
15.8%
50.0%
0-10 (n=68)
10-20 (n=1443)
20-30 (n=330)
30-40 (n=104)
40-50 (n=32)
50-60 (n=19)
60-70 (n=4)
nSPS (n = total number of molecules)

### 2.3.2 One-Shot Model C

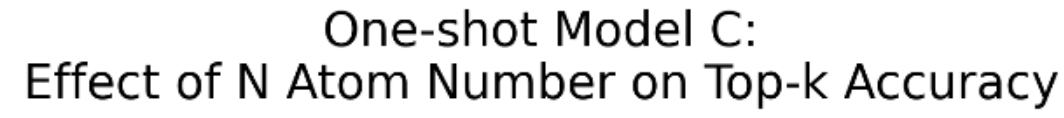

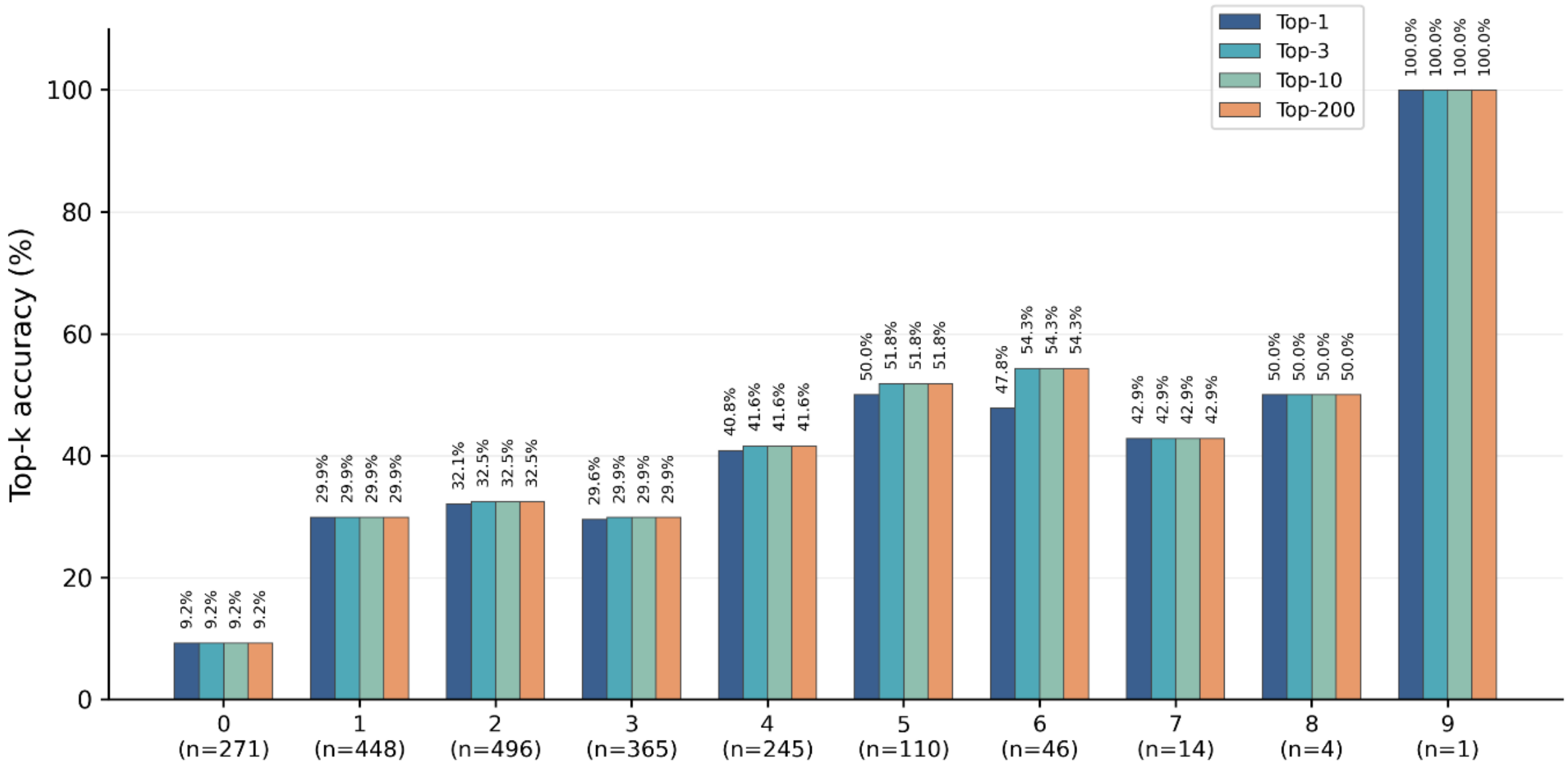

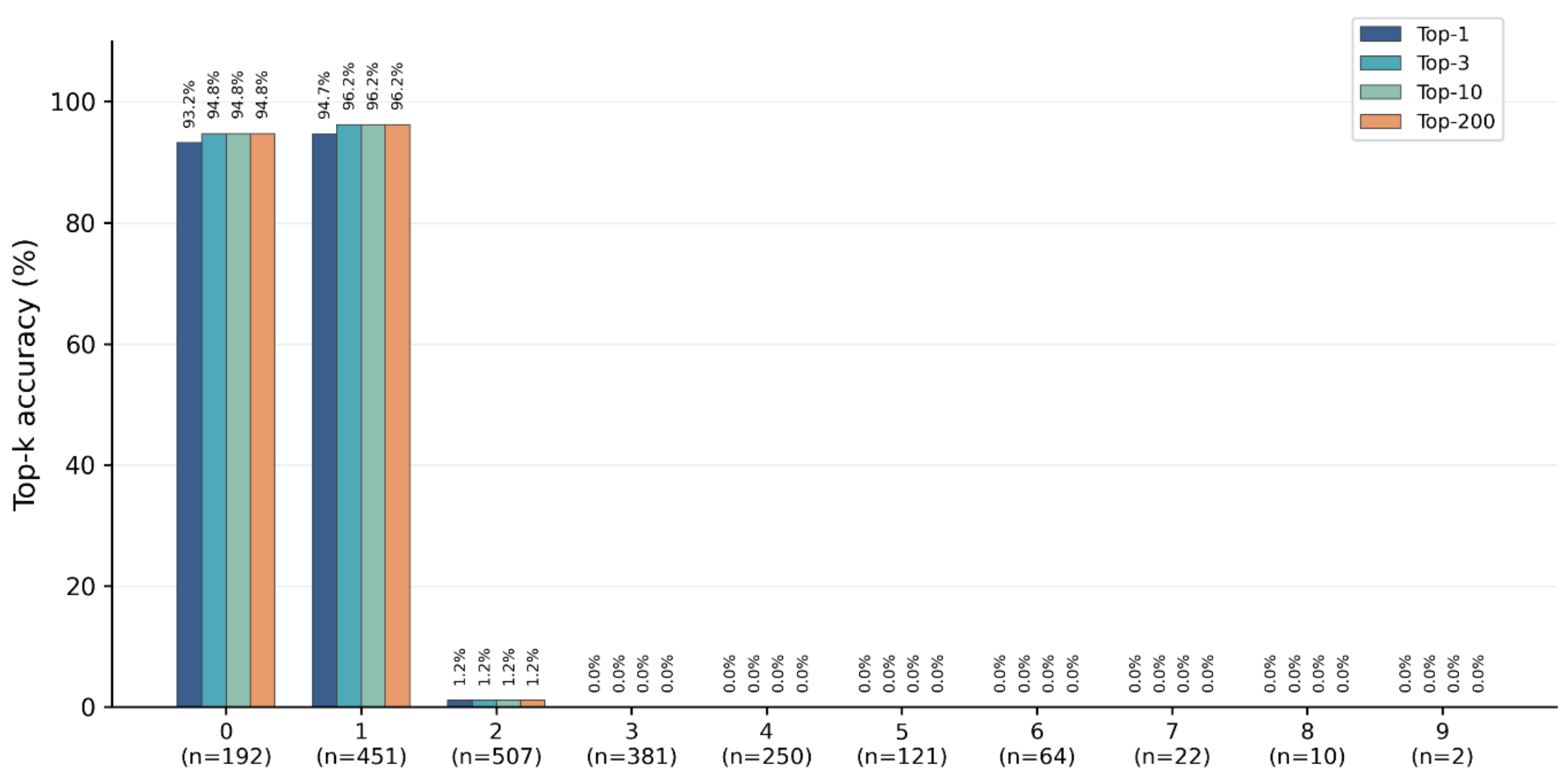

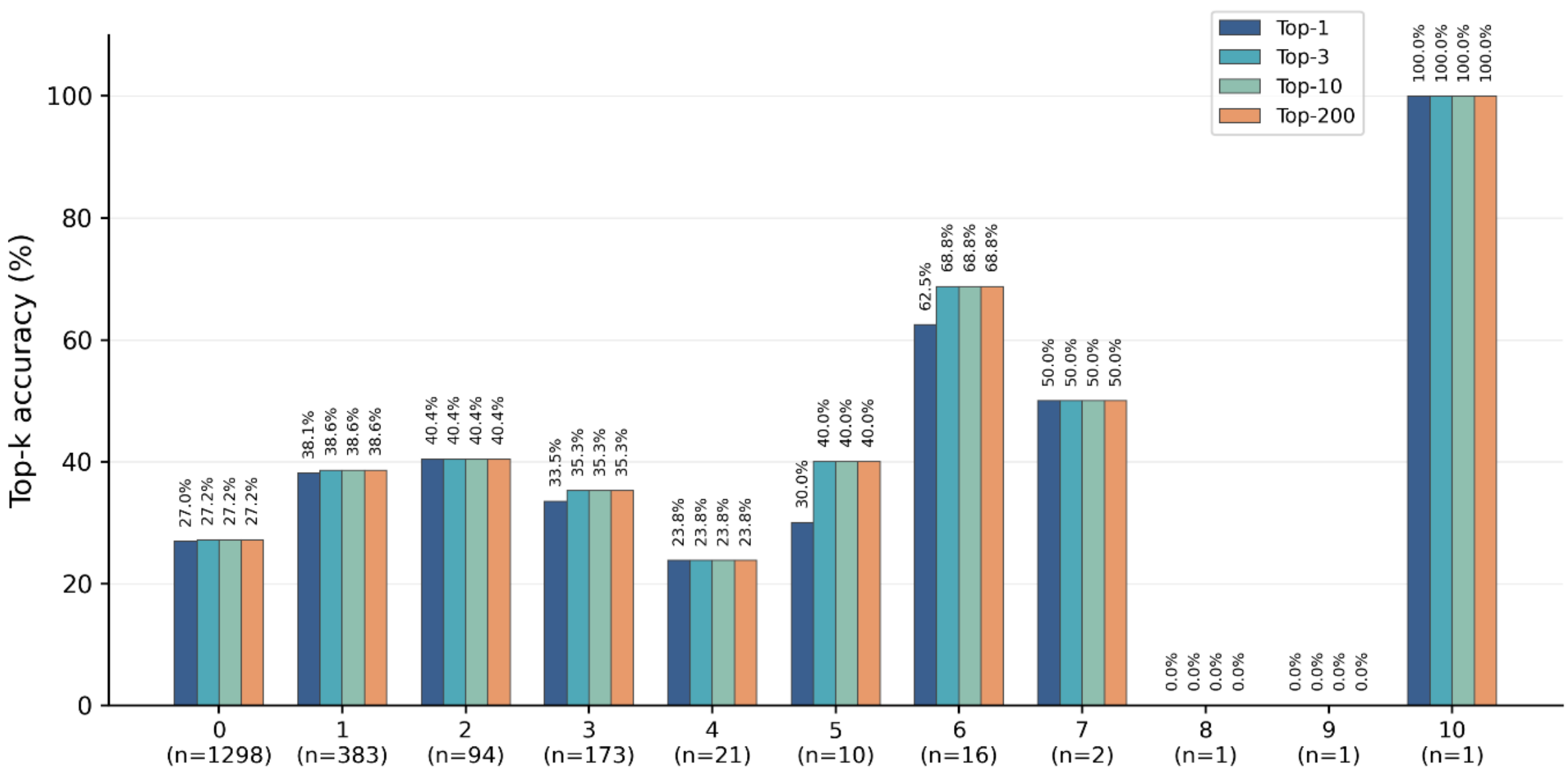
One-shot Model C:
Effect of F Atom Number on Top-k Accuracy
Top-1
Top-3
Top-10
Top-200
Top-k accuracy (%)
Number of F atoms (n = total number of molecules)
0 (n=1298)
1 (n=383)
2 (n=94)
3 (n=173)
4 (n=21)
5 (n=10)
6 (n=16)
7 (n=2)
8 (n=1)
9 (n=1)
10 (n=1)
27.0% 27.2% 27.2% 27.2%
38.1% 38.6% 38.6% 38.6%
40.4% 40.4% 40.4% 40.4%
33.5% 35.3% 35.3% 35.3%
23.8% 23.8% 23.8% 23.8%
30.0% 40.0% 40.0% 40.0%
62.5% 68.8% 68.8% 68.8%
50.0% 50.0% 50.0% 50.0%
0.0% 0.0% 0.0% 0.0%
0.0% 0.0% 0.0% 0.0%
100.0% 100.0% 100.0% 100.0%

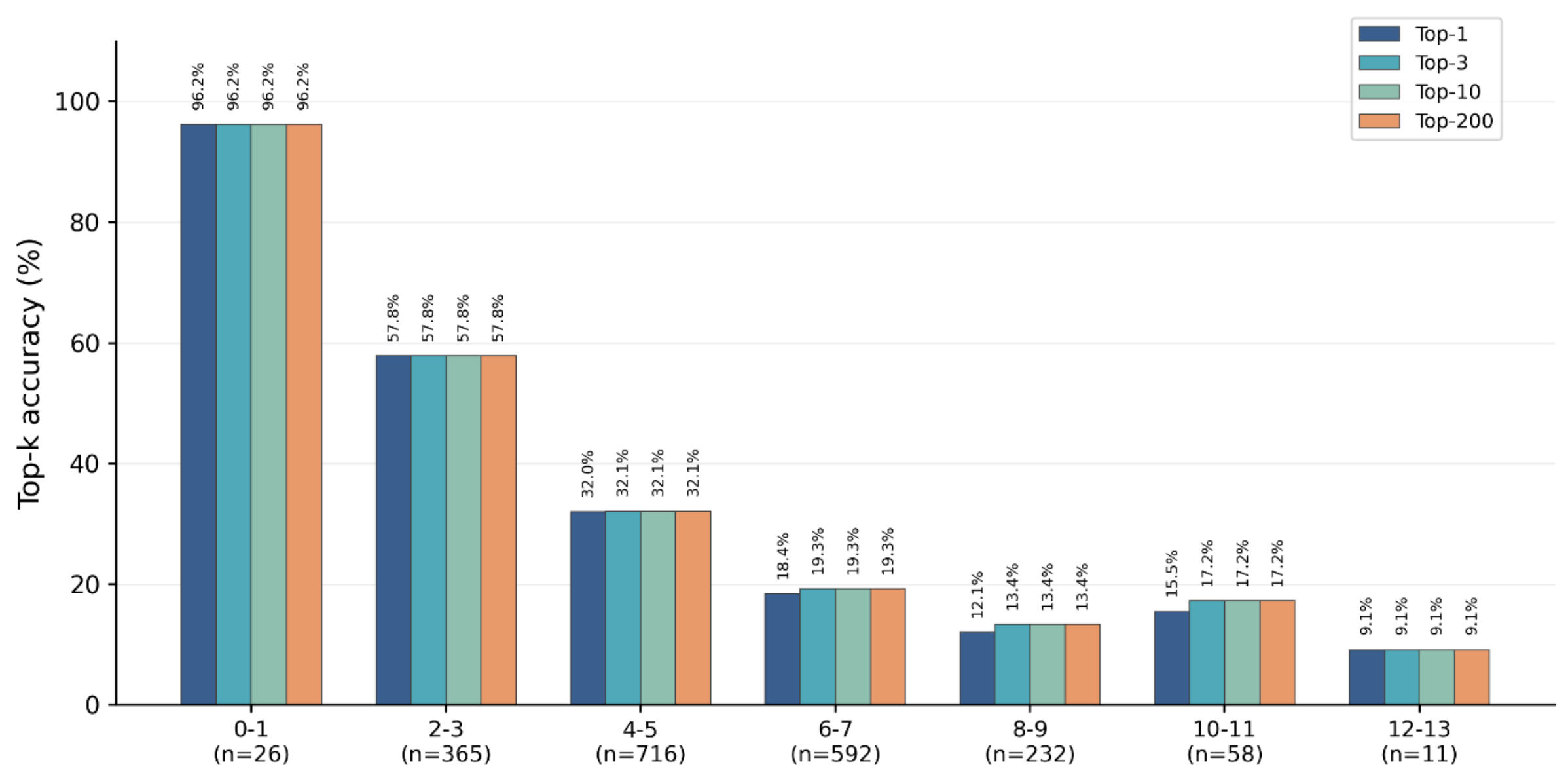
One-shot Model C:
Effect of Heteroatom (N, O and F) Number on Top-k Accuracy
Top-1
Top-3
Top-10
Top-200
Top-k accuracy (%)
Number of heteroatoms (n = total number of molecules)
0-1 (n=26)
2-3 (n=365)
4-5 (n=716)
6-7 (n=592)
8-9 (n=232)
10-11 (n=58)
12-13 (n=11)
96.2% 96.2% 96.2% 96.2%
57.8% 57.8% 57.8% 57.8%
32.0% 32.1% 32.1% 32.1%
18.4% 19.3% 19.3% 19.3%
12.1% 13.4% 13.4% 13.4%
15.5% 17.2% 17.2% 17.2%
9.1% 9.1% 9.1% 9.1%

One-shot Model C:
Effect of Heavy Atom (C, N, O and F) Number on Top-k Accuracy

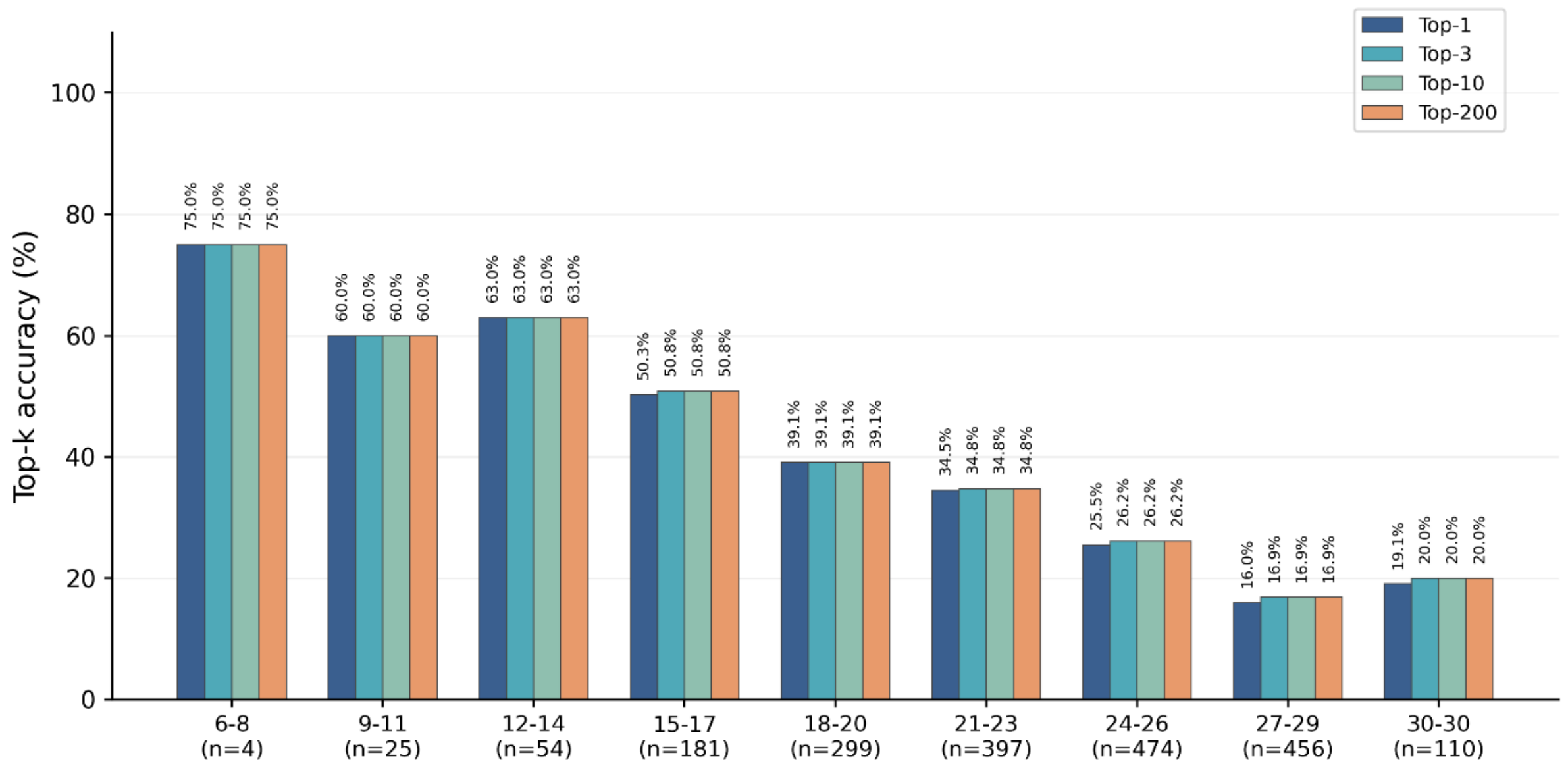

Number of heavy atoms (n = total number of molecules)

One-shot Model C:

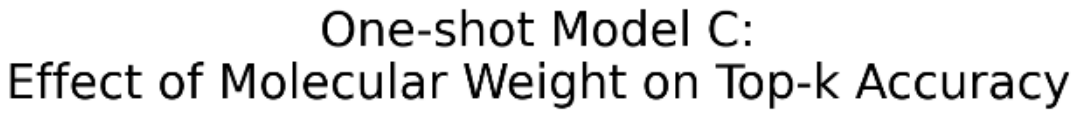

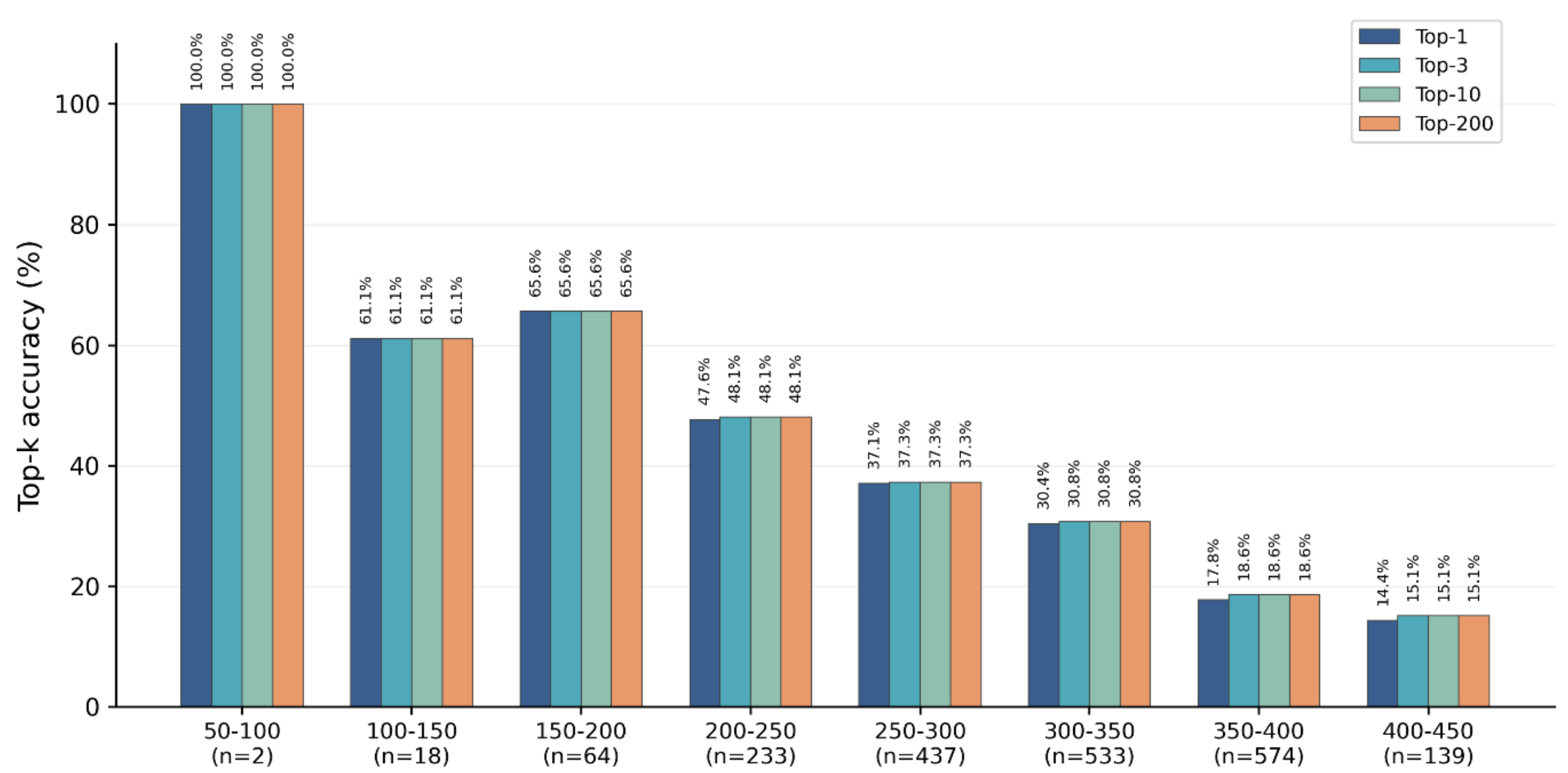

Molecular weight (Da, n = total number of molecules)

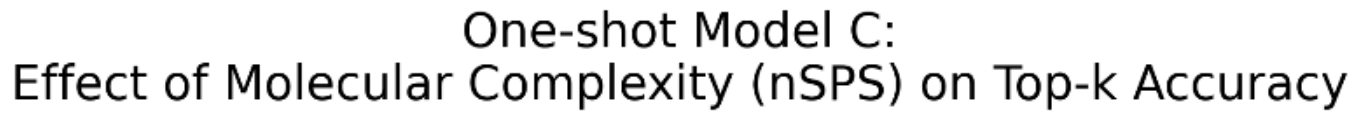
One-shot Model C:
Effect of Molecular Complexity (nSPS) on Top-k Accuracy

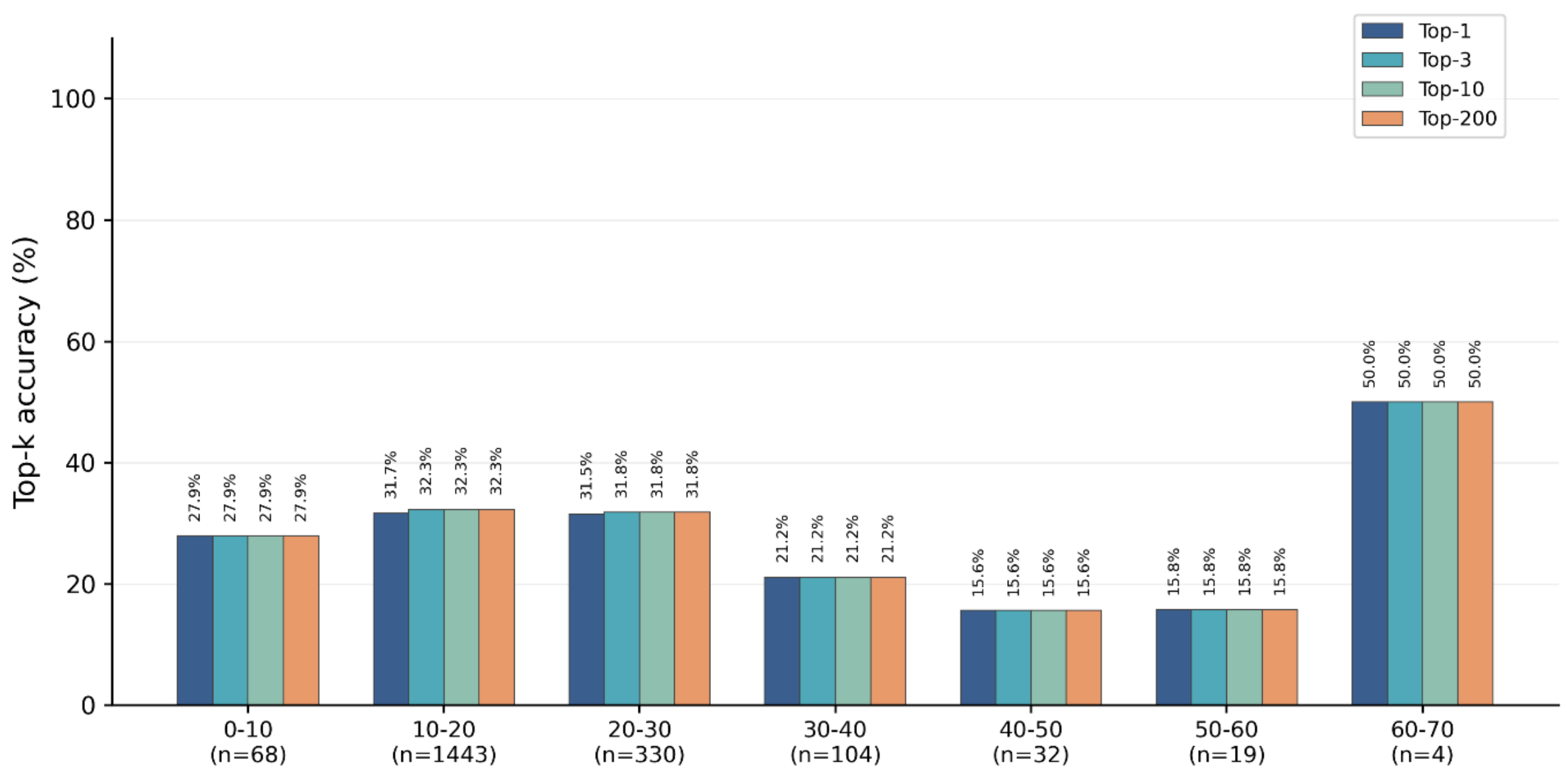
Top-1
Top-3
Top-10
Top-200
Top-k accuracy (%)
0
20
40
60
80
100
27.9% 27.9% 27.9% 27.9%
31.7% 32.3% 32.3% 32.3%
31.5% 31.8% 31.8% 31.8%
21.2% 21.2% 21.2% 21.2%
15.6% 15.6% 15.6% 15.6%
15.8% 15.8% 15.8% 15.8%
50.0% 50.0% 50.0% 50.0%
0-10 (n=68)
10-20 (n=1443)
20-30 (n=330)
30-40 (n=104)
40-50 (n=32)
50-60 (n=19)
60-70 (n=4)
nSPS (n = total number of molecules)

### 2.3.3 Stepwise Model B With Structure Correction

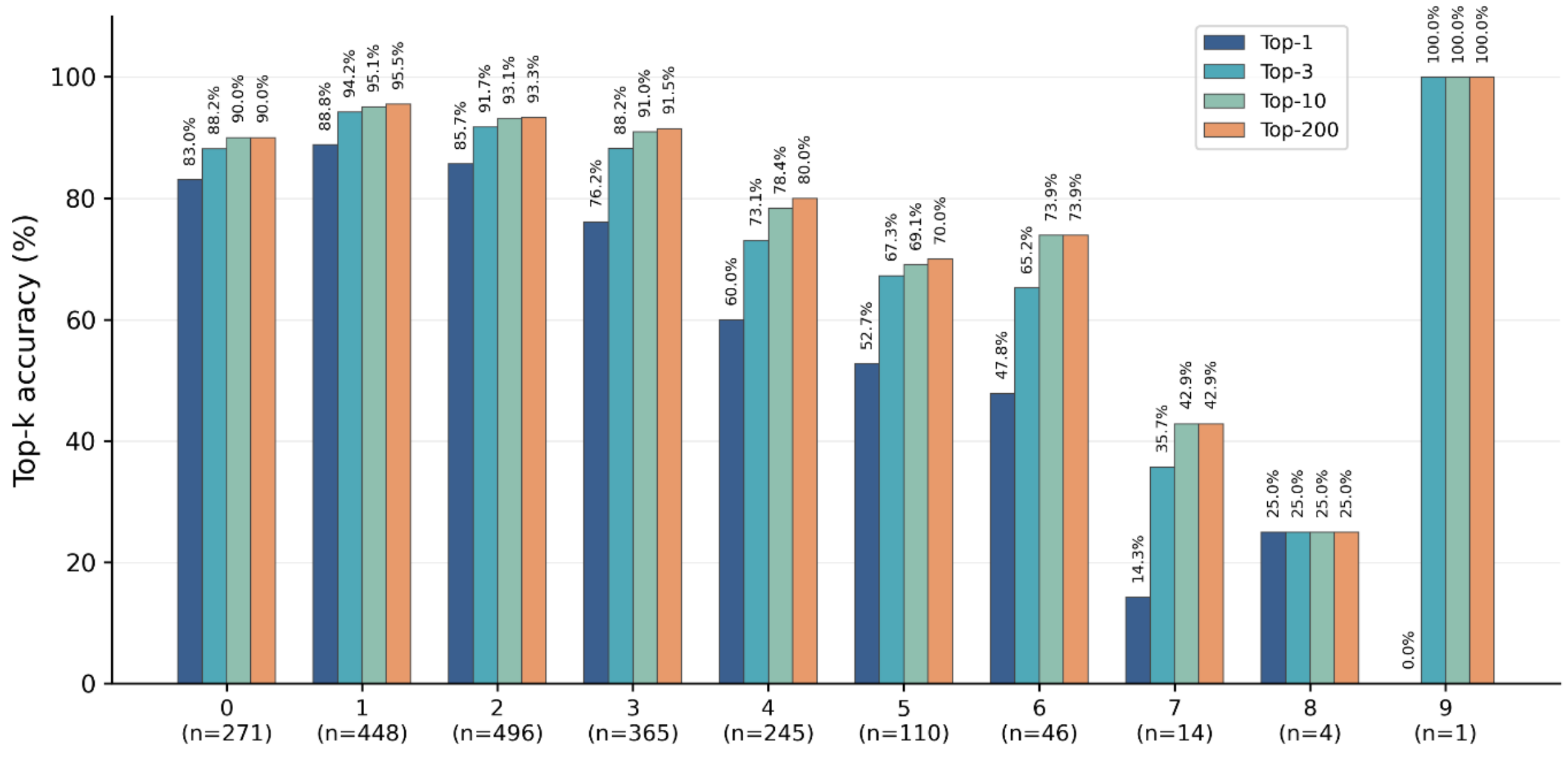

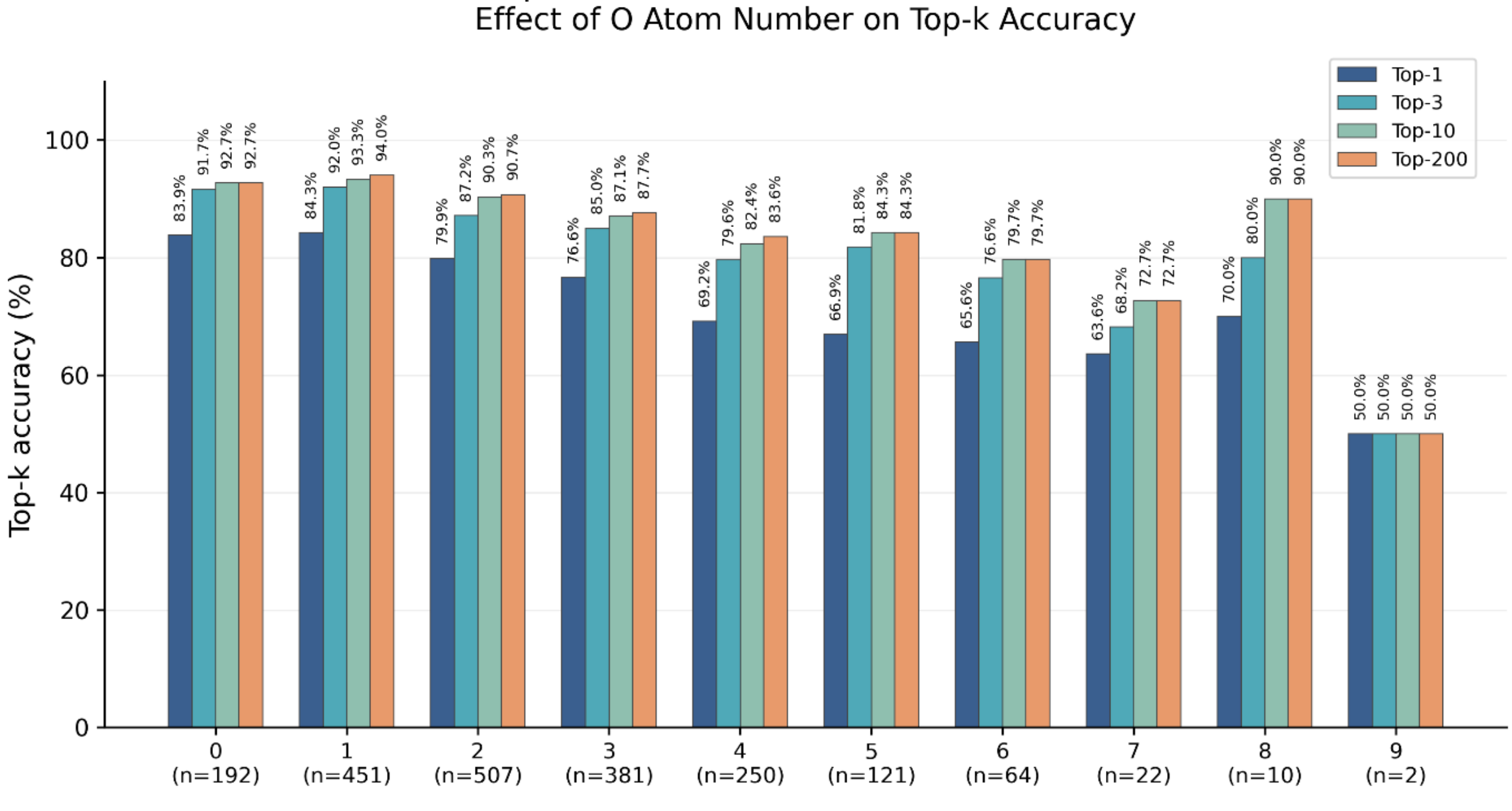

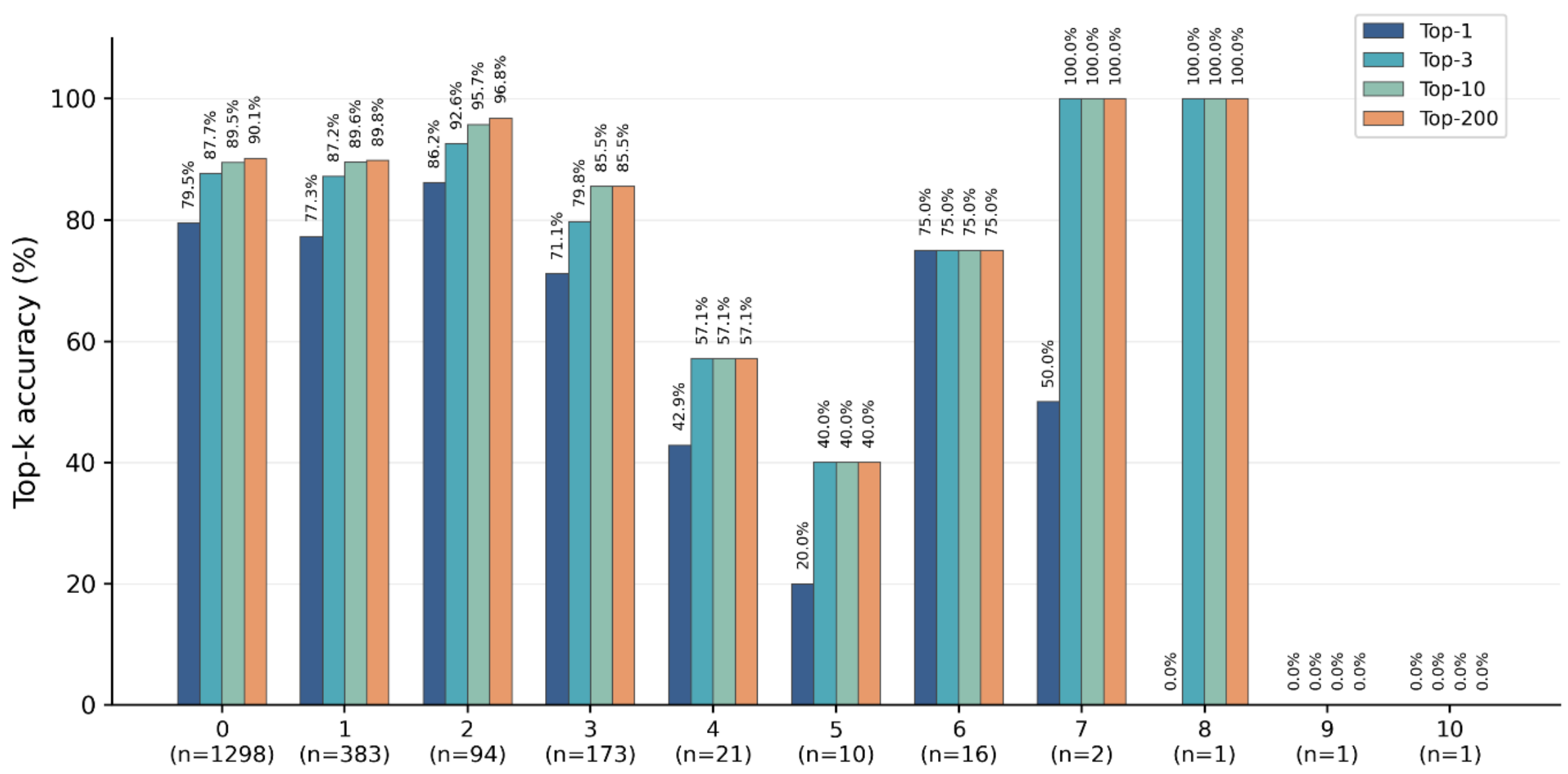
Stepwise Model B with Structure Correction:
Effect of F Atom Number on Top-k Accuracy
Top-1
Top-3
Top-10
Top-200
Top-k accuracy (%)
Number of F atoms (n = total number of molecules)
0 (n=1298)
1 (n=383)
2 (n=94)
3 (n=173)
4 (n=21)
5 (n=10)
6 (n=16)
7 (n=2)
8 (n=1)
9 (n=1)
10 (n=1)
79.5% 87.7% 89.5% 90.1%
77.3% 87.2% 89.6% 89.8%
86.2% 92.6% 95.7% 96.8%
71.1% 79.8% 85.5% 85.5%
42.9% 57.1% 57.1% 57.1%
20.0% 40.0% 40.0% 40.0%
75.0% 75.0% 75.0% 75.0%
50.0% 100.0% 100.0% 100.0%
0.0% 100.0% 100.0% 100.0%
0.0% 0.0% 0.0% 0.0%
0.0% 0.0% 0.0% 0.0%

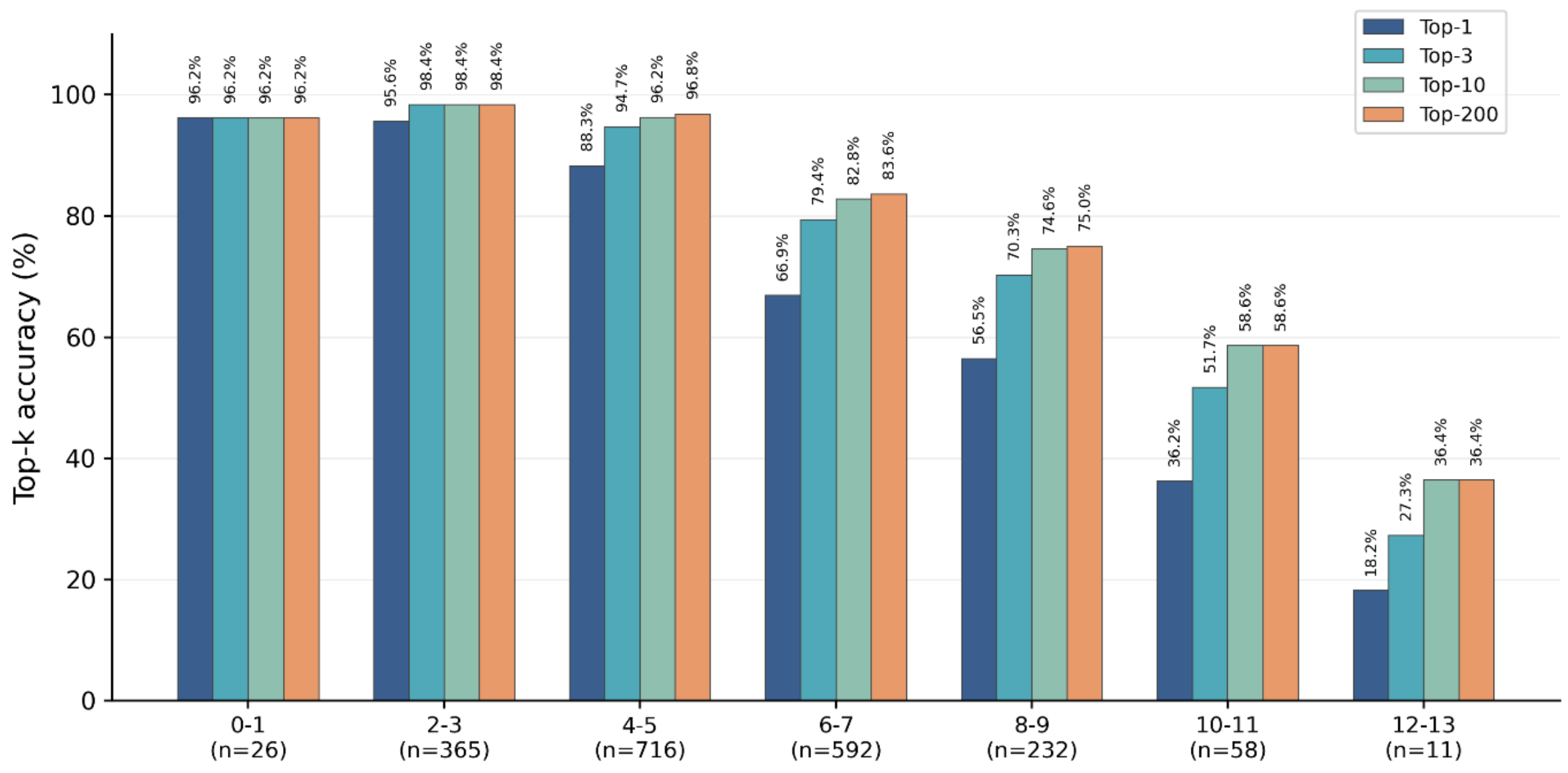
Stepwise Model B with Structure Correction:
Effect of Heteroatom (N, O and F) Number on Top-k Accuracy
Top-1
Top-3
Top-10
Top-200
Top-k accuracy (%)
Number of heteroatoms (n = total number of molecules)
0-1 (n=26)
2-3 (n=365)
4-5 (n=716)
6-7 (n=592)
8-9 (n=232)
10-11 (n=58)
12-13 (n=11)
96.2% 96.2% 96.2% 96.2%
95.6% 98.4% 98.4% 98.4%
88.3% 94.7% 96.2% 96.8%
66.9% 79.4% 82.8% 83.6%
56.5% 70.3% 74.6% 75.0%
36.2% 51.7% 58.6% 58.6%
18.2% 27.3% 36.4% 36.4%

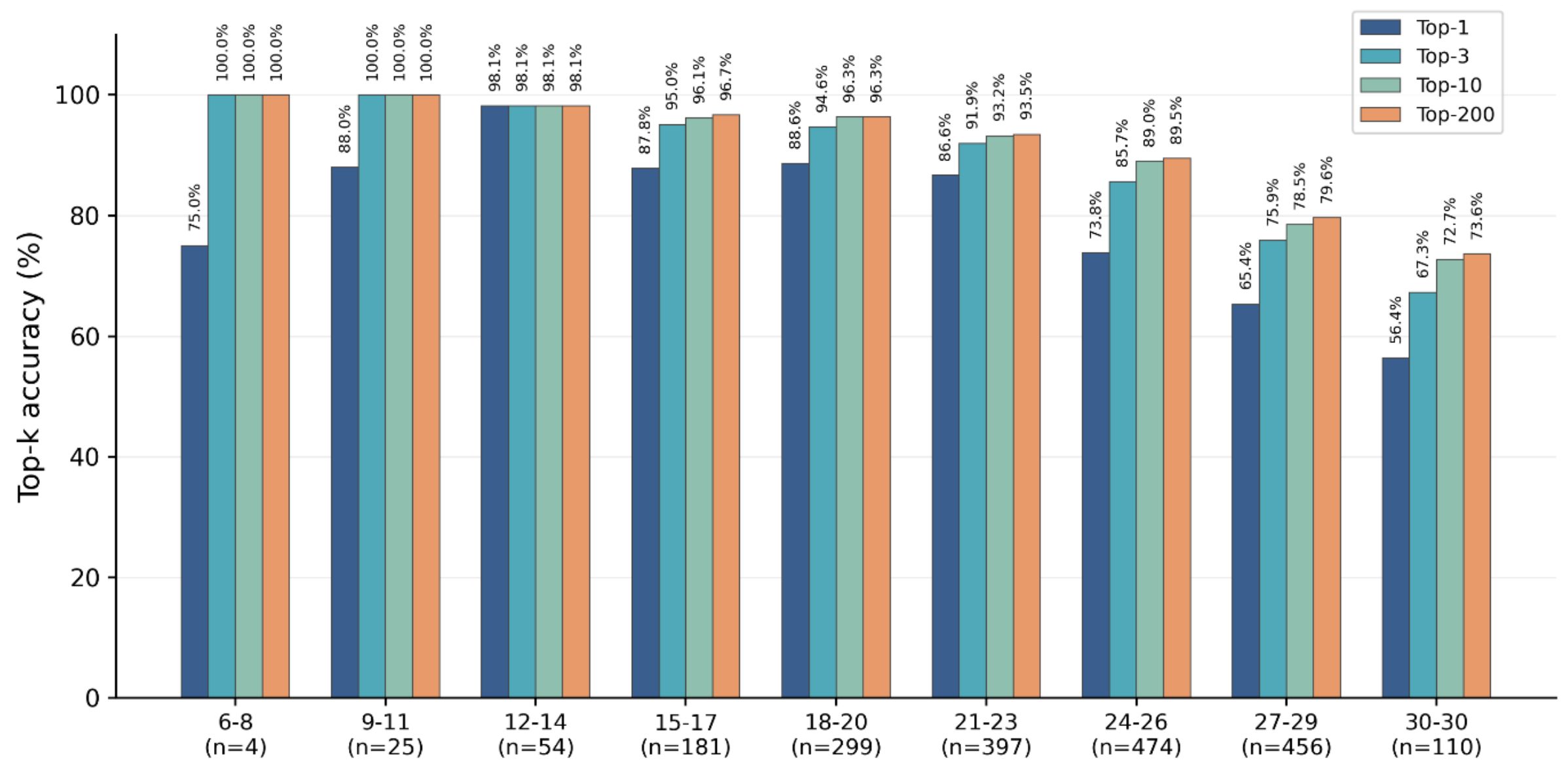
Stepwise Model B with Structure Correction:
Effect of Heavy Atom (C, N, O and F) Number on Top-k Accuracy
Top-1
Top-3
Top-10
Top-200
Top-k accuracy (%)
6-8 (n=4)
9-11 (n=25)
12-14 (n=54)
15-17 (n=181)
18-20 (n=299)
21-23 (n=397)
24-26 (n=474)
27-29 (n=456)
30-30 (n=110)
Number of heavy atoms (n = total number of molecules)

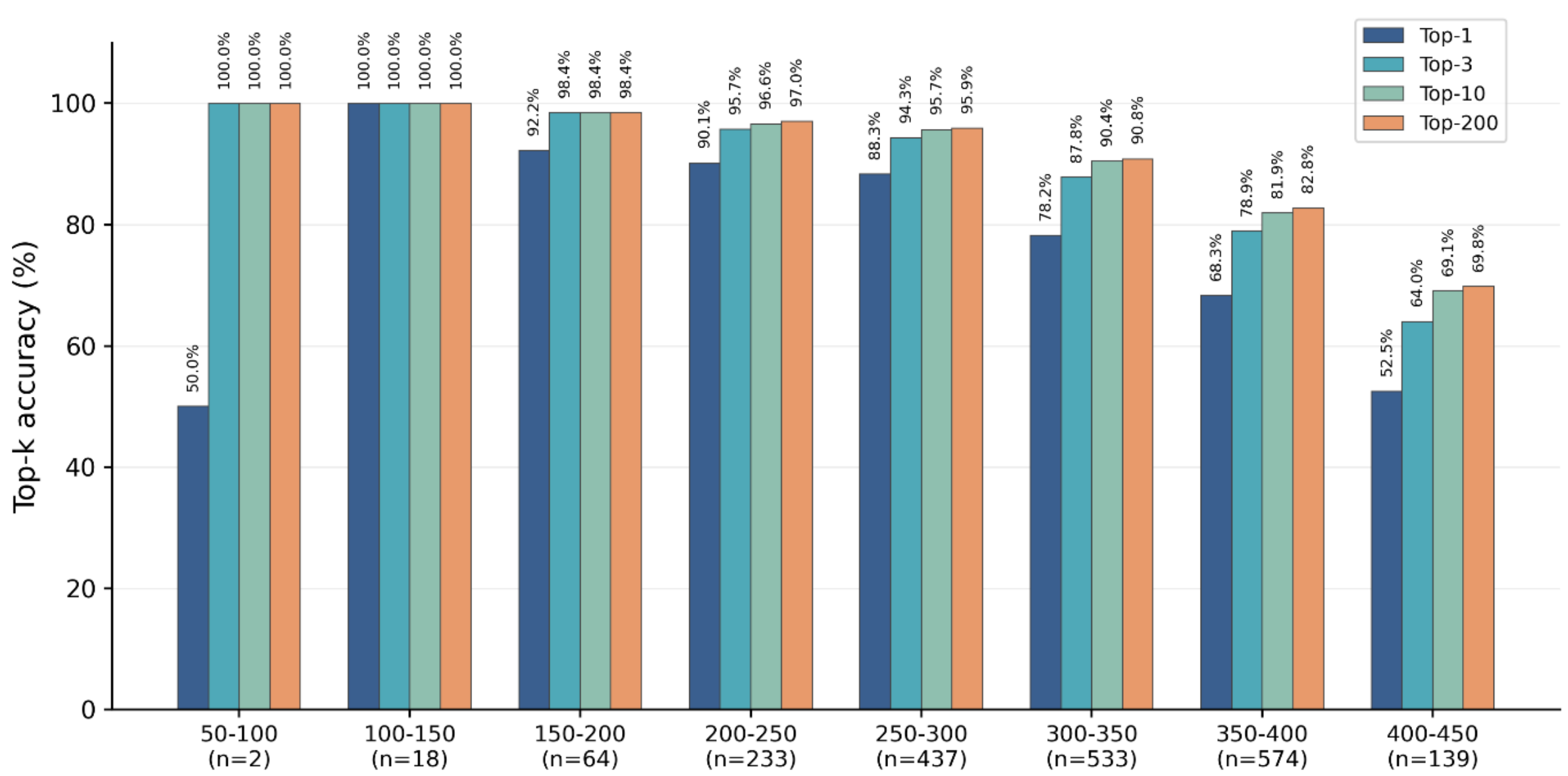
Stepwise Model B with Structure Correction:
Effect of Molecular Weight on Top-k Accuracy
Top-1
Top-3
Top-10
Top-200
Top-k accuracy (%)
50-100 (n=2)
100-150 (n=18)
150-200 (n=64)
200-250 (n=233)
250-300 (n=437)
300-350 (n=533)
350-400 (n=574)
400-450 (n=139)
Molecular weight (Da, n = total number of molecules)

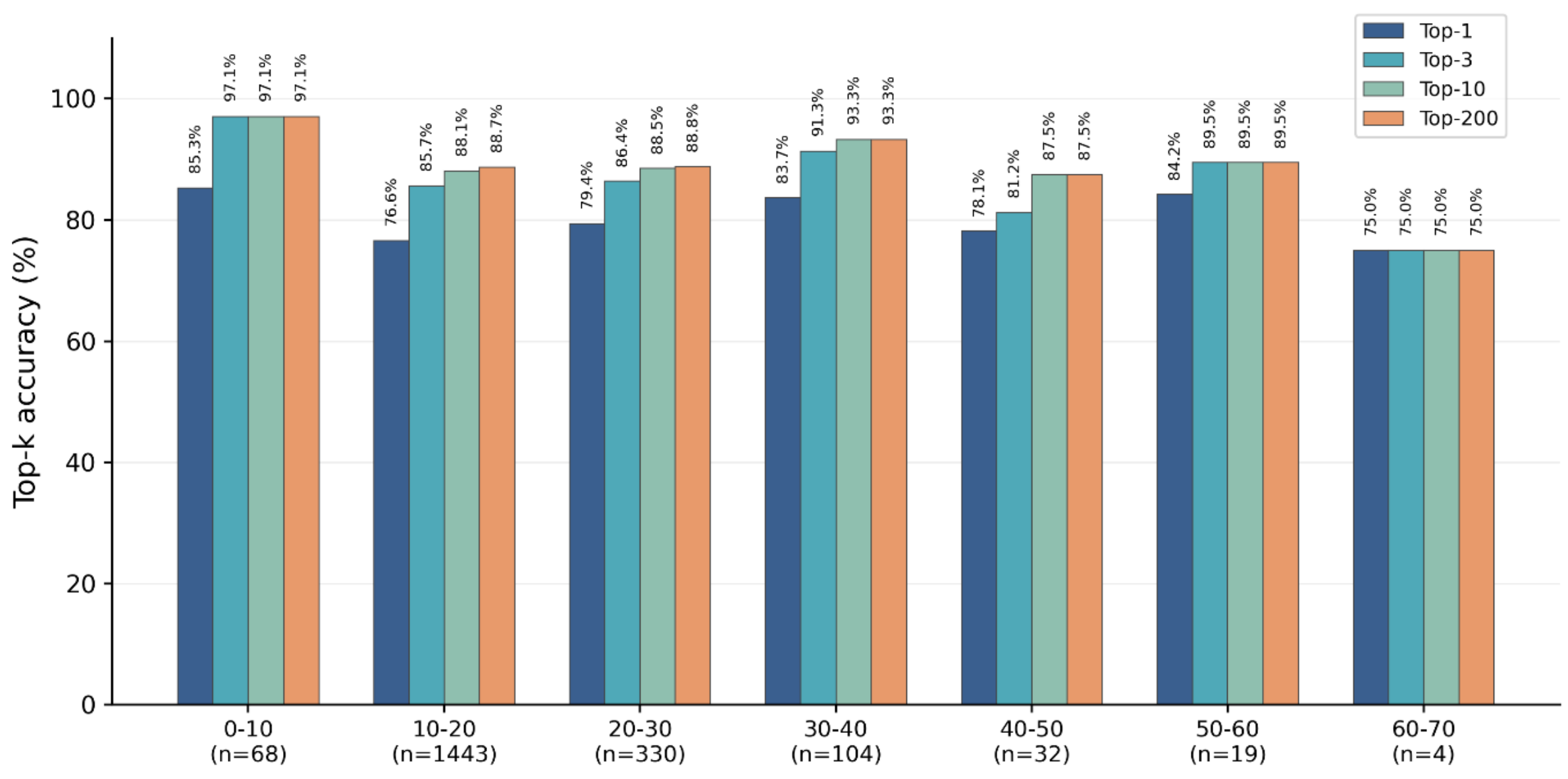
Stepwise Model B with Structure Correction:
Effect of Molecular Complexity (nSPS) on Top-k Accuracy
Top-1
Top-3
Top-10
Top-200
Top-k accuracy (%)
0
20
40
60
80
100
85.3%
97.1%
97.1%
97.1%
76.6%
85.7%
88.1%
88.7%
79.4%
86.4%
88.5%
88.8%
83.7%
91.3%
93.3%
93.3%
78.1%
81.2%
87.5%
87.5%
84.2%
89.5%
89.5%
89.5%
75.0%
75.0%
75.0%
75.0%
0-10 (n=68)
10-20 (n=1443)
20-30 (n=330)
30-40 (n=104)
40-50 (n=32)
50-60 (n=19)
60-70 (n=4)
nSPS (n = total number of molecules)

### 2.3.4 Stepwise Model C With Structure Correction

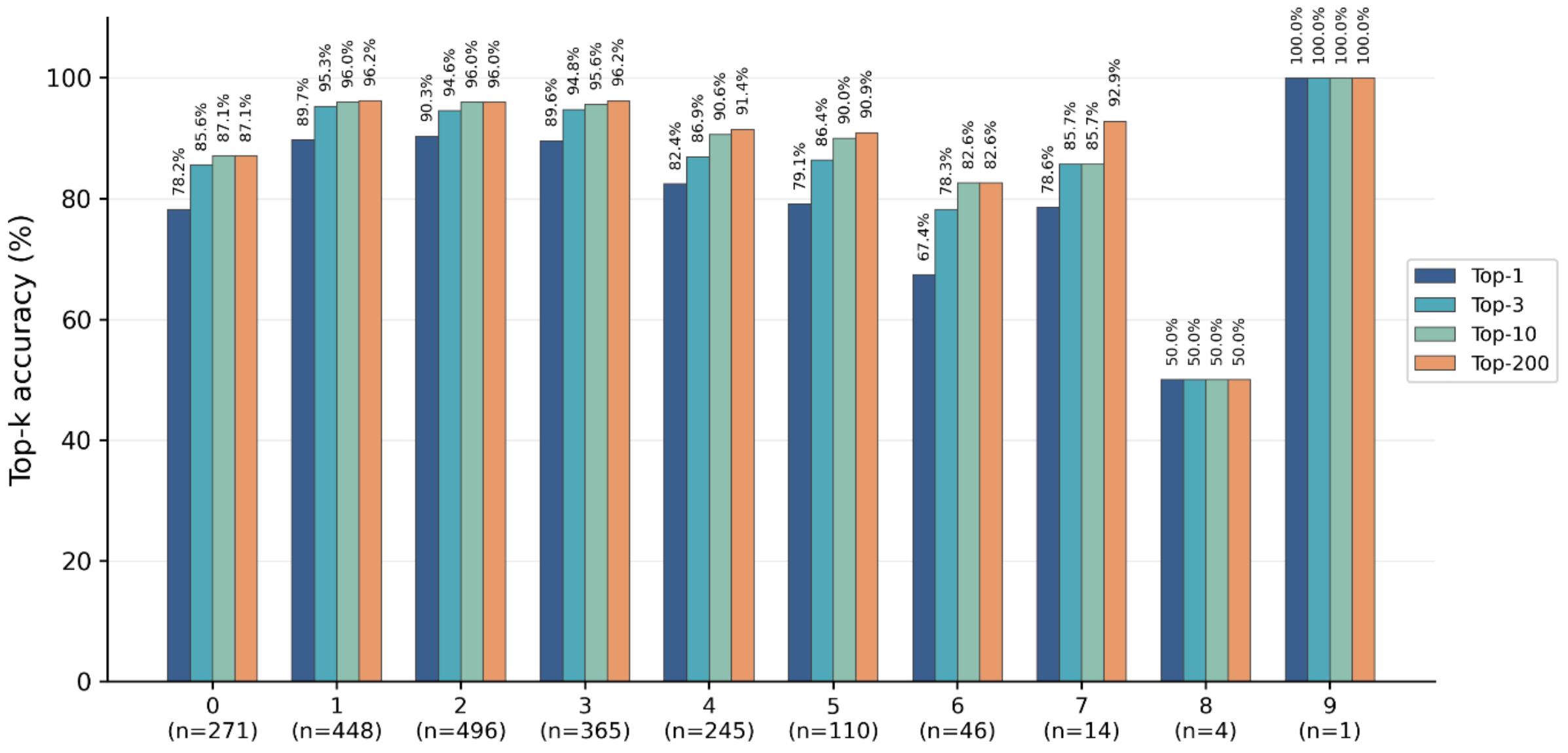

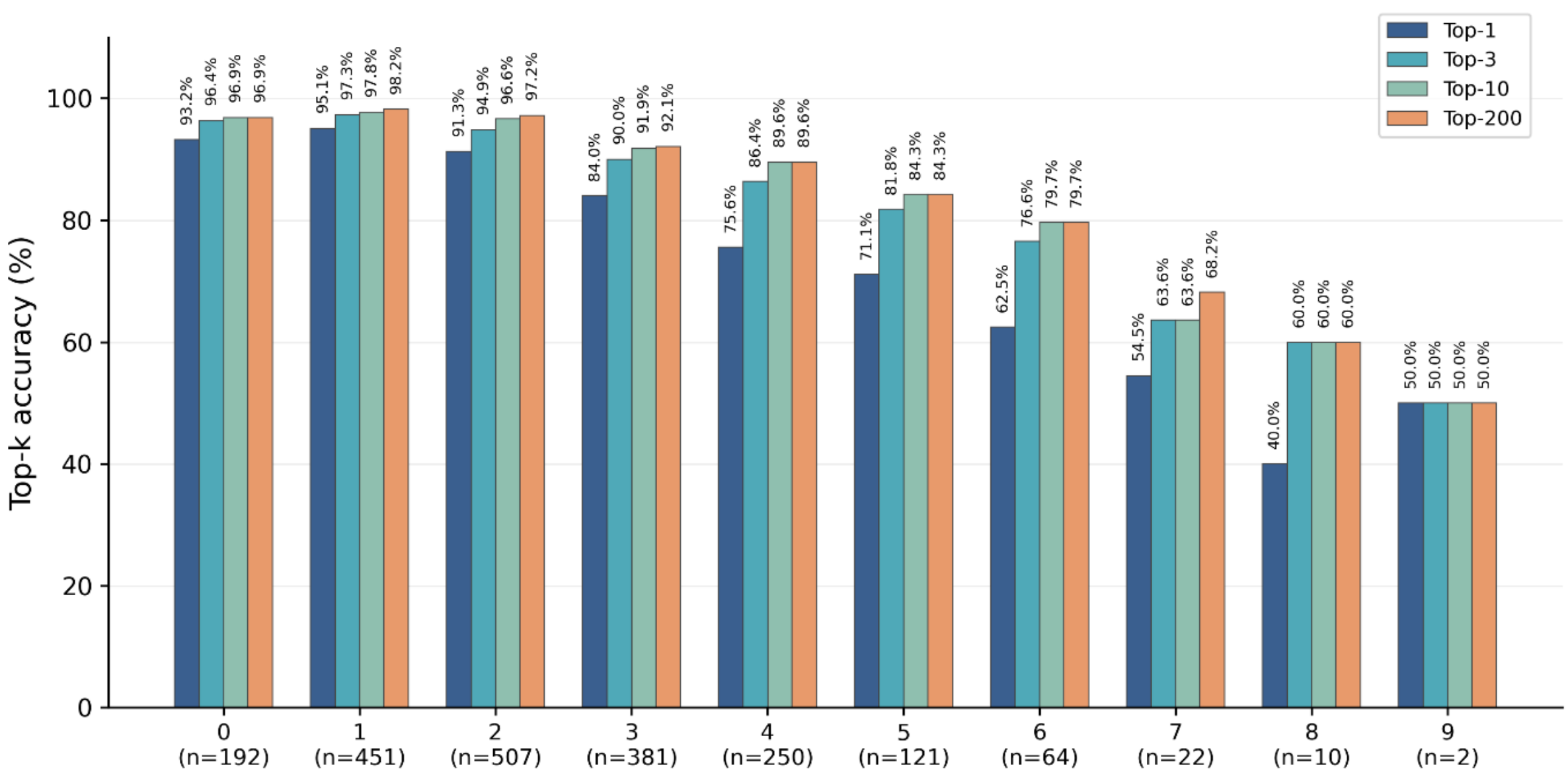

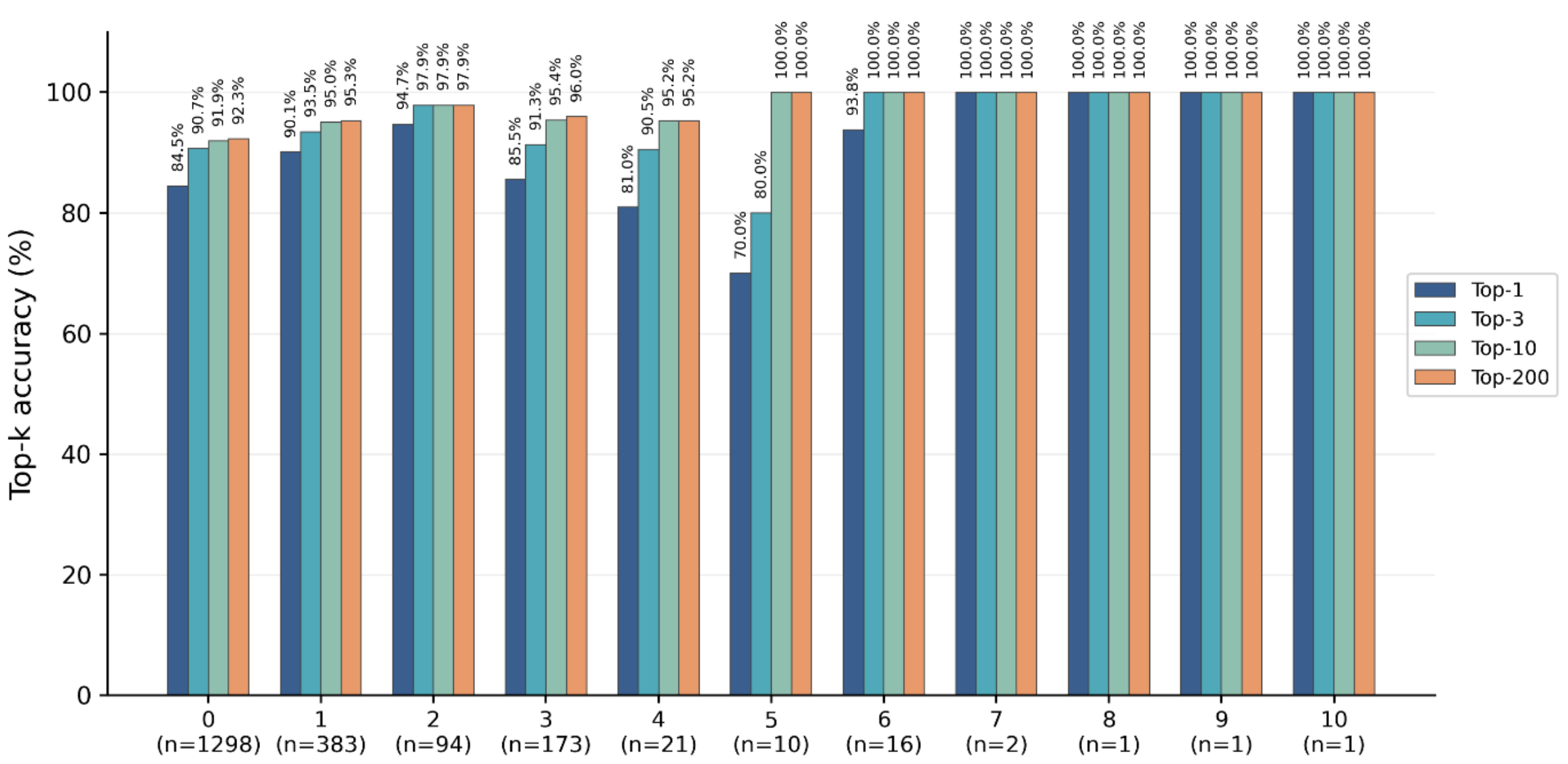
Stepwise Model C with Structure Correction:
Effect of F Atom Number on Top-k Accuracy
Top-k accuracy (%)
Number of F atoms (n = total number of molecules)
0 (n=1298)
1 (n=383)
2 (n=94)
3 (n=173)
4 (n=21)
5 (n=10)
6 (n=16)
7 (n=2)
8 (n=1)
9 (n=1)
10 (n=1)
Top-1
Top-3
Top-10
Top-200

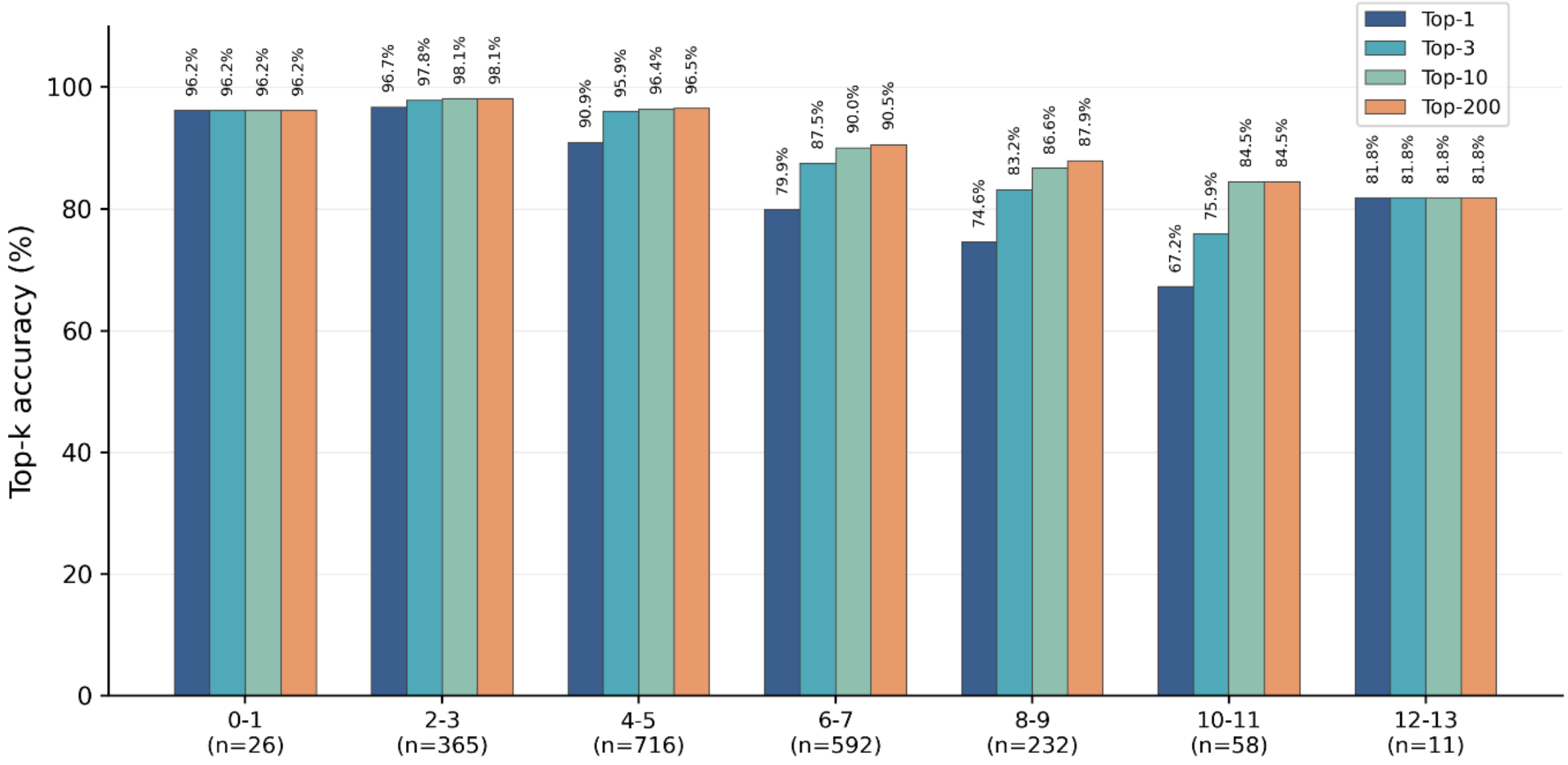
Stepwise Model C with Structure Correction:
Effect of Heteroatom (N, O and F) Number on Top-k Accuracy
Top-k accuracy (%)
Number of heteroatoms (n = total number of molecules)
0-1 (n=26)
2-3 (n=365)
4-5 (n=716)
6-7 (n=592)
8-9 (n=232)
10-11 (n=58)
12-13 (n=11)
Top-1
Top-3
Top-10
Top-200

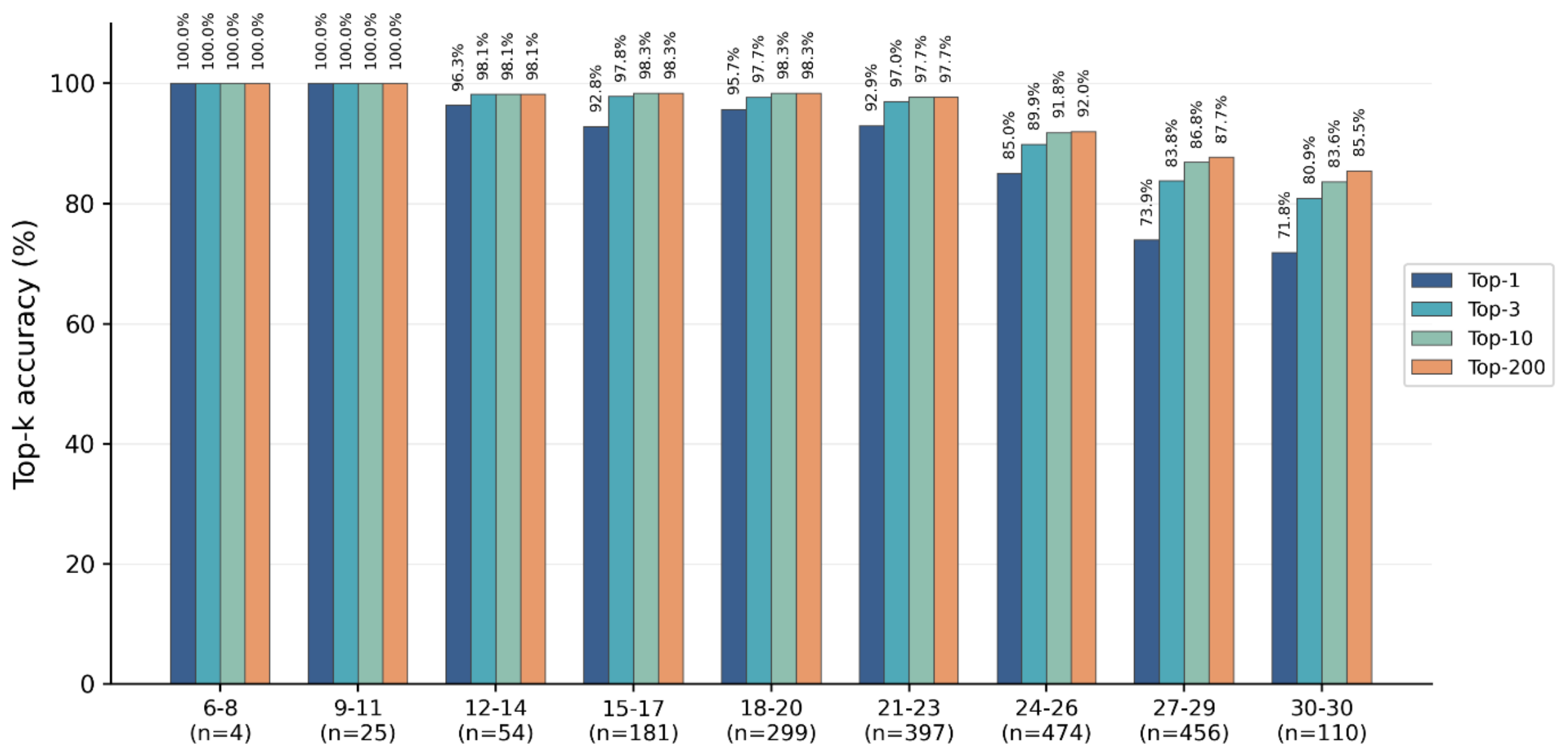
Stepwise Model C with Structure Correction:
Effect of Heavy Atom (C, N, O and F) Number on Top-k Accuracy
Top-k accuracy (%)
Number of heavy atoms (n = total number of molecules)
6-8 (n=4)
9-11 (n=25)
12-14 (n=54)
15-17 (n=181)
18-20 (n=299)
21-23 (n=397)
24-26 (n=474)
27-29 (n=456)
30-30 (n=110)
Top-1
Top-3
Top-10
Top-200

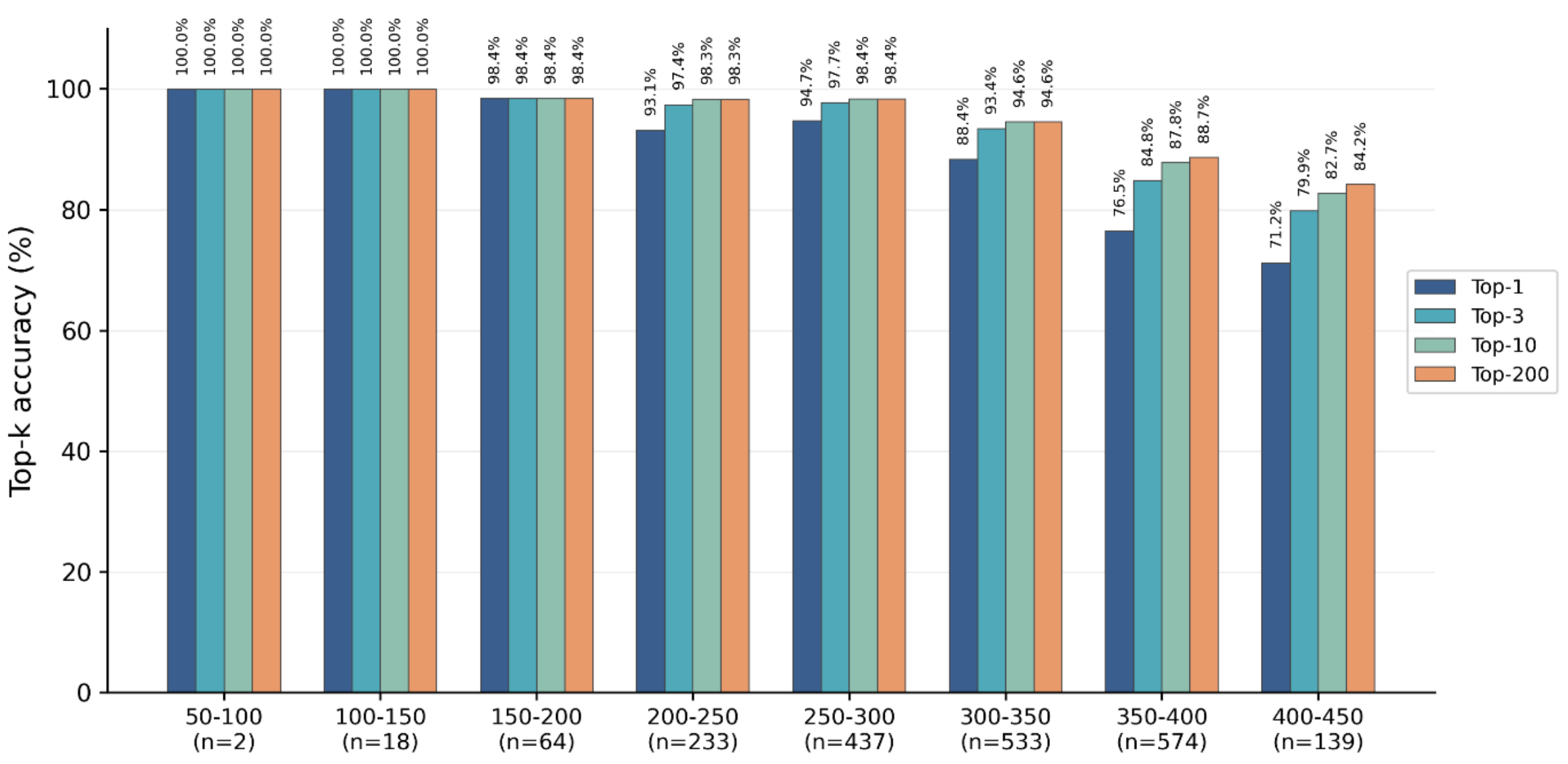
Stepwise Model C with Structure Correction:
Effect of Molecular Weight on Top-k Accuracy
Top-k accuracy (%)
Molecular weight (Da, n = total number of molecules)
50-100 (n=2)
100-150 (n=18)
150-200 (n=64)
200-250 (n=233)
250-300 (n=437)
300-350 (n=533)
350-400 (n=574)
400-450 (n=139)
Top-1
Top-3
Top-10
Top-200

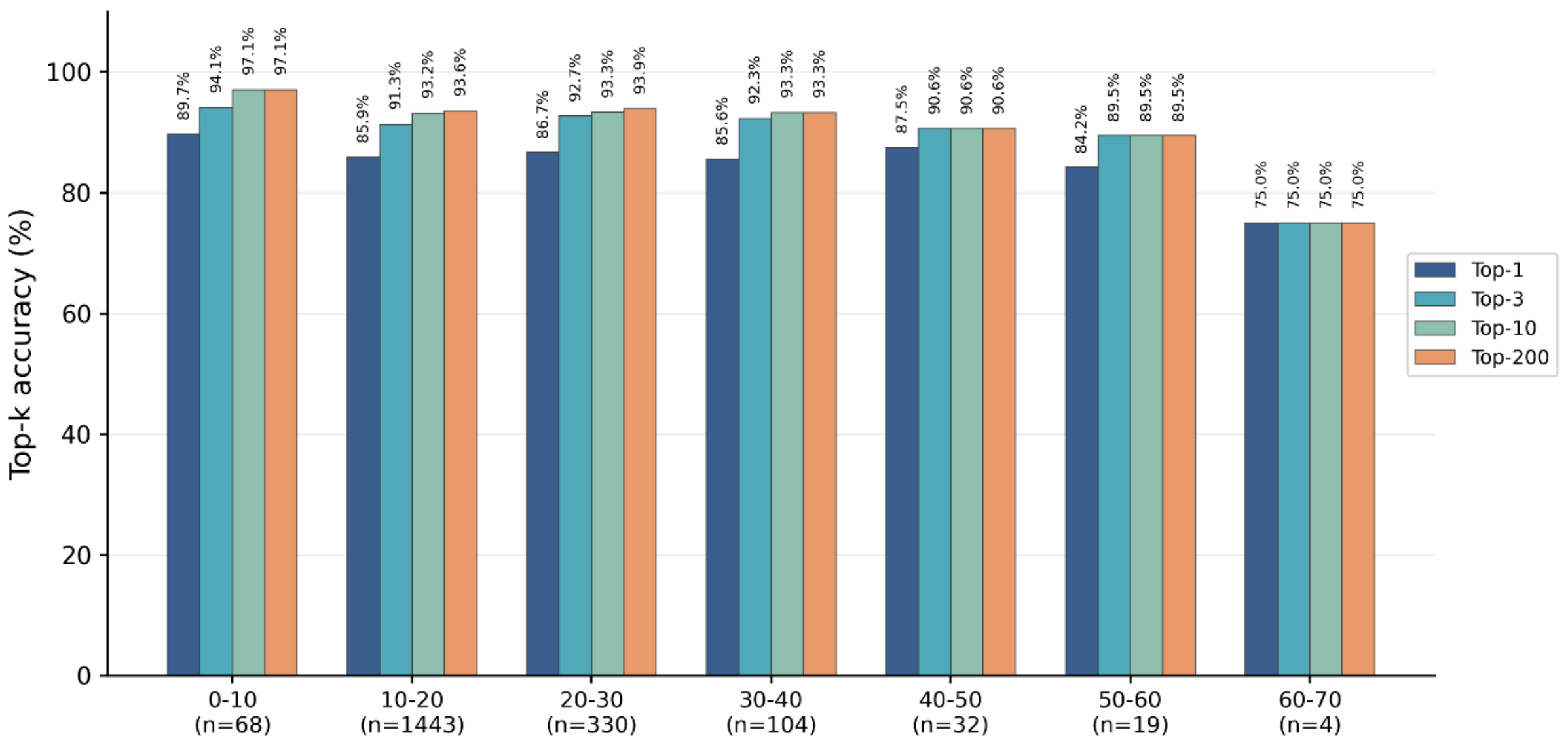
Stepwise Model C with Structure Correction:
Effect of Molecular Complexity (nSPS) on Top-k Accuracy
Top-k accuracy (%)
0
20
40
60
80
100
89.7%
94.1%
97.1%
97.1%
85.9%
91.3%
93.2%
93.6%
86.7%
92.7%
93.3%
93.9%
85.6%
92.3%
93.3%
93.3%
87.5%
90.6%
90.6%
90.6%
84.2%
89.5%
89.5%
89.5%
75.0%
75.0%
75.0%
75.0%
Top-1
Top-3
Top-10
Top-200
0-10 (n=68)
10-20 (n=1443)
20-30 (n=330)
30-40 (n=104)
40-50 (n=32)
50-60 (n=19)
60-70 (n=4)
nSPS (n = total number of molecules)

## 2.4 Application of Inverse-IMPRESSION on experimental molecules

The inverse-IMPRESSION platform was applied to predict the structures of 19 molecules using experimental $^{1}H$ and $^{13}C$ chemical shifts together with coupling constants obtained from COSY, HSQC, and HMBC spectra. Unless otherwise specified, 2000 sets of NMR noises were added to each molecule, with standard deviations of 0.3 ppm for $^{1}H$, 2.5 ppm for $^{13}C$, and 0.4 Hz for all coupling constants. The predicted candidate structures were ranked by frequency, defined as the proportion of times a structure was predicted among the 2000 candidates.

An additional ranking method based on the MAE of $^{13}C$ chemical shifts was also employed to improve performance. For each chemically valid candidate structure, 20 conformers were generated and optimised using the MMFF force field as implemented in RDKit. The $^{13}C$ chemical shifts of each conformer were then predicted using the IMPRESSION-G2 model. Boltzmann-weighted averaging of the conformers was performed to obtain the final predicted $^{13}C$ chemical shifts for each candidate structure. The mean absolute error (MAE) over all carbon atoms was subsequently calculated and used to rank the valid candidates.

The structures of the 19 molecules are shown below and classified into two categories: (1) correctly predicted structures ranked first by MAE; and (2) cases where no valid structures were predicted or where the top-ranked candidate is not the correct structure.

Supplementary Table 4 summarises the molecular properties associated with prediction accuracy: (1) Number of heteroatoms; (2) Number of quaternary-quaternary C-C bonds; (3) COSY ratio, defined as the proportion of 2JHH, 3JHH and 4JHH coupling constants among all H-H pairs separated by 2-4 bonds; (4) HSQC ratio, defined as the proportion of 1JCH coupling constants among all C-H bonds; and (5) HMBC ratio, defined as the proportion of 2JCH, 3JCH and 4JCH coupling constants among all C-H pairs separated by 2-4 bonds.

## 2.4.1 Correct Predictions (Top MAE Rank)

| **Structure** | | |
|---|---|---|
| **Name** | 1-(4-ethylphenyl)ethenone | aloperine |
| **Formula** | $C_{10}H_{12}O$ | $C_{15}H_{24}N_2$ |
| **Molecular weight** | 148.2 g/mol | 232.4 g/mol |
| **nSPS (SPS)** | 9.6 (106) | 45.8 (779) |
| **Number of NMR noise sets** | 2000 | 2000 |
| **Coupling constant noise** | 0.4 Hz | 0.4 Hz |
| **$^{13}C$ MAE** | 1.99 ppm | 2.94 ppm |
| **$^{13}C$ MAE rank** | 1 | 1 |
| **Frequency** | 99.85% | 80.5% |
| **Frequency rank** | 1 | 1 |

| Structure | | |
|---|---|---|
| **Name** | abiraterone | 4-piperidone ethylene acetal |
| **Formula** | $C_{24}H_{31}NO$ | $C_7H_{13}NO_2$ |
| **Molecular weight** | 349.5 g/mol | 143.2 g/mol |
| **nSPS (SPS)** | 44.4 (1154) | 31.2 (312) |
| **Number of NMR noise sets** | 2000 | 2000 |
| **Coupling constant noise** | 0.4 Hz | 0.4 Hz |
| **$^{13}$C MAE** | 2.99 ppm | 1.27 ppm |
| **$^{13}$C MAE rank** | 1 | 1 |
| **Frequency** | 1.5% | 0.85% |
| **Frequency rank** | 2 | 8 |

| Structure | | |
|---|---|---|
| **Name** | tadalafil | brucine |
| **Formula** | $C_{22}H_{19}N_3O_4$ | $C_{23}H_{26}N_2O_4$ |
| **Molecular weight** | 389.4 g/mol | 394.5 g/mol |
| **nSPS (SPS)** | 22.8 (661) | 41.0 (1189) |
| **Number of NMR noise sets** | 2000 | 5000 |
| **Coupling constant noise** | 0.4 Hz | 0.4 Hz |
| **$^{13}$C MAE** | 2.08 ppm | 4.31 ppm |
| **$^{13}$C MAE rank** | 1 | 1 |
| **Frequency** | 0.25 % | 0.05% |
| **Frequency rank** | 1 | 1 |

| Structure | | |
|---|---|---|
| **Name** | betulin | 2-ethoxy-5-methylpyridin-4-amine |
| **Formula** | $C_{30}H_{50}O_2$ | $C_8H_{12}N_2O$ |
| **Molecular weight** | 442.7 g/mol | 152.2 g/mol |
| **nSPS (SPS)** | 56.5 (1808) | 9.6 (106) |
| **Number of NMR noise sets** | 2000 | 2000 |
| **Coupling constant noise** | 0.4 Hz | 0.6 Hz |
| **$^{13}$C MAE** | 3.83 ppm | 1.71 ppm |
| **$^{13}$C MAE rank** | 1 | 1 |
| **Frequency** | 0.1% | 0.7% |
| **Frequency rank** | 1 | 7 |

| **Structure** | | |
|---|---|---|
| **Name** | 24-epibrassinolide | 1-Boc-4-(4-aminophenyl)piperazine |
| **Formula** | $C_{28}H_{48}O_6$ | $C_{15}H_{23}N_3O_2$ |
| **Molecular weight** | 480.7 g/mol | 277.4 g/mol |
| **nSPS (SPS)** | 48.1 (1635) | 16.2 (324) |
| **Number of NMR noise sets** | 2000 | 2000 |
| **Coupling constant noise** | 0.4 Hz | 0.6 Hz |
| **$^{13}$C MAE** | 3.66 ppm | 2.00 ppm |
| **$^{13}$C MAE rank** | 1 | 1 |
| **Frequency** | 0.35% | 0.05% |
| **Frequency rank** | 1 | 10 |

## 2.4.2 Incorrect Predictions

| Structure | | |
|---|---|---|
| **Name** | (R)-(+)-N-(1-phenylethyl)succinamic acid | Wrongly predicted candidate ranked #1 by $^{13}C$ MAE |
| **Formula** | $C_{12}H_{15}NO_3$ | $C_{12}H_{15}NO_3$ |
| **Molecular weight** | 221.3 g/mol | 221.3 g/mol |
| **nSPS (SPS)** | 11.8 (189) | - |
| **Number of NMR noise sets** | - | 2000 |
| **Coupling constant noise** | - | 0.4 Hz |
| **$^{13}C$ MAE** | - | 23.98 ppm |
| **$^{13}C$ MAE rank** | - | 1 |
| **Frequency** | - | 0.05% |
| **Frequency rank** | - | 1 |

| Structure | | - |
|---|---|---|
| **Name** | baccatin-III | - |
| **Formula** | $C_{31}H_{38}O_{11}$ | - |
| **Molecular weight** | 586.6 g/mol | - |
| **nSPS** | 40.4 (1697) | - |
| **Number of NMR noise sets** | - | 2000 |
| **Coupling constant noise** | - | 0.4 Hz |
| **$^{13}$C MAE** | - | - |
| **$^{13}$C MAE rank** | - | - |
| **Frequency** | - | - |
| **Frequency rank** | - | - |

| **Structure** | | |
|---|---|---|
| **Name** | enoxolone | Wrongly predicted candidate ranked #1 by $^{13}$C MAE |
| **Formula** | $C_{30}H_{46}O_4$ | $C_{30}H_{46}O_4$ |
| **Molecular weight** | 470.7 g/mol | 470.7 g/mol |
| **nSPS (SPS)** | 54.1 (1839) | - |
| **Number of NMR noise sets** | - | 2000 |
| **Coupling constant noise** | - | 0.4 Hz |
| **$^{13}$C MAE** | **-** | 5.75 ppm |
| **$^{13}$C MAE rank** | **-** | 1 |
| **Frequency** | **-** | 0.05% |
| **Frequency rank** | **-** | 1 |

| **Structure** | | |
|---|---|---|
| **Name** | evodiamine | Wrongly predicted candidate ranked #1 by $^{13}$C MAE |
| **Formula** | $C_{19}H_{17}N_3O$ | $C_{19}H_{17}N_3O$ |
| **Molecular weight** | 303.4 g/mol | 303.4 g/mol |
| **nSPS (SPS)** | 19.5 (449) | - |
| **Number of NMR noise sets** | - | 2000 |
| **Coupling constant noise** | - | 0.4 Hz |
| **$^{13}$C MAE** | - | 8.63 ppm |
| **$^{13}$C MAE rank** | - | 1 |
| **Frequency** | - | 0.05% |
| **Frequency rank** | - | 1 |

| **Structure** | | |
|---|---|---|
| **Name** | ginkgolide A | Wrongly predicted candidate ranked #1 by $^{13}C$ MAE |
| **Formula** | $C_{20}H_{24}O_9$ | $C_{20}H_{24}O_9$ |
| **Molecular weight** | 408.4 g/mol | 408.4 g/mol |
| **nSPS (SPS)** | 59.2 (1717) | - |
| **Number of NMR noise sets** | - | 2000 |
| **Coupling constant noise** | - | 0.4 Hz |
| **$^{13}C$ MAE** | **-** | 12.58 ppm |
| **$^{13}C$ MAE rank** | **-** | 1 |
| **Frequency** | **-** | 0.05% |
| **Frequency rank** | **-** | 1 |

| **Structure** | | - |
|---|---|---|
| **Name** | reserpine | - |
| **Formula** | $C_{33}H_{40}N_2O_9$ | - |
| **Molecular weight** | 608.7 g/mol | - |
| **nSPS (SPS)** | 26.1 (1148) | - |
| **Number of NMR noise sets** | - | - |
| **Coupling constant noise** | - | - |
| **$^{13}$C MAE** | - | - |
| **$^{13}$C MAE rank** | - | - |
| **Frequency** | - | - |
| **Frequency rank** | - | - |

| **Structure** | | |
|---|---|---|
| **Name** | tetrahydropalmatine | Wrongly predicted candidate ranked #1 by $^{13}C$ MAE |
| **Formula** | $C_{21}H_{25}NO_4$ | $C_{21}H_{25}NO_4$ |
| **Molecular weight** | 355.5 g/mol | 355.5 g/mol |
| **nSPS (SPS)** | 18.4 (478) | - |
| **Number of NMR noise sets** | - | 2000 |
| **Coupling constant noise** | - | 0.4 HZ |
| **$^{13}C$ MAE** | **-** | 16.15 ppm |
| **$^{13}C$ MAE rank** | **-** | 1 |
| **Frequency** | **-** | 0.05% |
| **Frequency rank** | **-** | 1 |

| **Structure** | | |
|---|---|---|
| **Name** | vindoline | Wrongly predicted candidate ranked #1 by $^{13}C$ MAE |
| **Formula** | $C_{25}H_{32}N_2O_6$ | $C_{25}H_{32}N_2O_6$ |
| **Molecular weight** | 456.5 g/mol | 456.5 g/mol |
| **nSPS (SPS)** | 38.4 (1267) | - |
| **Number of NMR noise sets** | - | 2000 |
| **Coupling constant noise** | - | 0.4 Hz |
| **$^{13}C$ MAE** | **-** | 11.51 ppm |
| **$^{13}C$ MAE rank** | **-** | 1 |
| **Frequency** | **-** | 0.05% |
| **Frequency rank** | **-** | 1 |

| **Structure** | | |
|---|---|---|
| **Name** | vinpocetine | Wrongly predicted candidate ranked #1 by $^{13}C$ MAE |
| **Formula** | $C_{22}H_{26}N_2O_2$ | $C_{22}H_{26}N_2O_2$ |
| **Molecular weight** | 350.5 g/mol | 350.5 g/mol |
| **nSPS (SPS)** | 27.2 (707) | - |
| **Number of NMR noise sets** | - | 2000 |
| **Coupling constant noise** | - | 0.4 Hz |
| **$^{13}C$ MAE** | **-** | 5.38 ppm |
| **$^{13}C$ MAE rank** | **-** | 1 |
| **Frequency** | **-** | 0.05% |
| **Frequency rank** | **-** | 1 |

## 2.4.3 Properties of Experimental molecules

**Supplementary Table 4** Molecular properties associated with prediction accuracy for 19 experimental molecules. Molecules highlighted in blue are correctly predicted. Values highlighted in red indicate deviations from the defined thresholds for successful prediction: (1) Heteroatom count ≤ 7; (2) Quaternary-quaternary C-C bond count ≤ 4; and (3) COSY/HMBC ratio ≥ 15%.

| Molecule name | $M_R$ (g/mol) | Heteroatom count | Quaternary-quaternary C-C bond count | COSY ratio (%) | HSQC ratio (%) | HMBC ratio (%) |
|---|---|---|---|---|---|---|
| 1-(4-ethylphenyl)ethenone | 148.2 | 1 | 1 | 38.1 | 100.0 | 29.4 |
| aloperine | 232.4 | 2 | 0 | 44.0 | 100.0 | 39.2 |
| abiraterone | 349.5 | 2 | 3 | 35.6 | 100.0 | 29.5 |
| 4-piperidone ethylene acetal | 143.2 | 3 | 0 | 23.5 | 100.0 | 52.5 |
| tadalafil | 389.4 | 7 | 4 | 50.0 | 100.0 | 35.4 |
| brucine | 394.5 | 6 | 3 | 30.9 | 100.0 | 34.9 |
| betulin | 442.7 | 2 | 1 | 22.1 | 100.0 | 19.0 |
| 2-ethoxy-5-methylpyridin-4-amine | 152.2 | 3 | 1 | 31.6 | 100.0 | 40.4 |
| 24-epibrassinolide | 480.7 | 6 | 0 | 25.6 | 100.0 | 20.7 |
| 1-Boc-4-(4-aminophenyl)piperazine | 277.4 | 5 | 0 | 15.4 | 100.0 | 40.7 |
| **(R)-(+)-N-(1-phenylethyl)succinamic acid** | 221.3 | 4 | 0 | **10.0** | 100.0 | **13.9** |
| **baccatin-III** | 586.6 | **11** | **5** | **12.5** | 100.0 | 32.7 |
| **enoxolone** | 470.7 | 4 | 3 | **13.5** | 100.0 | 39.1 |
| **evodiamine** | 303.4 | 4 | **5** | 61.5 | 100.0 | 30.2 |
| **ginkgolide A** | 408.4 | **9** | 4 | **13.0** | 100.0 | 20.4 |
| **reserpine** | 608.7 | **11** | **6** | 21.1 | 100.0 | 20.9 |
| **tetrahydropalmatine** | 355.5 | 5 | **5** | 27.8 | 100.0 | 21.3 |
| **vindoline** | 456.5 | **8** | 3 | 25.0 | 100.0 | 22.7 |
| **vinpocetine** | 350.5 | 4 | 4 | 49.3 | 100.0 | 31.4 |

# 3 Figures for dataset distributions

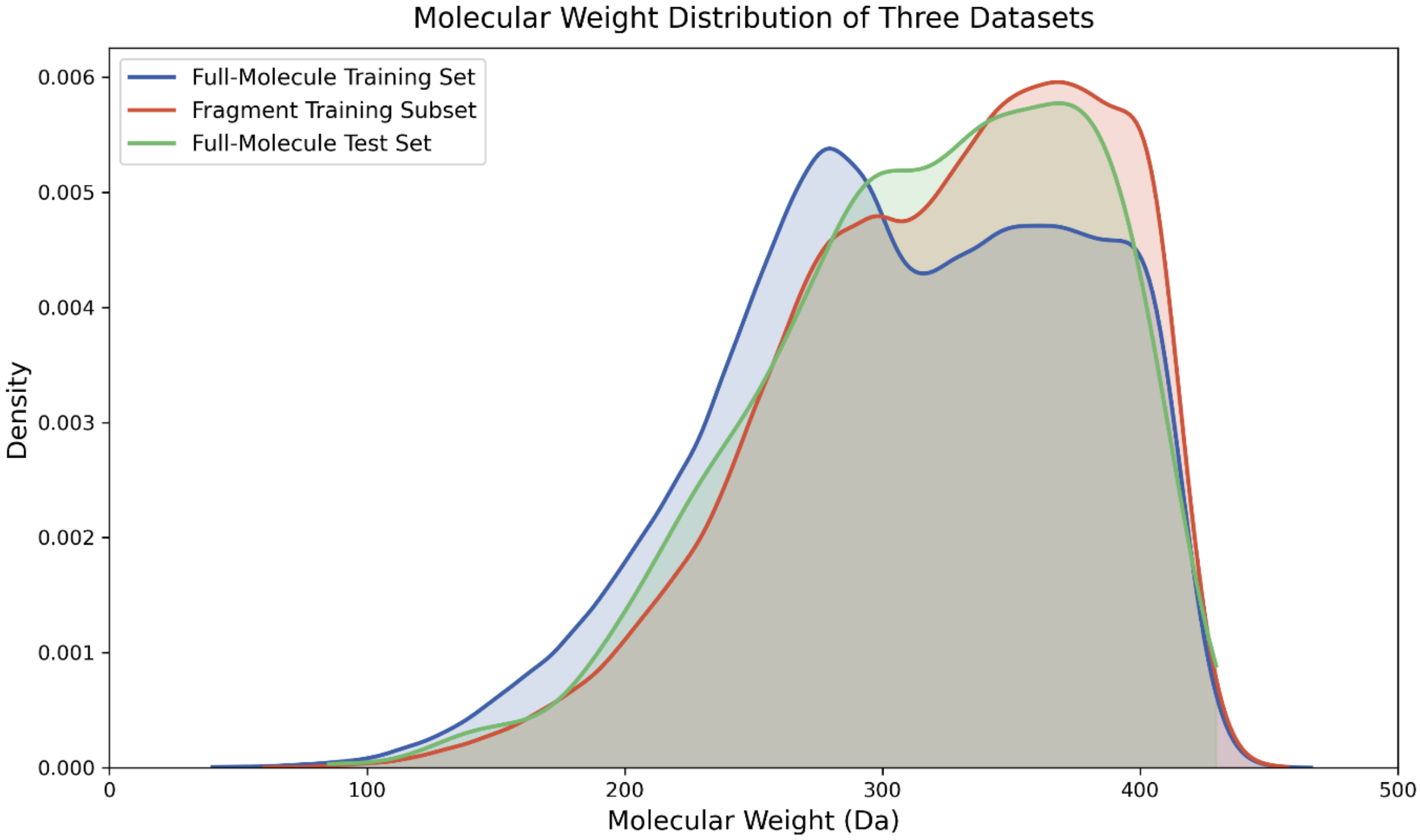

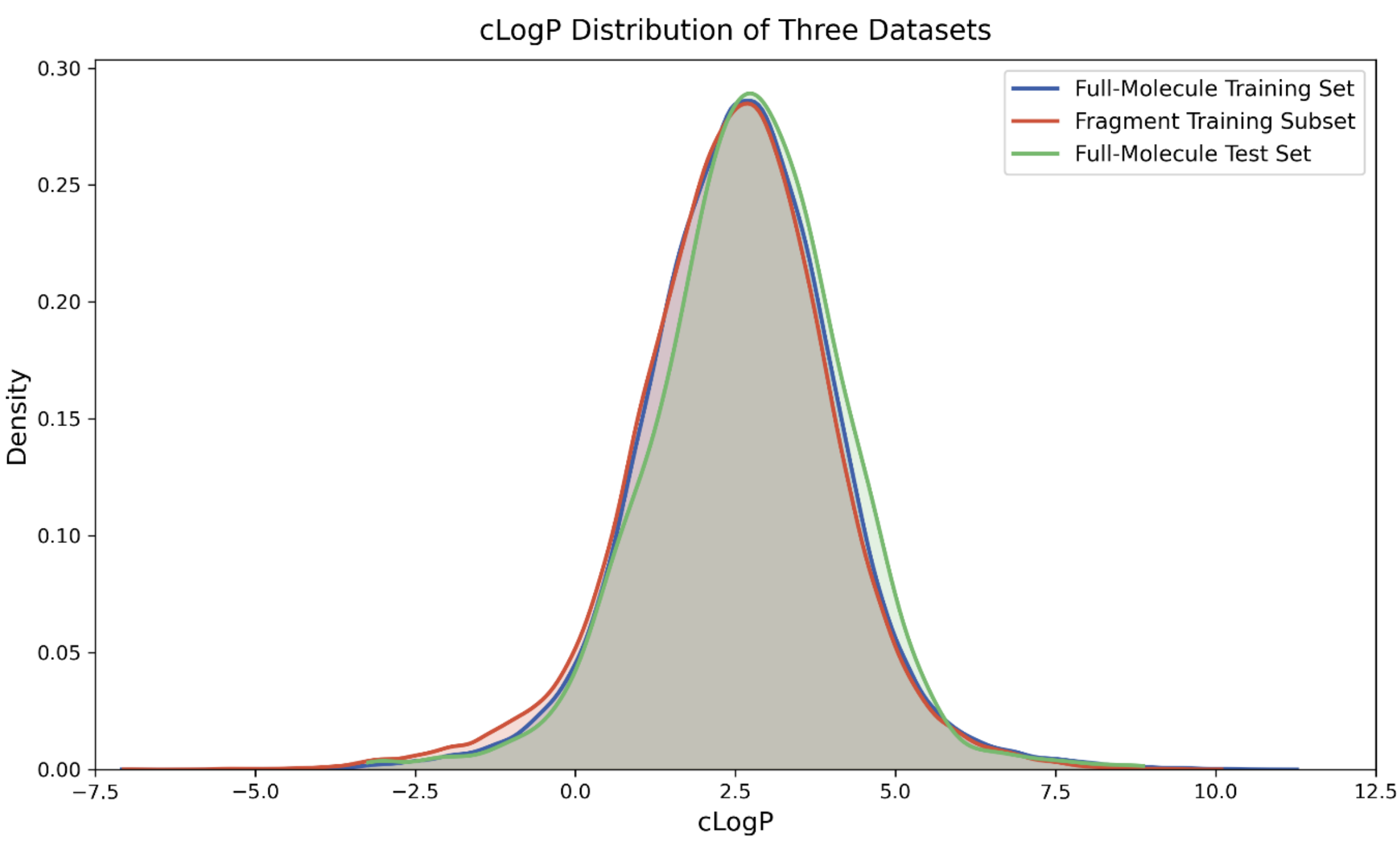

nSPS Distribution of Three Datasets

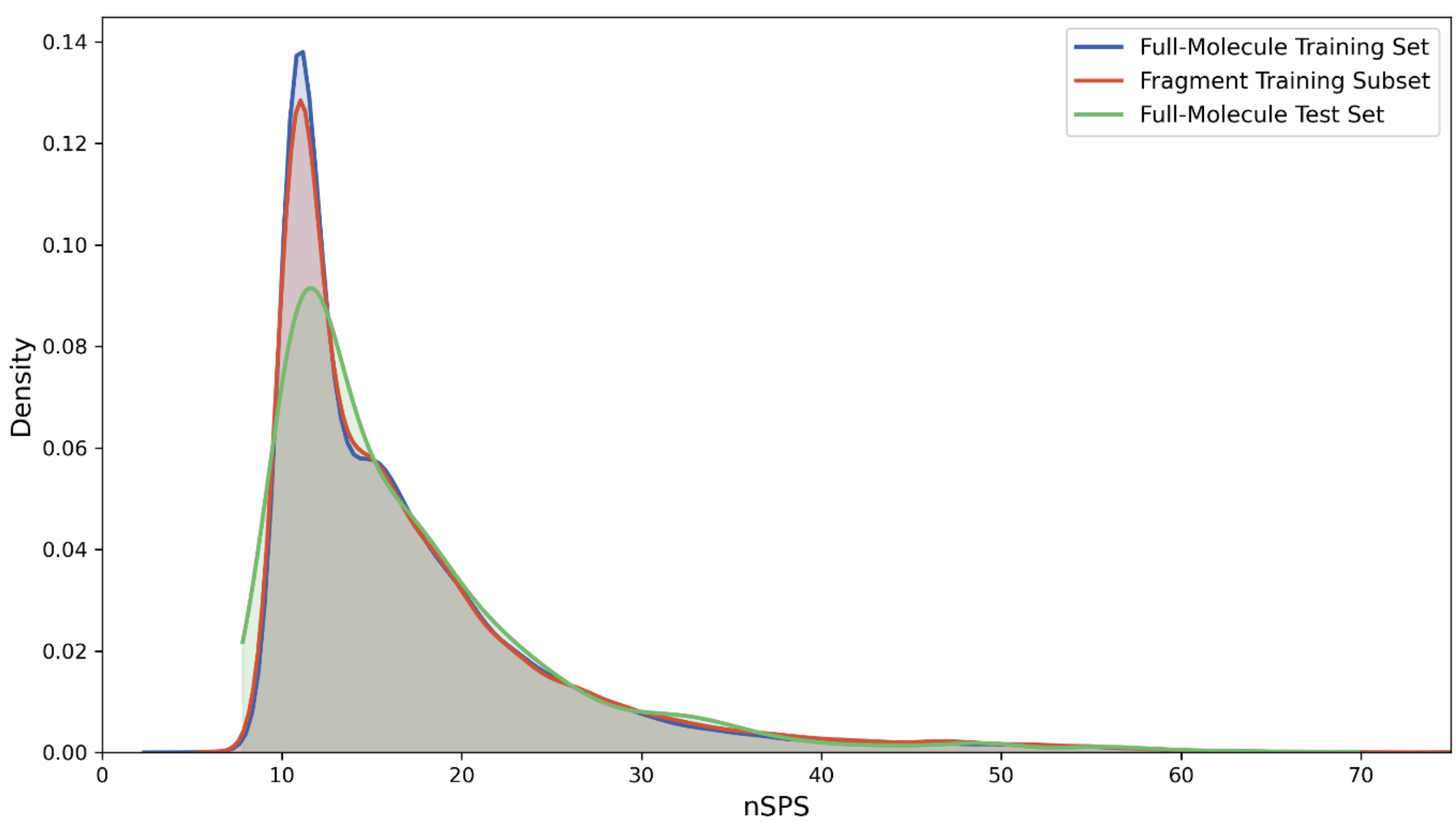

Heteroatom (N, O and F) Count Distribution of Three Datasets

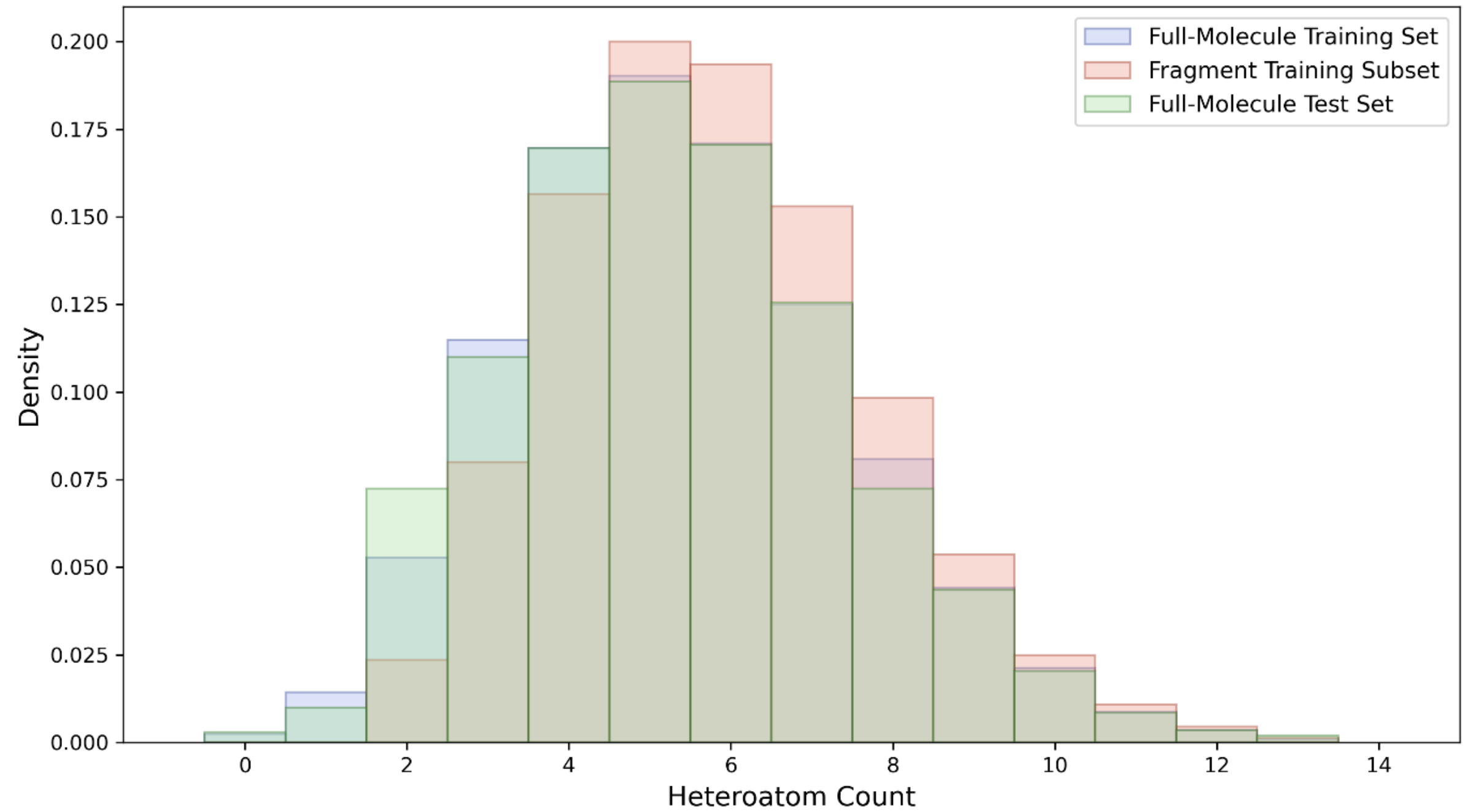

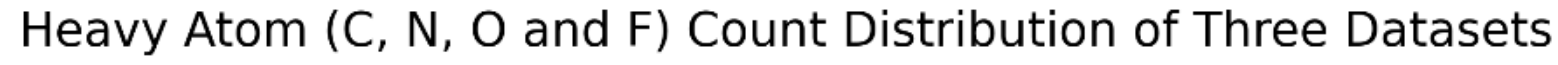
Heavy Atom (C, N, O and F) Count Distribution of Three Datasets

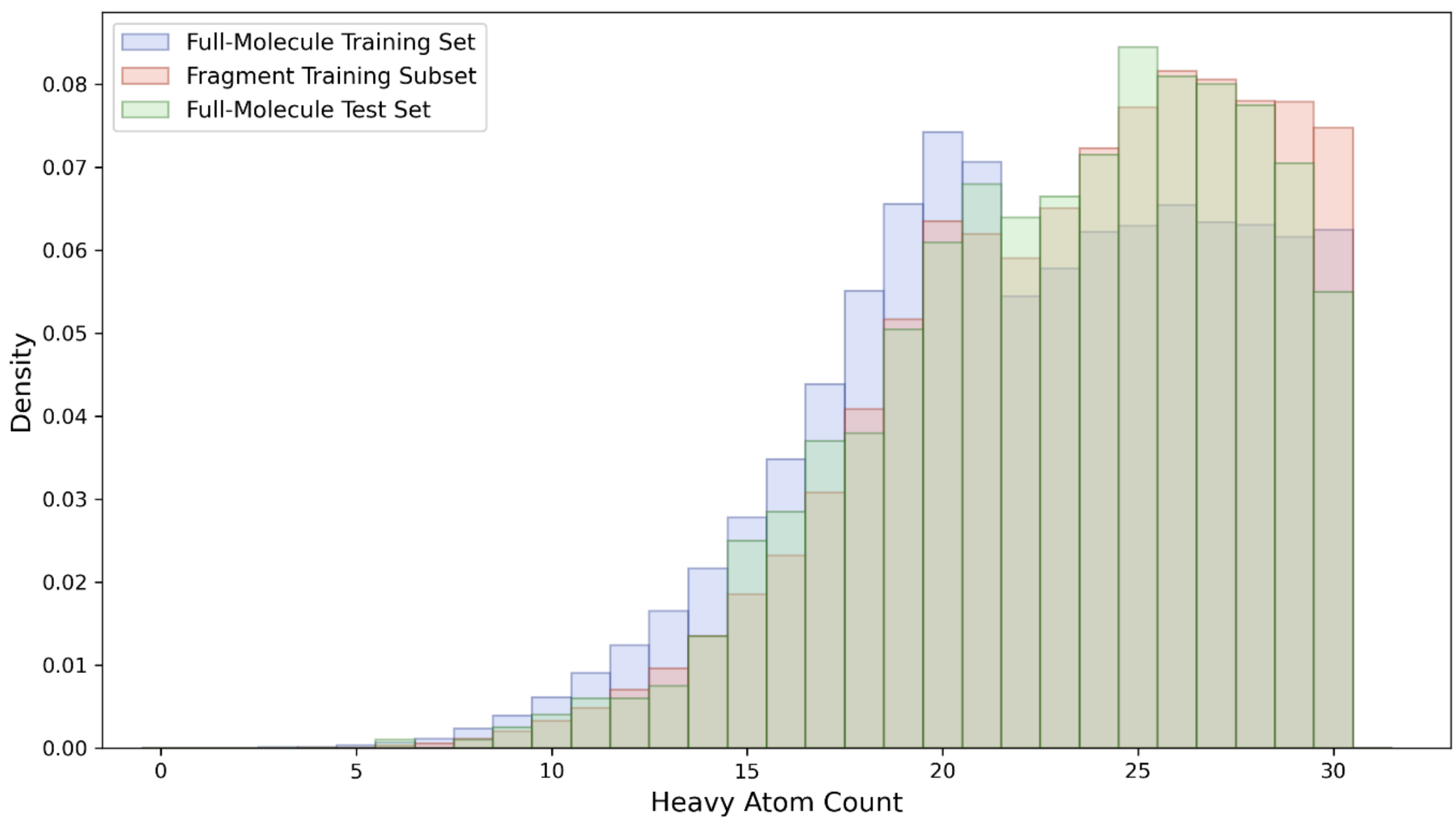
Full-Molecule Training Set
Fragment Training Subset
Full-Molecule Test Set
Density
Heavy Atom Count

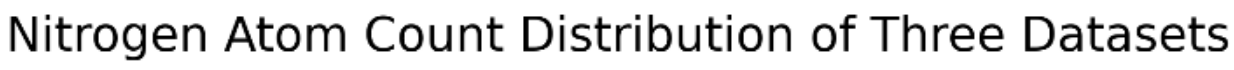
Nitrogen Atom Count Distribution of Three Datasets

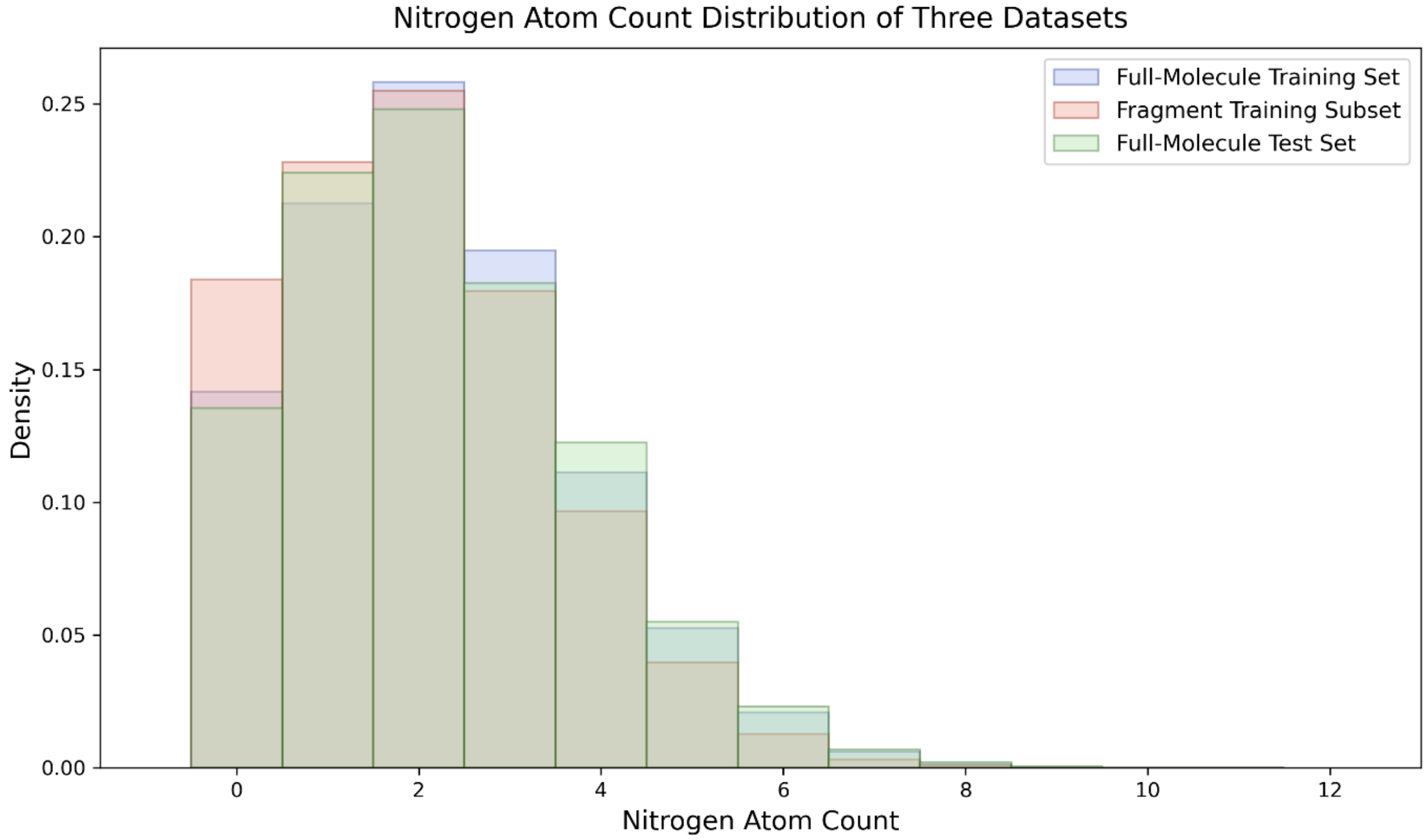
Full-Molecule Training Set
Fragment Training Subset
Full-Molecule Test Set
Density
Nitrogen Atom Count

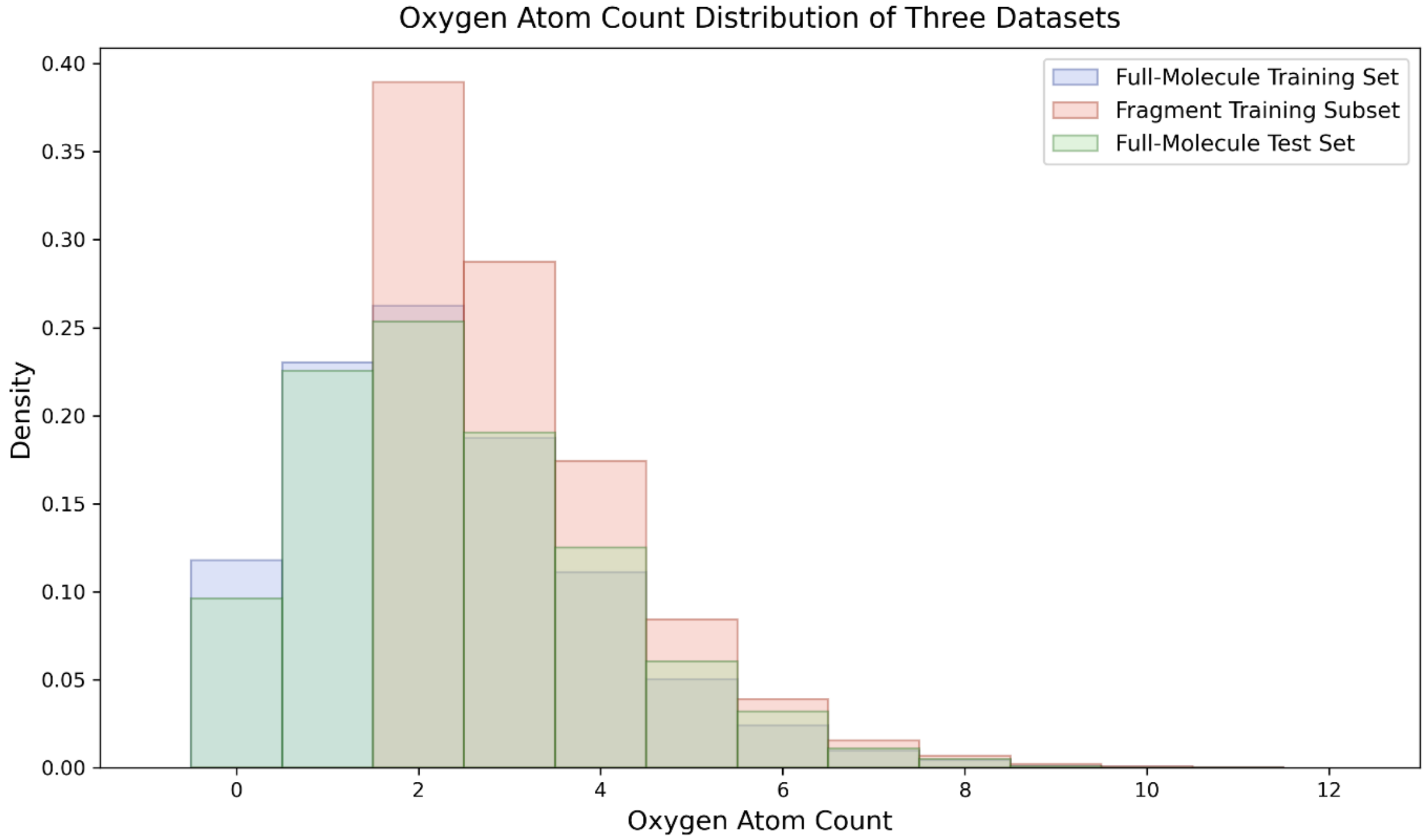
Oxygen Atom Count Distribution of Three Datasets
Full-Molecule Training Set
Fragment Training Subset
Full-Molecule Test Set
Density
Oxygen Atom Count
0.40
0.35
0.30
0.25
0.20
0.15
0.10
0.05
0.00
0
2
4
6
8
10
12

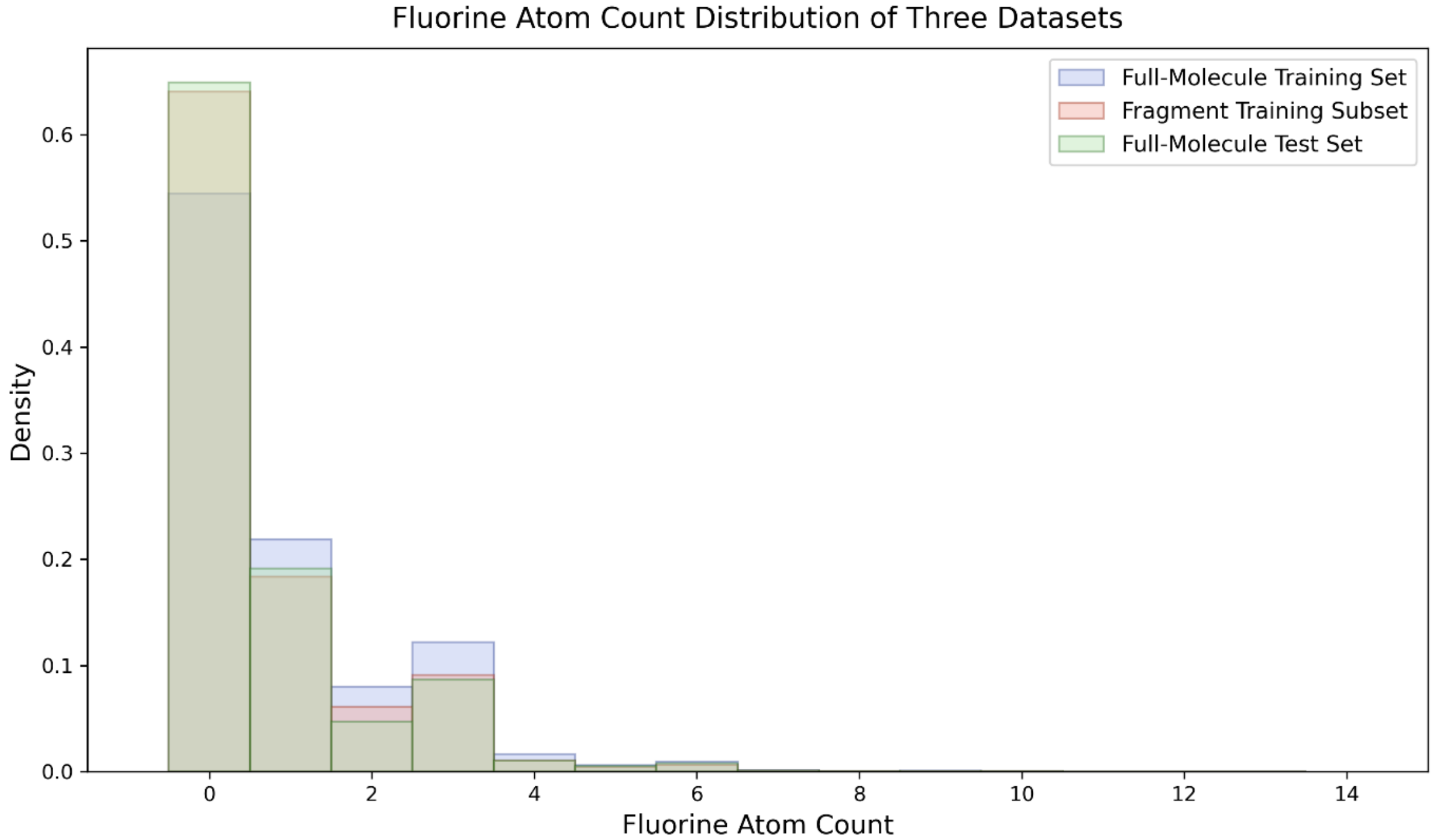
Fluorine Atom Count Distribution of Three Datasets
Full-Molecule Training Set
Fragment Training Subset
Full-Molecule Test Set
Density
Fluorine Atom Count
0.6
0.5
0.4
0.3
0.2
0.1
0.0
0
2
4
6
8
10
12
14

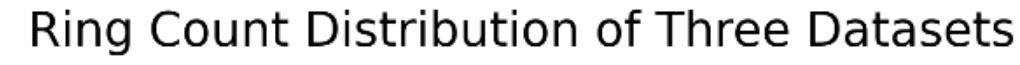
Ring Count Distribution of Three Datasets

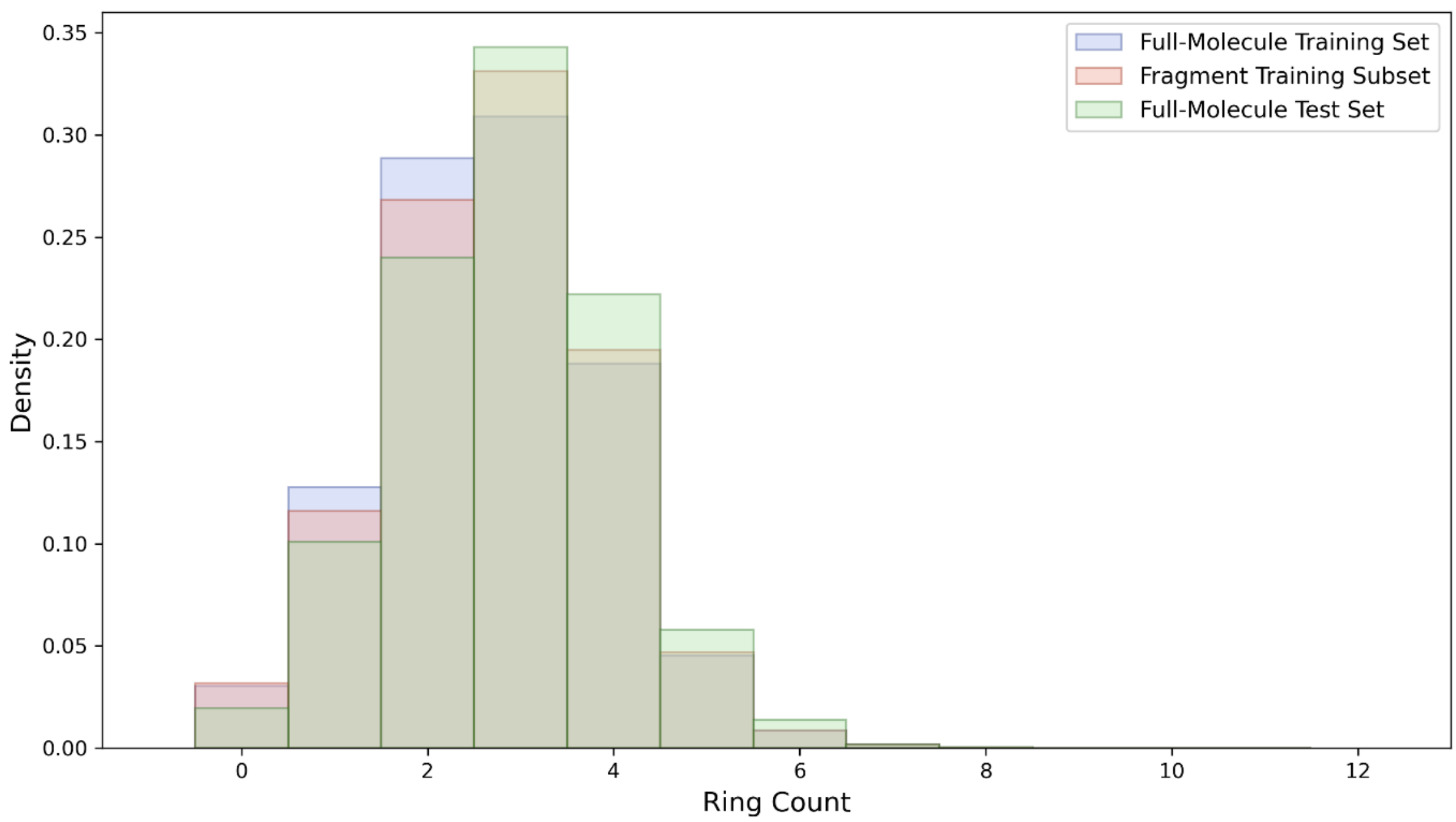
Full-Molecule Training Set
Fragment Training Subset
Full-Molecule Test Set
Density
0.00
0.05
0.10
0.15
0.20
0.25
0.30
0.35
0
2
4
6
8
10
12
Ring Count

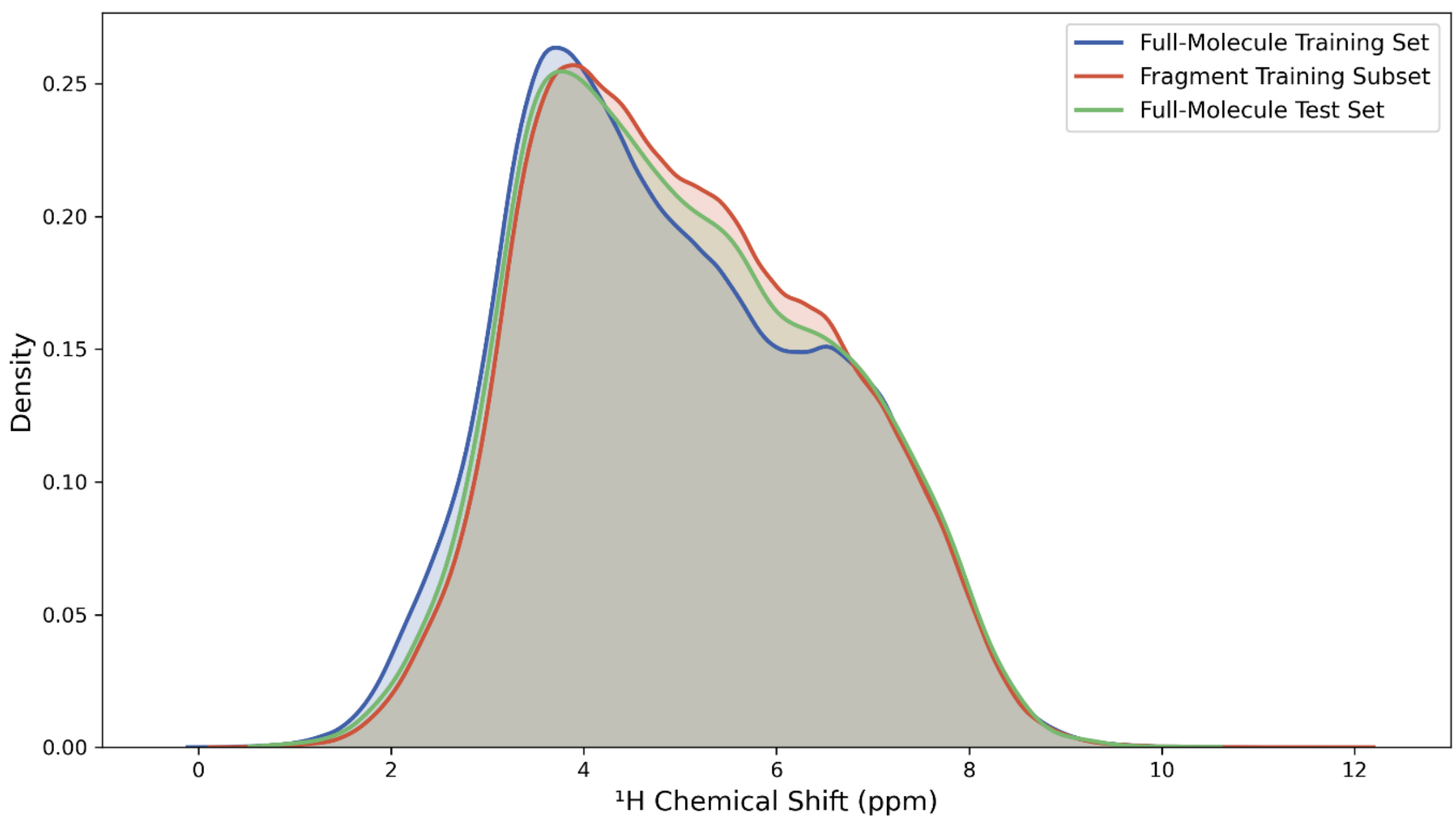
¹H Chemical Shift Distribution of Three Datasets
Full-Molecule Training Set
Fragment Training Subset
Full-Molecule Test Set
Density
0.00
0.05
0.10
0.15
0.20
0.25
0
2
4
6
8
10
12
¹H Chemical Shift (ppm)

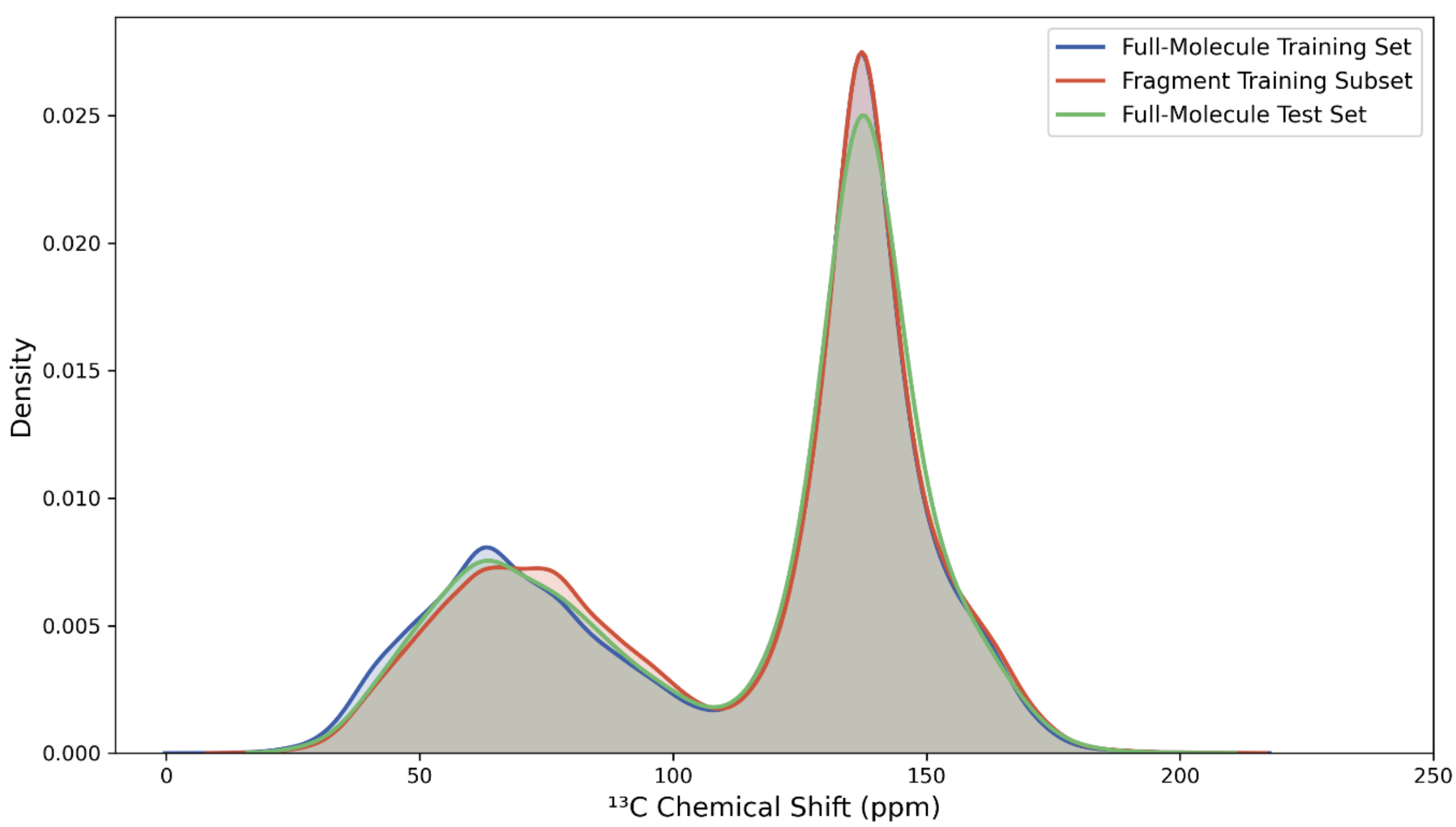

$^{13}$C Chemical Shift Distribution of Three Datasets
Full-Molecule Training Set
Fragment Training Subset
Full-Molecule Test Set
Density
0.000
0.005
0.010
0.015
0.020
0.025
0
50
100
150
200
250
$^{13}$C Chemical Shift (ppm)

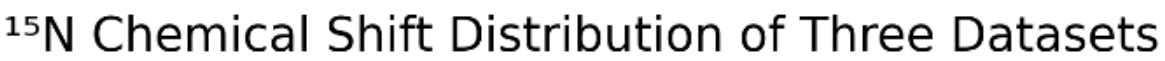

$^{15}$N Chemical Shift Distribution of Three Datasets

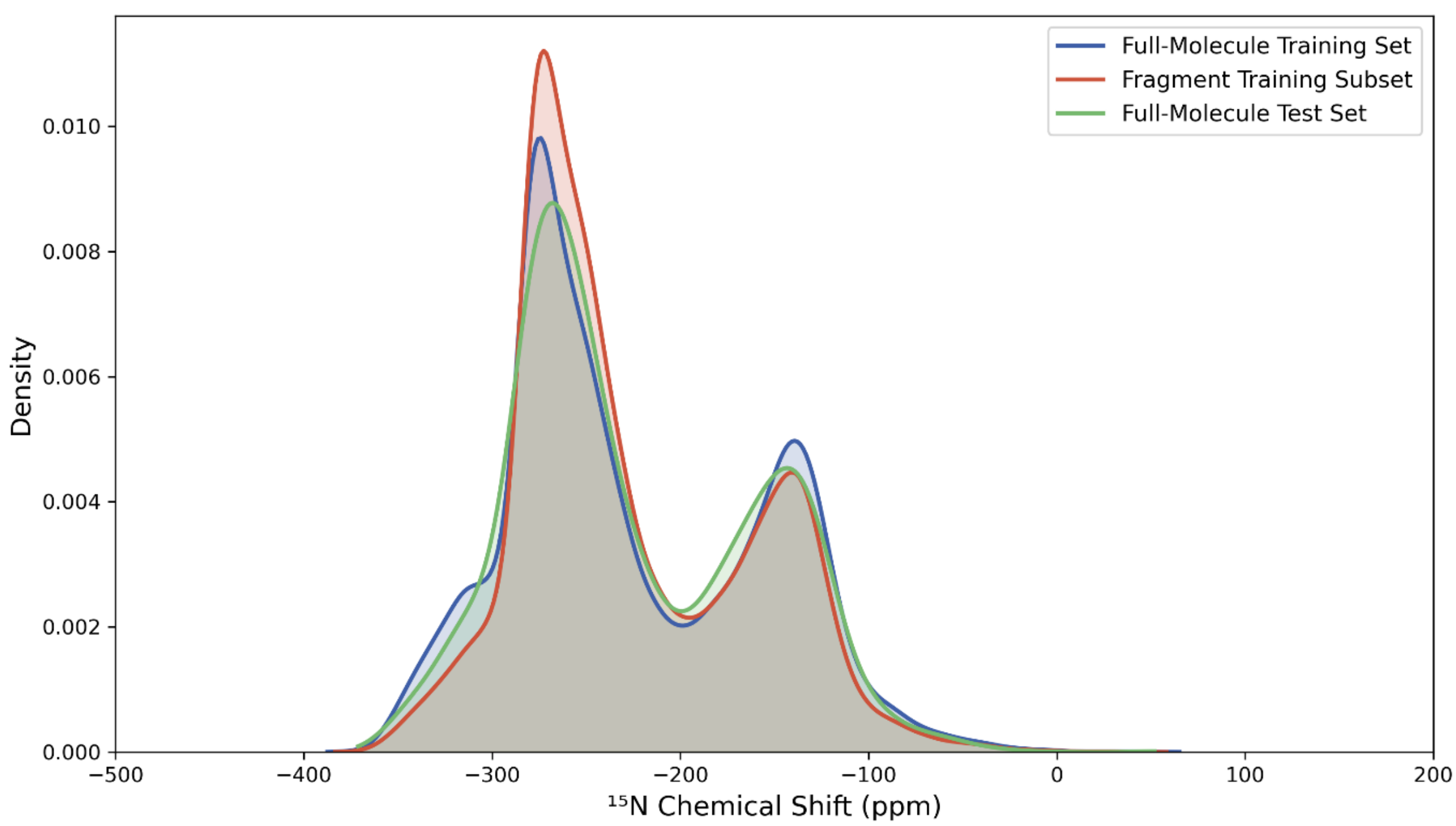

Full-Molecule Training Set
Fragment Training Subset
Full-Molecule Test Set
Density
0.000
0.002
0.004
0.006
0.008
0.010
−500
−400
−300
−200
−100
0
100
200
$^{15}$N Chemical Shift (ppm)

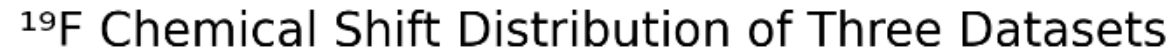

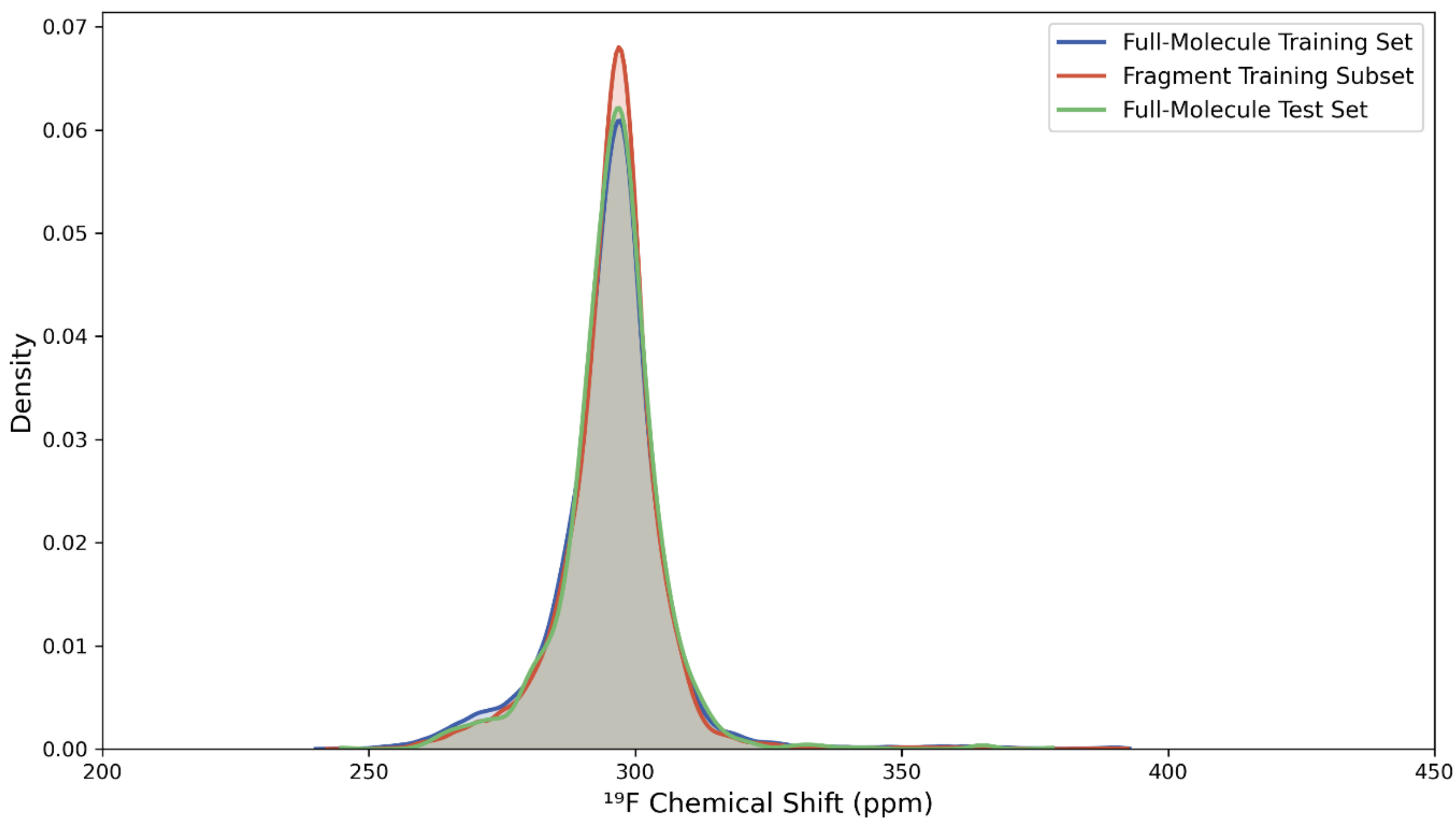

¹⁹F Chemical Shift Distribution of Three Datasets
Density
0.07
0.06
0.05
0.04
0.03
0.02
0.01
0.00
200
250
300
350
400
450
¹⁹F Chemical Shift (ppm)
Full-Molecule Training Set
Fragment Training Subset
Full-Molecule Test Set